\documentclass[11pt,letterpaper]{article}

\usepackage{geometry}
\usepackage{bbm}
\usepackage[toc,page]{appendix}
\usepackage{amsmath, amsthm, amsfonts, amssymb}
\usepackage{mathrsfs} 
\usepackage{hyperref}
\usepackage{cleveref}
\usepackage{slashed}
\usepackage{dsfont}
\usepackage{wrapfig}
\usepackage{minibox}
\usepackage{graphicx}
\usepackage{graphics}
\usepackage{epsfig}
\usepackage{xcolor}
\usepackage{array}
\usepackage{tikz}
\usepackage{tikz-cd}
\usetikzlibrary{arrows,calc}
\usepackage{relsize}
\usetikzlibrary{decorations.pathmorphing,calc}
\usepackage{authblk}
\usepackage[labelfont=bf]{caption}
\usepackage{caption}
\usepackage{subcaption}
\usepackage{placeins}

\def\a{\alpha}
\def\b{\beta}
\def\g{\gamma}

\def\d{\delta}
\def\D{\Delta}
\def\ve{\varepsilon}
\def\m{\mu}
\def\n{\nu}
\def\l{\lambda}
\def\L{\Lambda}
\def\r{\rho}
\def\s{\sigma}
\def\thintablerule{\hrule height0.4pt}

\def\bu{\bullet}

\def\pa{\partial}

\newcommand{\be}{\begin{equation}}
\newcommand{\ee}{\end{equation}}
\newcommand{\bea}{\begin{eqnarray}}
\newcommand{\eea}{\end{eqnarray}}
\newcommand{\bal}{\begin{aligned}}
\newcommand{\eal}{\end{aligned}}
\newcommand{\eq}[1]{Eq.~(\ref{#1})}
\newcommand{\fig}[1]{Fig.~\ref{#1}}

\newcommand{\sect}[1]{Section~\ref{#1}}
\newcommand{\tr}{\mathrm{tr}\,}

\numberwithin{equation}{section}

\allowdisplaybreaks

\begin{document}

\tikzset{
    photon/.style={decorate, decoration={snake}, draw=black},
    electron/.style={draw=black, postaction={decorate},
        decoration={markings,mark=at position .55 with {\arrow[draw=black]{>}}}},
    gluon/.style={decorate, draw=black,
        decoration={coil,amplitude=4pt, segment length=5pt}} 
}

\centerline{\LARGE RG flows in Non-Perturbative Gauge-Higgs Unification II. }
\vskip .2cm
\centerline{\LARGE Effective action for the Higgs phase  near the quantum phase transition}
\vskip .5cm

\vskip 2 cm
\centerline{\large Nikos Irges and Fotis Koutroulis}
\vskip 1cm
\centerline{\it Department of Physics}
\centerline{\it National Technical University of Athens}
\centerline{\it Zografou Campus, GR-15780 Athens, Greece}
\centerline{\it e-mail: irges@mail.ntua.gr, fkoutroulis@central.ntua.gr}

\vskip 2.2 true cm
\thintablerule
\vskip 2.0ex

\centerline{\bf Abstract}
We construct the zero temperature (no compact dimensions) effective action for an $SU(2)$ Yang-Mills theory in five dimensions, with boundary conditions that reduce the
symmetry on the four-dimensional boundary located at the origin to a $U(1)$-complex scalar system. In order to be sensitive to the Higgs phase,
we need to include higher dimensional operators in the effective action, which can be naturally achieved by generating it by expanding the corresponding lattice 
construction in small lattice spacing, taking the naive continuum limit and then renormalizing.
In addition, we build in the effective action non-perturbative information, related to a first order quantum phase transition known to exist.
As a result, the effective action acquires a finite cut-off that is low and the fine tuning of the scalar mass is rather mild.

\vskip 1.0ex\noindent
\vskip 2.0ex
\thintablerule

\newpage

\tableofcontents

\pagebreak


\section{Introduction}\label{Intro}

The target of the present work, a sequel to \cite{IrgesFotis2}, is the construction of a four-dimensional (4d) continuum
effective action for a five-dimensional (5d) model, originally constructed on the lattice \cite{IrgesK1, IrgesK2, IrgesK3} and dubbed as 
a model of Non-Perturbative Gauge-Higgs Unification (NPGHU).
The model in its simplest version has a pure $SU(2)$ gauge symmetry in 5d with orbifold boundary conditions that generate a 4d boundary on which a $U(1)$ gauge field coupled to a complex scalar survive.
The main novel property of this model is that, at the non-perturbative level, it exhibits spontaneous breaking of its gauge symmetry in infinite fifth dimension \cite{IrgesK2,IrgesK4,IrgesK5,IrgesK6,Roman},
which distinguishes it from extra-dimensional models where the scalar potential is of a finite temperature type, inversely proportional to the size of the extra dimension.
Its pure bosonic nature sets it apart also from similar mechanisms where the presence of fermions is necessary in order that a Higgs mechanism is triggered.
The absence of any polynomial terms in the (quantum) effective scalar potential distinguishes it from 
models with a classical potential but also from the Coleman-Weinberg model \cite{Coleman} and its generalizations, where at least a quartic potential operator appears at the classical level.
These features seem to point to a new class of Higgs-type mechanisms and for this reason it is worth investigating them in detail \cite{Corfu1,Corfu2}.
A further motivation is related to the fact that in the Higgs phase, the ratio of the scalar to the gauge boson mass near the 5d "bulk" or "zero-temperature" or "quantum" phase transition 
and not far from the triple point on the phase diagram turns out to be numerically close to the corresponding ratio of the Higgs to the $Z$ boson mass
in the Standard Model \cite{IrgesK6}. This regime is near the line of first order phase transitions that separates the Higgs phase from the "Hybrid" phase, where in the bulk
the system decomposes into a weakly interacting array of 4d, confined hyperplanes. The third phase is just a 5d Confined phase with which we will not be concerned here.
Finally, because of the fact that the quantum phase transition of interest is of first order, the effective action must be constructed with a finite cut-off.
It is non-trivial that such an effective action exists at all in a perturbatively non-renormalizable theory but if it does and the associated cut-off is low, it may give us a possible 
resolution to the Higgs mass fine tuning problem. Being able to draw Lines of Constant Physics (LCP), that is lines on the phase diagram ending on the phase transition along which the 
mass spectrum remains constant, supports such a conclusion.

In part I of this work \cite{IrgesFotis2}, we outlined the strategy for building a continuum effective action starting from the lattice construction,
which we briefly review. Start from the lattice plaquette action and expand it in small lattice spacing. It is known that this process generates
an infinite tower of operators of increasing classical dimension, with the ones of dimension larger than $d$, 
typically called Higher Dimensional Operators (HDO).\footnote{At a fixed order in the power of the lattice spacing 
there may also be generated gauge variant operators (lattice artefacts) that we must drop 
in a continuum approach.} Truncating the expansion at a given order and then taking the naive continuum limit, gives us a classical, continuum effective action that may be quantized.
In part I we truncated this expansion at the leading order (LO) in the lattice spacing, while here we will truncate it at next to leading order (NLO), 
thus including the dominant HDO.\footnote{In a lattice language such a process is called Symanzik's improvement 
and it was conducted for the first time, for a pure Yangs-Mills gauge theory, in \cite{Weisz1}. For its application to lattice QCD, see \cite{Weisz2}.}
The reason is that the LO effective action can reproduce several of the non-perturbative properties of the system seen on the lattice but not the ones 
associated with the Higgs mechanism. As we will see, the additional presence of the HDO at NLO, will also unlock the physical properties of the Higgs phase.
In Appendix \ref{review} there is a review of the lattice model and the construction of the classical continuum action from it, that we wish to quantize here.
At the end of the process described there, a continuum action enhanced with HDO, for both the 4d boundary and the 5d bulk, is obtained.
Then we focus on the boundary action which we renormalize diagrammatically at 1-loop order and obtain its quantum effective version.
Subsequently, we analyze the renormalized scalar potential in order to expose the Higgs mechanism on the four-dimensional boundary.
It is important to notice that the boundary effective action, even though naively decoupled from the bulk,
carries information of its 5d origin, hidden inside its couplings and the constrained
way it can move on the phase diagram that contains genuinely 5d structures, such quantum phase transitions.
We point out to this effect a subtle constraint that the boundary RG flows are subject to. Since the phase transition is a place
where the effective cut-off assumes its maximum possible value (apart from the trivial Gaussian fixed point), 
RG flows in the respective phases that the line of phase transitions separates and that terminate
at the same point on the line of phase transitions, are necessarily correlated. Taking into account the fact that some of the phase transitions are of a bulk origin
and that the system near them is dimensionally reduced via localization, results in non-trivial constraints that the most general, unconstrained 4d effective action
with the same field and operator content, would not see.

We do not have to do any explicit new calculations regarding the bulk, as we can safely use known results for both the case where the system is dimensionally
reduced to 4d planes, in which case the RG flow along the 4d planes is just that of an asymptotically free 4d $SU(2)$ coupling and for the 5d bulk 
in the absence of dimensional reduction we can use results from \cite{IrgesFotis2}
when HDO are absent and from \cite{Wise,Grinstein,Schuster,Casarin} when HDO are present.

The non-perturbative phase diagram is determined by two dimensionless couplings, $\beta_4$ and $\beta_5$, or equivalently by $\beta$ and $\g$,
(see \eq{b45}) with $\g$ referred to as the anisotropy parameter. 
As implied by the terminology, this amounts to introducing an anisotropy in the fifth dimension 
without disrupting the four-dimensional Euclidean/Lorentz invariance in the naive continuum limit, which disappears when $\g=1$.
The dimensionful quantities out of which these couplings are constructed, are the lattice spacings $a_4$ and $a_5$ along the 4d and extra dimension respectively
and the 5d gauge coupling $g_5$ that has dimension $-1/2$ in 5d. Practically on the lattice the extent of the fifth dimension is always finite,
introducing in principle an extra dimensionful parameter $R$, the physical length of the fifth dimension. In \cite{IrgesFotis2} it is explained how it can 
be removed from the continuum effective action: since the lattice has a reflection symmetry about the middle point along the extra dimension,
it can be folded about it and then the limit of infinite points in the fifth dimension can be taken. As a result, we end up with 
a semi-infinite dimension at the origin of which the 4d boundary sits and the phase diagram is truly 2-dimensional, parametrized by $\beta_4$ and $\beta_5$ only.
%
\begin{figure}[!htbp]
\centering
\includegraphics[width=10cm]{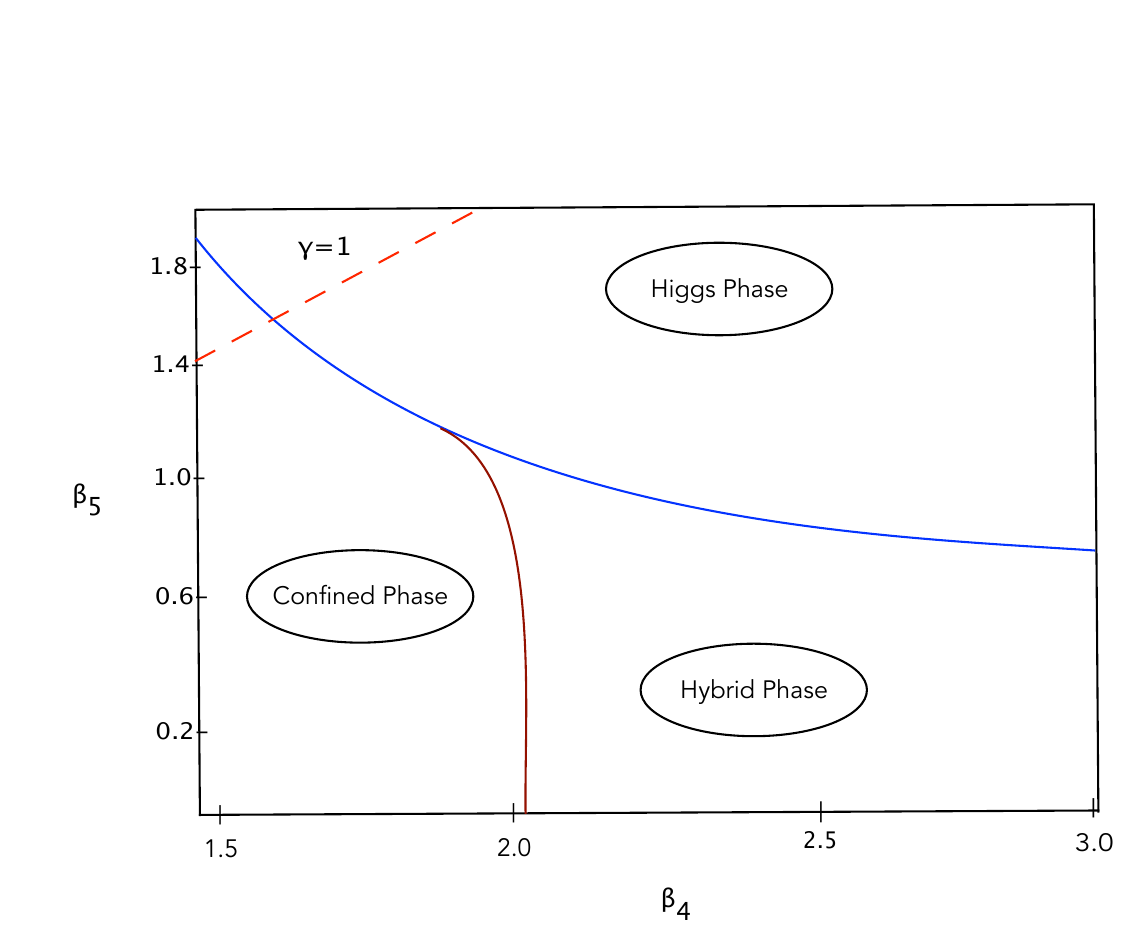}
\caption{\small The Phase Diagram of the anisotropic orbifold lattice \cite{IrgesK5,IrgesK6}.
It exhibits three distinct phases, the Confined, the Higgs and the Hybrid phase, separated by either bulk-driven (blue) or by boundary-driven (red) first-order phase transitions.
The dashed line corresponds to $\g=1$. \label{npPD}}
\end{figure}
\FloatBarrier
%

A crucial step in our construction of a continuum effective action is relating lattice to continuum parameters.
In the continuum, the boundary theory will have a dimensionless 4d coupling and at the quantum level it will develop
two dimensionful scales: a regularization scale, say $\m$ in Dimensional Regularization (DR) and a vacuum expectation value (vev) $v$ in the Higgs phase.
The structure of the phase diagram \cite{IrgesK5,IrgesK6} will guide us in this respect and it is useful to review it in some more detail.
For  general $\g$, the model exhibits three distinct phases, separated by first-order phase transitions.
These are a Higgs phase, a Hybrid phase and a Confined phase, see \fig{npPD}.
We point out here certain features of this phase diagram which will have to be incorporated in, or reproduced by, the continuum effective action.
Actually, the first feature will be reproduced by it while the second will have to be input, as perturbation theory seems to be blind to it.
The first feature is the fact that the line that separates the Higgs from the other two phases (blue line in \fig{npPD}) is bulk driven.
This just means that it is unaffected by the boundary conditions and it is present even on a fully periodic, infinite lattice.
For the effective action this implies that the presence of this phase transition should be detectable by 5d equations only.
This was done in \cite{IrgesFotis2} using the $\ve$-expansion,
according to which the bulk driven phase transition may be re-constructed as a line of 5d Wilson-Fisher (WF) fixed points.
The subtle issue with this is that a WF fixed point is usually interpreted as the sign of a second order phase transition, whereas here
we are after a first order transition. 
Such a distinction while not important in \cite{IrgesFotis2} where the LO expansion could not distinguish first from second order phase transitions,
here with HDO developing in a NLO expansion, becomes necessary.
The second non-perturbative property that has been observed on the lattice is that the entire Hybrid phase, as well as the Higgs phase 
but only near the Higgs-Hybrid phase transition, are layered. This seems to be a fully non-perturbative property \cite{Fu}, not seen by the $\ve$-expansion
and it must be built in by hand in the effective action. The way to do it is to set for the $d=4-\ve$ parameter $\ve=-1$ when locating the bulk driven phase transition
as a line of WF points, but use $\ve=0$ when computing the spectrum and the RG flows in the dimensionally reduced regimes.
When locating the phase transition as a first order transition, we can use instead purely $d=4$ language, which however should
result in small deviations from the second order, WF line. This would imply that the first order transition is weak.
The phase transition that separates the Hybrid and Confined phases on the other hand, is boundary driven.
Indeed, its presence is a non-trivial consistency fact of dimensional reduction, as the boundary of the system has the 
degrees of freedom of an Abelian-Higgs model, where such a phase transition is indeed present (for Higgs charge 2).
The 5d Confined phase will not concern us here much.
Finally, there is the above mentioned link between the lattice and continuum parameters that we need.
As explained in detail in \cite{IrgesFotis2}, this comes down to a relation of the form 
\be\label{ma4}
\m = \frac{F(\b_4,\b_5)}{a_4}
\ee
which relates the DR regularization scale $\m$ to the lattice spacing $a_4$. In general and especially in a non-perturbative regime of a 
non-renormalizable, spontaneously broken theory, $F(\b_4,\b_5)$ may be a complicated function. Its perturbative effect will be taken into account 
by promoting the fixed numerical factors that lattice spacing expansion generates in front of operators to general couplings, to be determined by the renormalization process.
As far as its non-perturbative effects are concerned built in the effective action, as argued in \cite{IrgesFotis2}, near the phase transition it can be safely approximated by a constant.

We finally point out that even though we carry out our analysis for a specific model, analogous considerations are expected to apply for 
any quantum gauge theory with boundaries of reduced gauge symmetry and a phase diagram of similar structure. 
This is a rather broad class of models whose zero temperature properties have not been yet sufficiently investigated.

\section{Quantization with higher derivative operators}\label{quan.bh}

Here we become more specific of the action to be quantized.
The starting point is the lattice orbifold action $S^{\rm orb}$ defined in \cite{IrgesK1,IrgesK3} and reproduced in Appendix \ref{review}:
\be\label{Sor}
S^{\rm orb} = S^{\rm b-h} + S^{B} \, ,
\ee 
with $S^{\rm b-h}$ the Boundary-Hybrid action and $S^{B}$ the bulk action, given by \eq{SbhA} and \eq{SBA}. These are
\be
S^{\rm b-h} = \frac{1}{2 N} \sum_{n_\m} \Biggl[ \frac{\b_4}{2} \sum_{\m<\n} \tr \Bigl \{  1- U^{b}_{\m\n}(n_\m,0) \Bigr\}  + \b_5 \sum_{\m} \tr \Bigl \{  1- U^{h}_{\m5} (n_\m,0) \Bigr\}  \Biggr] \nonumber
\ee
and
\be\label{SBA}
S^{B} = \frac{1}{2 N} \sum_{n_\m,n_5} \Biggl[ \b_4 \sum_{\m<\n} \tr \Bigl \{  1- U_{\m\n}(n_\m,n_5) \Bigr\}  + \b_5 \sum_{\m} \tr \Bigl \{  1- U_{\m5}(n_\m,n_5) \Bigr\}  \Biggr] \, , \nonumber
\ee
respectively. $N=2$ for $SU(2)$, $\b_4$ and $\b_5$ are the lattice couplings, $n_\m, n_5$ the discrete coordinates of the nodes and
$U_{MN}$ is the plaquette lying in the $MN$ directions, with $M,N=\m, 5$. 
The Boundary-Hybrid action represents plaquettes lying on the boundary and plaquettes that are orthogonal to it with one of their sides only on the boundary.
The Bulk action represents all other plaquettes.
The lattice spacings in which the above actions are to be expanded 
are in the definitions
\be\label{b4b5def}
\b_4 = \frac{4}{g_4^2} \, , \hskip 1cm \b_5 = \frac{4 a_4^2}{ a_5^2 g_4^2} 
\ee
with $g_4$ a dimensionless derived coupling, defined in terms of the 5d gauge coupling as $g_4^2=g_5^2/a_5$.
From these definitions it is clear that the model has three raw dimensionful parameters ($a_4$, $a_5$ and $g_5$), or two dimensionless ($\b_4$ and $\b_5$).
Expanding now in small $a_4$ and $a_5$ and truncating at next to leading order in the expansion, yields 
\be\label{Sbhf}
S^{\rm b-h} = \sum_{n_\m} a_4^4 \sum_{\m} \Biggl [ \sum_{\n} \Biggl (  \frac{1}{4} F^3_{\m\n} F^3_{\m\n} + 
\frac{1}{16}  a_4^2 ( \hat\D_\m F^3_{\m\n} ) ( \hat\D_\m F^3_{\m\n} )  \Biggr ) +   | \hat D_\m \phi |^2 +  \frac{a_4^2}{4} | \hat D_\m \hat D_\m \phi |^2   \Biggr ] \, 
\ee
for the Boundary-Hybrid part of the action, on which we will mainly concentrate. For the details of this step, as well as for the analogous step 
for the Bulk part, see Appendix \ref{review}.

Next, we have to take the naive continuum limit to obtain a continuum action.
For that purpose we exploit \eq{a4to0} along with $\hat \D_\m \to \partial_\m$, $\hat D_\m \to D_\m$ and $\hat p_M = (2/a_M) \sin ( a_M p_M/2) \to p_M$.
Moreover we move to Minkowski space
with metric $\eta_{\m\n} \equiv (+,-,-,-)$. 
These are standard operations and they are also shown in detail in Appendix \ref{review}. Here we only comment on the handling of the dimension 6 operators
multiplied by $a_4^2$, for which we use \eq{ma4}.
After these steps, we arrive at 
\be\label{SBH1}
S^{\rm b-h} = \int d^4x \Biggl [ - \frac{1}{4} F^3_{\m\n} F^{3,\m\n}  +  | D_\m \phi |^2  + \frac{ c_\a^{(6)} }{2 \m^2}  ( \partial^\m F^3_{\m\n} ) ( \partial_\m F^{3,\m\n} )  -  \frac{ c_2^{(6)} }{\m^2} |  D^\m D_\m \phi |^2   \Biggr ] \, ,
\ee
where $\phi$ is a complex scalar field, $A^3_\m$ is the photon field and $F^3_{\m\n} = \pa_\m A^3_\n - \pa_\n A^3_\m $.
The couplings $c_\a^{(6)}$ and $c_2^{(6)}$ are introduced for the HDO of the gauge and scalar field respectively, absorbing the unknown function
$F$ in \eq{ma4}. The final step before the quantization process starts is to interpret $\m$ in \eq{SBH1}. In principle we could leave it as it is (see \cite{FotisLetter2}),
however we can rewrite it in a more convenient form that resembles usual Effective Field Theory (EFT) treatments. Multiplying and dividing by a constant
scale $\L^2$ and absorbing $\L^2/\m^2$ in the couplings, we can replace $\m\to\L$ in  \eq{SBH1}.
Now $\L$ can be regarded as the cut-off of the EFT. In fact, we will see that in our case $\L$ is not an external scale that must be introduced 
at this point by hand. It is rather an internal scale, given by the value of the regulating scale at the phase transition, $\m_*$, where it assumes its maximum value.
Notice that we could have arrived at \eq{SBH1} directly by using gauge invariance. One reason we went through the painful process of 
generating it by expanding the lattice action is because of the anisotropy factor $\g$, hiding in the covariant derivative $D_\m = \partial_\m - i g_4 A^3_\m $, where
\be
g_4^2 = \frac{g_5^2}{a_5} = \frac{g_5^2}{a_4} \frac{a_4}{a_5} = \frac{g_5^2}{a_4} \g = g^2 \g\, .
\ee
In the above we have used the bare value of the anisotropy originating from \eq{b4b5def} and defined another convenient dimensionless coupling, $g^2=g_5^2/a_4=g_4^2/\g$.
The presence of $\g$ is non-trivial since it opens a second dimension in the phase diagram, where new phases and a triple point appear, among others.
But there is another, equally important reason.
Notice that in the presence of quadratic and quartic potential terms for the complex scalar, \eq{SBH1} would be just
the Lee-Wick Scalar QED whose 1-loop renormalization was extensively studied in \cite{Simmons}.
However, the above effective action does not have a scalar potential since the lattice does not produce polynomial terms for the boundary effective action at any order 
in the expansion in the lattice spacings. 
Using standard jargon, our Higher Dimensional Operators are exclusively Higher Derivative Operators.
This is due to the 5d origin of the boundary theory and it is a crucial characteristic of our model which distinguishes it from other models of the sort.

\subsection{The ghost-free basis of the gauge-fixed classical action}\label{Red.action}

Now we are almost ready to renormalize the Lagrangean at 1-loop level, obtain its $\b$-functions and through them determine the Renormalization Group (RG) flows.
We have to tackle one more obstacle though, associated with the scalar HDO in \eq{SBH1} which contains an Ostrogradsky instability.
We will deal with this immediately, but first we will fix the gauge.
The gauge-fixing term for the bulk action is $ \pa_M A^A_M $ while for the boundary, using the boundary conditions as in \cite{IrgesFotis2}, 
it is $\pa_\m A^3_\m $.
The same is true for the Faddeev-Popov ghosts which in the bulk are $\bar c^A$, $c^A$ and on the boundary $\bar c^3$, $c^3$.
In the latter case recall that the Faddeev-Popov ghosts are decoupled from the spectrum.
Given the above, the gauge-fixed $S^{\rm b-h}$ reads
\bea\label{SBH2}
S^{\rm b-h} &=& \int d^4x \Biggl [ - \frac{1}{4} F^3_{\m\n} F^{3,\m\n} - \frac{1}{2\xi} (\pa^\m A^3_\m)^2 +  | D_\m \phi |^2  \nonumber\\
&+& \frac{ c_\a^{(6)} }{2\L^2} ( \partial^\m F^3_{\m\n} ) ( \partial_\m F^{3,\m\n} ) -  \frac{ c_2^{(6)} }{\L^2}  |  D^\m D_\m \phi |^2 + \pa^\m \bar c^3 \pa_\m c^3  \Biggr ] \, .
\eea
The instability is exposed by expanding the covariant derivative and rearranging terms up to total partial derivatives.
By doing this, we arrive at the bare Boundary-Hybrid action 
\bea\label{SBH3}
S_0^{\rm b-h} &=& \int d^4x \Biggl [ - \frac{1}{4} F^3_{\m\n,0} F_0^{3,\m\n} + \frac{1}{2\xi} A^3_{\m,0} \pa^\m \pa_\n A_0^{3,\n} - \bar \phi_0 \Box \phi_0  - \frac{ c_{\a,0}^{(6)} }{2\L^2}  F^3_{\m\n,0} \Box F_0^{3,\m\n}  - \frac{ c_{2,0}^{(6)} }{\L^2} \bar \phi_0 \Box^2 \phi_0 - \bar c_0^3 \Box c_0^3   \nonumber\\
&+& i g_0 \sqrt{\g_0} A^3_{\m,0} \Biggl(  \Bigl\{  \bar \phi_0 \pa^\m \phi_0 - \phi_0 \pa^\m \bar \phi_0 \Bigr \} + \frac{  c_{2,0}^{(6)} }{\L^2} \Bigl \{ \Box \bar \phi_0 \pa^\m \phi_0 - \Box \phi_0 \pa^\m \bar \phi_0 \Bigr \} \Biggr )  +  g_0^2 \g_0 (A^3_{\m,0})^2 \bar \phi_0 \phi_0   \nonumber\\
&+& g_0^2 \g_0 \Biggl(  \frac{  c_{2,0}^{(6)} }{\L^2} \Bigl \{  \phi_0 \Box^2 \bar \phi_0 + \bar \phi_0 \Box^2 \phi_0 \Bigr \}  -  \frac{  c_{2,0}^{(6)}  }{\L^2}  \pa^\m ( A^3_{\m,0} \bar \phi_0 )  \pa_\m ( A_0^{3,\m} \phi_0 )  \Biggr)    \nonumber\\
&-& \frac{ i g_0^3 \g_0^{3/2} c_{2,0}^{(6)}  }{\L^2} (A^3_{\r,0})^2  A^3_{\m,0}  \Bigl( \bar \phi_0  \pa^\m \phi_0  - \phi_0 \pa^\m \bar \phi_0  \Bigr)  - \frac{ g_0^4 \g_0^2 c_{2,0}^{(6)} }{\L^2}  (A^3_{\r,0} )^4 \bar \phi_0 \phi_0   \Biggr ]  \nonumber\\
&=& S^{\rm b-h}_{\rm Kin, 0} + S^{\rm b-h}_{\rm Int, 0} \, ,
\eea
where $S^{\rm b-h}_{\rm Kin, 0} $ is the kinetic part of the action
\be
S^{\rm b-h}_{\rm Kin, 0}  = \int d^4x \Biggl [  - \frac{1}{4} F^3_{\m\n,0} F_0^{3,\m\n} + \frac{1}{2\xi} A^3_{\m,0} \pa^\m \pa_\n A_0^{3,\n} - \bar \phi_0 \Box \phi_0  - \frac{ c_{\a,0}^{(6)} }{2\L^2}  F^3_{\m\n,0} \Box F_0^{3,\m\n}  - \frac{ c_{2,0}^{(6)} }{\L^2} \bar \phi_0 \Box^2 \phi_0 - \bar c_0^3 \Box c_0^3   \Biggr ] \nonumber
\ee
while $S^{\rm b-h}_{\rm Int, 0} $ is the interaction part
\bea
S^{\rm b-h}_{\rm Int, 0} &=& \int d^4x \Biggl [  i g_0 \sqrt{\g_0} A^3_{\m,0} \Biggl(  \Bigl\{  \bar \phi_0 \pa^\m \phi_0 - \phi_0 \pa^\m \bar \phi_0 \Bigr \} + \frac{  c_{2,0}^{(6)} }{\L^2} \Bigl \{ \Box \bar \phi_0 \pa^\m \phi_0 - \Box \phi_0 \pa^\m \bar \phi_0 \Bigr \} \Biggr )     \nonumber\\
&+&  g_0^2 \g_0 (A^3_{\m,0})^2 \bar \phi_0 \phi_0 + g_0^2 \g_0 \Biggl(  \frac{  c_{2,0}^{(6)} }{\L^2} \Bigl \{  \phi_0 \Box^2 \bar \phi_0 + \bar \phi_0 \Box^2 \phi_0 \Bigr \}  -  \frac{  c_{2,0}^{(6)}  }{\L^2}  \pa^\m ( A^3_{\m,0} \bar \phi_0 )  \pa_\m ( A_0^{3,\m} \phi_0 )  \Biggr)    \nonumber\\
&-& \frac{ i g_0^3 \g_0^{3/2} c_{2,0}^{(6)}  }{\L^2} (A^3_{\r,0})^2  A^3_{\m,0}  \Bigl( \bar \phi_0  \pa^\m \phi_0  - \phi_0 \pa^\m \bar \phi_0  \Bigr)  - \frac{ g_0^4 \g_0^2 c_{2,0}^{(6)} }{\L^2}  (A^3_{\r,0} )^4 \bar \phi_0 \phi_0    \Biggr ]  \, . \nonumber
\eea
The subscript $0$ denotes the bare fields and couplings. 
Looking at the kinetic part we notice that each of the two higher derivative operators may impose non-physical degrees of freedom on the spectrum.
These are the Ostrogradsky ghosts (the O-ghosts) \cite{Ost.Wood.} and a possible way to describe their effect can be found in \cite{Simmons}.
There, these ghosts correspond to extra poles in the gauge and scalar propagators, reducing the divergence level of the loop diagrams.
However, this observation is not sufficient to fully deal with them at the quantum level, since if O-ghosts remain in the spectrum, the instability remains.
In \cite{FotisLetter2,FotisLetter1} an algorithm was developed so as to get a ghost free basis when, after a general field redefinition, the Jacobean 
of the transformation is properly taken into account.
According to this algorithm, another ghost-field must be introduced, the Reparameterization ghost (R-ghost), which cancels the pole due to the O-ghost.
Here we do not get into the details of these operations and just use the result of \cite{FotisLetter2,FotisLetter1} to eliminate the O-ghosts, after performing the field redefinition 
\bea\label{fred.}
\phi_0 &\to& \hat \phi_0 =  \phi_0 + \frac{x}{\L^2} D^2 \phi_0 + \frac{y}{\L^2} (\bar \phi_0 \phi_0) \phi_0 \nonumber\\
\bar \phi_0 &\to& \bar{\hat \phi}_0 = \bar \phi_0 + \frac{x}{\L^2} \bar D^2 \bar \phi_0 + \frac{y}{\L^2} (\bar \phi_0 \phi_0) \bar \phi_0 \nonumber\\
A^3_{\m,0} &\to& \hat A^3_{\m,0} =  A^3_{\m,0} + \frac{x_\a}{\L^2}( \eta_{\m\r} \Box - \pa_\m \pa_\r )A^{3,\r}_0 
\nonumber\\
F^3_{\m\n,0} &\to& \hat F^3_{\m\n,0} = F^3_{\m\n,0} + \frac{x_\a}{\L^2} \Box F^3_{\m\n,0} \, ,
\eea
with $D^2$ standing for $D^\m D_\m$.
Now these field redefinitions raise two related questions.
The first regards the generality of the redefinition (for example the gauge-field redefinition seems to be incomplete, as we could have added the term $(A^3_\r)^2 A_\m^3/\L^2$)
and the second is concerned about the fate of gauge invariance of the redefined action.
Actually these questions are related and the answer to both of them is contained in the analysis of Appendix \ref{FRGI}, 
according to which a properly redefined field should transform covariantly, in such a way that leaves gauge invariance intact.
This is the case for \eq{fred.} and this is made clear if we gauge transform the scalar field as $\phi'_0 = e^{i \a(x)} \phi_0 $ to get 
\bea
\hat \phi'_0 &=& \phi'_0 + \frac{x}{\L^2} D'^2 \phi'_0 + \frac{y}{\L^2} (\bar \phi'_0 \phi'_0) \phi'_0 \nonumber\\
 &=& e^{i \a(x)} \phi_0 + \frac{x}{\L^2} D'^\m (D'_\m e^{i \a(x)} \phi_0 ) + \frac{y}{\L^2} (\bar \phi_0  e^{- i \a(x)} e^{i \a(x)} \phi_0) e^{i \a(x)} \phi_0 \nonumber\\
 &=& e^{i \a(x)} \{ \phi_0 + \frac{x}{\L^2} D^2 \phi_0  + \frac{y}{\L^2} (\bar \phi_0  \phi_0) \phi_0 \} \equiv e^{i \a(x)} \hat \phi_0 
\eea
and the gauge field as $(A^3_{\m,0})' = A^3_{\m,0} + \pa_\m \a(x) $ to get
\bea
(\hat A^3_{\m,0})'  &=& (A^3_{\m,0})'  + \frac{x_\a}{\L^2}( \eta_{\m\r} \Box - \pa_\m \pa_\r )(A^{3,\r}_0)' \nonumber\\
 &=& A^3_{\m,0} + \pa_\m \a(x) + \frac{x_\a}{\L^2}( \eta_{\m\r} \Box - \pa_\m \pa_\r )A^{3,\r}_0 + \frac{x_\a}{\L^2} \{ \Box \pa_\m \a(x) - \pa_\m \pa_\r \pa^\r \a(x) \} \nonumber\\
 &=& \{ A^3_{\m,0} + \frac{x_\a}{\L^2}( \eta_{\m\r} \Box - \pa_\m \pa_\r )A^{3,\r}_0 \} + \pa_\m \a(x) \equiv \hat A^3_{\m,0} + \pa_\m \a(x) \, ,
\eea
with $\a(x)$ a gauge tranformation function and then under a gauge transformation $D_\m \phi \to e^{i \a(x)} D_\m \phi $.
Note also that the R-ghosts which are inherited in the spectrum due to the field redefinitions are in accordance with \eq{FphiFA}.
Hence the redefined action remains gauge invariant.
Then, \eq{SBH3} becomes
\bea
S_0^{\rm b-h} &=&  \int d^4x \Biggl [ - \frac{1}{4} F^3_{\m\n,0} F_0^{3,\m\n} + \frac{1}{2\xi} A^3_{\m,0} \pa^\m \pa_\n A_0^{3,\n} - \bar \phi_0 \Box \phi_0  - \frac{ c_{\a,0}^{(6)} + x_\a }{2\L^2}  F^3_{\m\n,0} \Box F_0^{3,\m\n}  \nonumber\\
&-& \frac{ c_{2,0}^{(6)} + 2 x }{\L^2} \bar \phi_0 \Box^2 \phi_0 - \frac{2 y }{\L^2}  (\bar \phi_0 \phi_0) \bar \phi_0 \Box \phi_0 - \bar c_0^3 \Box c_0^3  \nonumber\\
&+& i g_0 \sqrt{\g_0} A^3_{\m,0} \Biggl( \Bigl \{ 1 + \frac{2y (\bar \phi_0 \phi_0)}{\L^2} \Bigr \} \Bigl(  \bar \phi_0 \pa^\m \phi_0 - \phi_0 \pa^\m \bar \phi_0 \Bigr ) + \frac{  c_{2,0}^{(6)} + 2 x }{\L^2} \Bigl( \Box \bar \phi_0 \pa^\m \phi_0 - \Box \phi_0 \pa^\m \bar \phi_0 \Bigr ) \Biggr )   \nonumber\\
&+& i g_0 \sqrt{\g_0} x_\a \frac{ ( \eta_{\m\r} \Box - \pa_\m \pa_\r ) A^{3,\r}_0 }{\L^2}  \Bigl(  \bar \phi_0 \pa^\m \phi_0 - \phi_0 \pa^\m \bar \phi_0 \Bigr ) + g_0^2 \g_0 (A^3_{\m,0})^2 \bar \phi_0 \phi_0  \Bigl( 1 +  \frac{2 y }{\L^2} \bar \phi_0 \phi_0 \Bigr)  \nonumber\\
&+& g_0^2 \g_0 \Biggl(  \frac{  c_{2,0}^{(6)} + 2 x }{\L^2} \Bigl \{  \phi_0 \Box^2 \bar \phi_0 + \bar \phi_0 \Box^2 \phi_0 \Bigr \} - \frac{ x }{\L^2} A^3_{\m,0} \Bigl \{ \pa^\r ( A^3_{\r,0} \bar \phi_0 ) \pa^\m \phi_0 + \pa^\r ( A^3_{\r,0} \phi_0 ) \pa^\m \bar \phi_0  \Bigr \}  \nonumber\\
&-&  \frac{  c_{2,0}^{(6)} + 2 x }{\L^2}  \pa^\m ( A^3_{\m,0} \bar \phi_0 )  \pa_\m ( A_0^{3,\m} \phi_0 )  \Biggr) +  2 g_0^2 \g_0 x_\a \frac{ A^{3,\m}_0 ( \eta_{\m\r} \Box - \pa_\m \pa_\r ) A^{3,\r}_0   }{\L^2} \bar \phi_0 \phi_0   \nonumber\\
&-& \frac{ i g_0^3 \g_0^{3/2} ( c_{2,0}^{(6)}+2x) }{\L^2} (A^3_{\r,0})^2  A^3_{\m,0}  \Bigl( \bar \phi_0  \pa^\m \phi_0  - \phi_0 \pa^\m \bar \phi_0  \Bigr)  - \frac{ g_0^4 \g_0^2 ( c_{2,0}^{(6)}+2x) }{\L^2}  (A^3_{\r,0} )^4 \bar \phi_0 \phi_0    \Biggr ]  \nonumber
\eea
and there is indeed extra freedom from the redefinition so as to eliminate the two higher derivative operators.
In particular choosing $x_\a = - c_{\a,0}^{(6)} $, $x = - c_{2,0}^{(6)}/2$ and $2y = c_{1,0}^{(6)}/4 $ the redefined bare action becomes
\bea\label{SBHrd}
S_0^{\rm b-h} &=& \int d^4x \Biggl [ - \frac{1}{4} F^3_{\m\n,0} F_0^{3,\m\n} + \frac{1}{2\xi} A^3_{\m,0} \pa^\m \pa_\n A_0^{3,\n} - \bar \phi_0 \Box \phi_0  - \frac{c_{1,0}^{(6)}}{4\L^2}  (\bar \phi_0 \phi_0) \bar \phi_0 \Box \phi_0 - \bar c_0^3 \Box c_0^3  \nonumber\\
&+& i g_{4,0}  \Bigl \{  \eta_{\m\r} - \frac{  \eta_{\m\r} \Box - \pa_\m \pa_\r  }{\L^2} \Bigr \} A^{3,\r}_0 \Bigl(  \bar \phi_0 \pa^\m \phi_0 - \phi_0 \pa^\m \bar \phi_0 \Bigr )  +  g^2_{4,0} (A^3_{\m,0})^2 \bar \phi_0 \phi_0   \nonumber\\
&+& \frac{  g^2_{4,0} }{2 \L^2}  \Bigl ( A^3_{\m,0} A^3_{\r,0} \pa^\r \bar \phi_0  \pa^\m \phi_0 + A^3_{\m,0} \pa^\r A^3_{\r,0} \pa^\m ( \bar \phi_0  \phi_0) \Bigr ) -  2 g^2_{4,0} \frac{ A^{3,\m}_0 ( \eta_{\m\r} \Box - \pa_\m \pa_\r ) A^{3,\r}_0   }{\L^2} \bar \phi_0 \phi_0 \nonumber\\
&+&  i \frac{ g_{4,0} \, c_{1,0}^{(6)} }{4 \L^2}A^3_{\m,0} \bar \phi_0 \phi_0 \Bigl(  \bar \phi_0 \pa^\m \phi_0 - \phi_0 \pa^\m \bar \phi_0 \Bigr ) + \frac{g^2_{4,0} \, c_{1,0}^{(6)} }{4 \L^2} (A^3_{\m,0})^2 ( \bar \phi_0 \phi_0 )^2  \Biggr ]  \, ,
\eea
where $c_{1,0}^{(6)} $ is a dimensionless coupling which is undetermined at present.
For simplicity of notation, we have turned back to our original notation $g_4 = g \sqrt{\g}$ and normalized the undefined couplings as $c_{2,0}^{(6)} = c_{\a,0}^{(6)} \equiv 1$. 
Notice that the gauge-fixing term is untouched since it is an arbitrary function and can be redefined to its original form.
Another way to see this is that since the redefinition commutes\footnote{For more details see \cite{FotisLetter2} and references therein.} 
with renormalization it could have been performed before gauge fixing.

Comparing now \eq{SBHrd} to its original form \eq{SBH3}, the interesting point to notice is that the 
former includes, after the field redefinition, the scalar quartic-like term $(\bar \phi \phi)\bar \phi \Box \phi$, instead of the original higher derivative term.
In fact, apart from the modified vertices, $S_0^{\rm b-h}$ now resembles an effective version of the Coleman-Weinberg (CW) \cite{Coleman} model.
What happened is that a term like $\bar \phi \Box^2  \phi $ has dual nature since it could be both part of the kinetic Lagrangian and a mass term of the scalar field.
Then performing the field redefinition we threw away the O-ghost whose nature as a mass term was left implicit in the theory through the appeared potential.
The Feynman rules for \eq{SBHrd} are given in Appendix \ref{FRBHA}.

\subsection{One loop corrections}\label{QuCo}

Now that a consistent basis for the Boundary-Hybrid action has been developed we are finally ready to initiate the renormalization program.
We choose to work in the Feynman gauge, where $\xi=1$.
We first set some notation used in the following.
The 1-loop corrections to the 2-, 3-, 4-, 5- and 6-point functions of the fields are denoted by ${\cal M}_F$, ${\cal K}_F$, ${\cal B}_{4,F}$, ${\cal B}_{5,F}$ and ${\cal B}_{6,F}$ respectively.
The subscript $F$ corresponds to the different combinations that $\phi$, $\bar \phi$ and $A^3_\m$ can form as external fields.
As an example consider the correction to the scalar-scalar-gauge vertex which is represented by ${\cal K}_{\bar \phi A \phi}$ where $A$ in the subscript represents  $A^3_\m$.
The symmetry factor which is associated with a given diagram will be denoted as $S_{{\cal G}}$ with ${\cal G} \equiv {\cal M}_F, {\cal K}_F, {\cal B}_{4,F}, {\cal B}_{5,F}, {\cal B}_{6,F}$.  
Finally if the scalar and gauge fields are external, then the notation of the momentum is $p_\m$ and $q_\m$  respectively.
Otherwise, when they run in the loop, the vector $k_\m$ is used.
For the finite parts of the diagrams we use the notation $[\,\,\,]_f$.

Now looking at the Feynman rules of \eq{SBHrd} note that there is only one possible 1-leg Tadpole, a correction to the gauge field, 
which is however forbidden due to gauge and Lorentz symmetry.
Hence the loop calculation starts with the 2-point functions of $\phi$ and $A^3_\m$.
More specifically for the scalar case there are three contributions, ${\cal M}^1_\phi$, ${\cal M}^2_\phi$ and ${\cal M}^3_\phi$ with equal symmetry factors $S_{{\cal M}^1_\phi}=S_{{\cal M}^2_\phi}=S_{{\cal M}^3_\phi}=1$.
The first diagram is a two-leg Tadpole which includes only scalar fields and is given by
\vskip .5cm
\begin{center}
\begin{tikzpicture}[scale=0.7]
\draw [dashed] (0,0)--(1.8,0);
\draw [dashed] (0,0.9) circle [radius=0.9];
\draw [dashed] (-1.8,0)--(0,0);
\draw [->]  (-1.9,0.3)--(-1.2,0.3);
\node at (-1.7,-0.4) {$p$};
\draw [<-, very thick] (-0.1,1.8)--(0.1,1.8);
\node at (0,1.3) {$k$};
\node at (1.7,-0.4) {$p$};
\draw [<-, very thick] (1,0)--(0.8,0);
\draw [->, very thick] (-1,0)--(-0.8,0);
\draw [<-]  (1.9,0.3)--(1.2,0.3);
\node at (4,0) {$=  \hskip .1 cm i {\cal M}^1_\phi$};
\end{tikzpicture}
\end{center}
whose evaluation gives
\bea\label{M1fi}
i {\cal M}^1_\phi &=&  i c_{1}^{(6)} \frac{p^2} {\L^2} S_{{\cal M}^1_\phi} \int \frac{d^4 k }{(2 \pi)^4}  \frac{i }{ k^2} \Rightarrow \nonumber\\
{\cal M}^1_\phi &=& - c_{1}^{(6)}  \frac{p^2} {\L^2} \int \frac{d^4 k }{(2 \pi)^4 i }  \frac{1 }{ k^2}  \nonumber\\
 &=& - c_{1}^{(6)} \frac{p^2} {\L^2} A_0 (0)  \, . 
\eea
In the last line we have used the standard Passarino-Veltman notation, which we will be using throughout.
The next diagram in line is another two-leg Tadpole with gauge contribution:
\vskip .5cm
\begin{center}
\begin{tikzpicture}[scale=0.7]
\draw [dashed] (0,0)--(1.8,0);
\draw [photon] (0,0.9) circle [radius=0.9];
\draw [dashed] (-1.8,0)--(0,0);
\node at (-1.7,-0.4) {$p$};
\draw [->]  (-1.9,0.3)--(-1.2,0.3);
\draw [<-]  (1.9,0.3)--(1.2,0.3);
\node at (0,1.3) {$k$};
\node at (1.7,-0.4) {$p$};
\draw [<-, very thick] (1,0)--(0.8,0);
\draw [->, very thick] (-1,0)--(-0.8,0);
\node at (4,0) {$= i {\cal M}^2_{\phi}$};
\end{tikzpicture}
\end{center}
and it is equal to
\bea\label{M2fi}
i {\cal M}^2_\phi &=& 2 i g_4^2 S_{{\cal M}^2_\phi} \int{ \frac{d^4 k}{(2 \pi)^4}}  \frac{-i \eta^{\m\n}}{ k^2} Q_{\m\n}(p,k) \Rightarrow \nonumber\\
{\cal M}^2_\phi &=& 2 g_4^2  \int \frac{d^4 k }{(2 \pi)^4 i }  \frac{ \eta^{\m\n}}{ k^2} \left[  \eta_{\m\n} + \frac{ \eta_{\m\n} p^2 } {2 \L^2}  +  \frac{2 ( \eta_{\m\n} k^2- k_\m k_\n )} { \L^2}   \right]  \nonumber\\
 &=& 2 g_4^2  \int \frac{d^4 k }{(2 \pi)^4 i }  \left[  d ( 1+ \frac{ p^2 } {2 \L^2} ) + \frac{ 2(d-1)k^2 } { \L^2} \right] \frac{1}{ k^2} \nonumber\\
 &=&  8 g_4^2 \, ( 1+ \frac{ p^2 } {2 \L^2} ) A_0(0) \, ,
\eea
where \eq{QmQmn} and the fact that
\be
\int \frac{d^d k }{(2 \pi)^d } \equiv 0 \nonumber 
\ee
in DR were used.
Up to now we have faced two massless tadpoles which vanish in dimensional regularization.
A useful relation based on that is obtained if we convert \eq{M2fi} to a usual 2-point function of the form
\be\label{M2fio}
{\cal M}^2_{\phi} = 8 g_4^2 \,  ( 1+ \frac{ p^2 } {2 \L^2} ) \int \frac{d^4 k }{(2 \pi)^4 i }  \frac{k^2}{ k^2(k+p)^2} = 8 g_4^2 \,  ( 1+ \frac{ p^2 } {2 \L^2} ) \eta_{\m\n} B^{\m\n}(k,k+p) \, .
\ee
Combining \eq{M2fi} with \eq{M2fio} shows that $\eta_{\m\n} B^{\m\n}$, when scaleless, is analogous to $A_0$ and vanishes in DR.
Therefore in the following only $B_0$'s will contribute a divergent part to the calculation, as we will see.
The last contribution of the current category comes from the square of a vertex and yields
\vskip .5cm
\begin{center}
\begin{tikzpicture}[scale=0.7]
\draw [dashed] (-2.3,0)--(-1,0);
\draw [dashed] (1,0)--(2.3,0);
\draw [dashed] (-1,0)--(1,0);
\node at (-1.7,-0.5) {$p$};
\node at (0,1.6) {$k$};
\node at (0,-0.5) {$k+p$};
\node at (1.7,-0.5) {$p$};
\draw [<-, very thick] (1.6,0)--(1.4,0);
\draw [->, very thick] (-1.6,0)--(-1.4,0);
\draw [->, very thick] (-0.1,0)--(0.1,0);
\node at (4.5,0) {$=\, i {\cal M}^3_\phi $\, . };
\draw  [photon] (-1,0) .. controls (-1,0.555) and (-0.555,1) .. (0,1)
.. controls (0.555,1) and (1,0.555) .. (1,0);
\end{tikzpicture}
\end{center}
Its explicit form is given by
\bea
i {\cal M}^3_\phi &=& - g_4^2 S_{{\cal M}^3_\phi} \int{ \frac{d^4 k}{(2 \pi)^4}}  \frac{-i \eta^{\m\n}}{ k^2 }  \frac{i Q_{\m}(p,k) Q_{\n}(p,k) }{ (k+p)^2 }   \Rightarrow \nonumber\\
{\cal M}^3_\phi &=& - g_4^2  \int \frac{d^4 k }{(2 \pi)^4 i }  \frac{ Q_{\m}(p,k) Q^{\m}(p,k) }{ k^2 (k+p)^2 } \nonumber\\
 &=& - g_4^2  \int \frac{d^4 k }{(2 \pi)^4 i }  \left[  (2p + k)^2 + 2 (2p + k)^2 \frac{ k^2 } { \L^2}  -  2\frac{ ( k^2 + 2p \cdot k)^2 } { \L^2} \right] \frac{ 1 }{ k^2 (k+p)^2 }  + {\cal O}(\frac{1}{\L^4}) \, , \nonumber
\eea
using again \eq{QmQmn}.
As indicated in the above relation there are terms of order higher than ${\cal O}(1/\L^2)$ contributing to ${\cal M}^3_\phi$.
Nevertheless, these cannot be renormalized unless operators with dimension higher that 6 appear in the action.
Since \eq{SBHrd} includes only dim-4 and -6 operators, in the following we neglect such contributions without loss of consistency.
To move on let us clarify that our calculating algorithm is to reduce the integrals until they reach the scalar form 
$A_0$ and $B_0$ following the Appendix of \cite{IrgesFotis1}, using the massless limit of the formulae.
When necessary the relation 
\be
\int \frac{d^4 k }{(2 \pi)^4 i } \frac{k_\m}{k^2} \equiv 0 \, , \nonumber
\ee
due to the anti-symmetry under $k \to -k$, is exploited.
So with the above in our hand ${\cal M}^3_{\phi}$ becomes 
\be\label{M3fi}
{\cal M}^3_\phi = - g_4^2 \left[  \frac{ 4 p^2 } { \L^2} A_0(0) + 2 p^2 ( 1- \frac{ p^2 } { \L^2} ) B_0(k,k+p) + \eta_{\m\n}  B^{\m\n}(k,k+p)  \right] 
 \ee
and collectively the 1-loop correction to the scalar field propagator reads
\bea\label{Mfi}
{\cal M}_\phi &=& {\cal M}^1_{\phi} + {\cal M}^2_{\phi} + {\cal M}^3_{\phi} \nonumber\\
 &=&  - c_1^{(6)} \frac{p^2} {\L^2}   A_0 (0) +  8 g_4^2  A_0(0) - 2 g_4^2 p^2 ( 1- \frac{ p^2 } { \L^2} ) B_0(k,k+p) - g_4^2 \eta_{\m\n}  B^{\m\n}(k,k+p) \, . \nonumber\\
\eea
The next set of 2-point functions regards the quantum corrections to the gauge-field propagator.
Here there are two possible diagrams at 1-loop level, ${\cal M}^1_{A,\m\n}$ and ${\cal M}^2_{A,\m\n}$, again with equal symmetry factors $S_{{\cal M}^1_A}=S_{{\cal M}^2_A}=1$.
The first diagram is the only two-leg Tadpole left given by  
\vskip .5cm
\begin{center}
\begin{tikzpicture}[scale=0.7]
\draw [photon] (0,0)--(1.8,0);
\draw [dashed] (0,0.9) circle [radius=0.9];
\draw [photon] (-1.8,0)--(0,0);
\node at (-1.7,-0.4) {$q$};
\draw [->]  (-1.9,0.3)--(-1.2,0.3);
\draw [<-]  (1.9,0.3)--(1.2,0.3);
\node at (0,1.3) {$k$};
\node at (1.7,-0.4) {$q$};
\draw [<-, very thick] (-0.1,1.8)--(0.1,1.8);
\node at (4,0) {$= i {\cal M}^1_{A,\m\n}$};
\end{tikzpicture}
\end{center}
and is equal to
\bea
i {\cal M}^1_{A,\m\n} &=& 2 i g_4^2 S_{{\cal M}^1_A} \int{ \frac{d^4 k}{(2 \pi)^4}}  \frac{i Q_{\m\n}(k,q)}{ k^2}  \Rightarrow \nonumber\\
{\cal M}^1_{A,\m\n} &=& - 2 g_4^2  \int \frac{d^4 k }{(2 \pi)^4 i }  \frac{1}{ k^2} \left[  \eta_{\m\n} + \frac{ \eta_{\m\n} k^2 } {2 \L^2}  +  \frac{2 ( \eta_{\m\n} q^2- q_\m q_\n )} { \L^2}   \right]  \nonumber\\
 &=& - 2 g_4^2  \left[  \eta_{\m\n}  + \frac{2 ( \eta_{\m\n} q^2- q_\m q_\n )} { \L^2}   \right] A_0(0) 
\eea
while its contracted version gives
\be\label{M1A}
{\cal M}^1_A = \frac{1}{3} \left( - \eta^{\m\n} + \frac{q^\m q^\n}{q^2} \right) {\cal M}^1_{A,\m\n} = 2 g_4^2  \left[  1  + \frac{2 q^2} { \L^2}   \right] A_0(0) \, .
\ee
The second contribution to the gauge field propagator comes from the $\bar \phi$-$\phi$-$A^3_\m$ interaction and yields
\vskip .5cm
\begin{center}
\begin{tikzpicture}[scale=0.7]
\draw [photon] (1,0)--(3,0);
\draw [dashed] (0,0) circle [radius=1];
\draw [photon] (-3,0)--(-1,0);
\node at (-2,-0.4) {$q$};
\draw [->]  (-2.2,0.3)--(-1.5,0.3);
\draw [<-]  (2.2,0.3)--(1.5,0.3);
\node at (0,1.5) {$k$};
\node at (0,-1.5) {$k+q$};
\node at (2,-0.4) {$q$};
\draw [<-, very thick] (-0.1,1)--(0.1,1);
\draw [->, very thick] (-0.1,-1)--(0.1,-1);
\node at (4.5,0) {$= i {\cal M}^2_{A,\m\n} $};
\end{tikzpicture}
\end{center}
while its evaluation gives
\bea
i {\cal M}^2_{A,\m\n} &=& - g_4^2 S_{{\cal M}^2_A} \int{ \frac{d^4 k}{(2 \pi)^4}}  \frac{i Q_{\m}(k,q)}{ k^2 }  \frac{i  Q_{\n}(k,q) }{ (k+q)^2 }   \Rightarrow \nonumber\\
{\cal M}^2_{A,\m\n} &=&  g_4^2  \int \frac{d^4 k }{(2 \pi)^4 i }  \frac{ Q_{\m}(k,q) Q_{\n}(k,q) }{ k^2 (k+q)^2 } \nonumber\\
 &=& g_4^2  \int \frac{d^4 k }{(2 \pi)^4 i }  \Bigl [  (2k + q)_\m(2k + q)_\n ( 1 + \frac{ 2 q^2 } { \L^2}  )  - \frac{ q_\n q^\r } { \L^2} (2k + q)_\m(2k + q)_\r  \nonumber\\
 &-& \frac{ q_\m q^\r } { \L^2} (2k + q)_\n(2k + q)_\r  \Bigr] \frac{ 1 }{ k^2 (k+q)^2 }  + {\cal O}(\frac{1}{\L^4}) \, , \nonumber
\eea
where again terms of higher than ${\cal O}(1/\L^2)$ are neglected.
Expanding the parentheses and performing the appropriate reduction the above expression becomes
\bea\label{M2Amn}
{\cal M}^2_{A,\m\n} &=&  g_4^2 ( 1 + \frac{ 2 q^2 } { \L^2}  ) \left[ q_\m q_\n B_0(k,k+p) + 2 q_\m B_\n(k,k+p) + 2 q_\n B_\m(k,k+p) + 4 B_{\m\n}(k,k+p)   \right]  \nonumber\\
&-& \frac{ q_\n q^\r } { \L^2}  \left[ q_\m q_\r B_0(k,k+p) + 2 q_\m B_\r(k,k+p) + 2 q_\r B_\m(k,k+p) + 4 B_{\m\r}(k,k+p)   \right]  \nonumber\\
&-& \frac{ q_\m q^\r } { \L^2}  \left[ q_\n q_\r B_0(k,k+p) + 2 q_\n B_\r(k,k+p) + 2 q_\r B_\n(k,k+p) + 4 B_{\n\r}(k,k+p)   \right] \, ,
\eea
and its contracted version
\bea\label{M2A}
{\cal M}^2_A =  \frac{1}{3} \left( - \eta^{\m\n} + \frac{q^\m q^\n}{q^2} \right) {\cal M}^2_{A,\m\n}  =  \frac{ g_4^2 q^2}{3} ( 1 + \frac{ 2 q^2 } { \L^2}  )  B_0(k,k+p) - \frac{4 g_4^2}{3} ( 1 + \frac{ 2 q^2 } { \L^2}  ) \eta_{\m\n} B^{\m\n}(k,k+p) \, . \nonumber\\
\eea
Finally the complete, contracted, contribution of the current set of diagrams is
\bea\label{MAm}
{\cal M}_A &=& {\cal M}^1_A  + {\cal M}^2_A  \nonumber\\
 &=& \frac{2 g_4^2}{3}  (  1  + \frac{2 q^2} { \L^2}  ) A_0(0) +  \frac{ g_4^2 q^2}{3} ( 1 + \frac{ 2 q^2 } { \L^2}  )  B_0(k,k+p) \, ,\nonumber\\
\eea
where we took advantage of \eq{M2fio}.
The quantum corrections of the propagators, from which the anomalous dimensions of the fields will be constructed, are now finished.

The next step is to compute the corrections to the vertices.
Recall that \eq{SBHrd} contains the couplings $g_4$ and $c^{(6)}_1$ whose desired running will be revealed if we correct the vertices $\bar \phi$-$\phi$-$A^3_\m$ and $(\bar \phi \phi) \bar \phi \Box \phi$ respectively.
Nevertheless the explicit calculation of the 3-point vertex corrections is not necessary for the model which we are looking at.
In particular note that in SQED gauge invariance forces the overall counterterm of the 3-point vertex and the counterterm of the scalar field to be equal, at least at the divergent level, as it was shown in \cite{IrgesFotis2}.
Actually this is a well known fact in both scalar and regular QED due to the universality of the electromagnetic coupling. 
So for the action $S^{\rm b-h}$ which is an extended, with gauge invariant HDO, version of SQED there is no reason to expect a different conclusion.
Therefore, as it will be revealed in the renormalization program, the $\b$-function of $g_4$ is determined only by the gauge-field counter-term.
So in the following we focus on the quantum corrections of the four-scalar vertex.

Let us first perform a qualitative study of the possible one-loop contributions to the four point function, usually called Boxes. 
Recall that Box diagrams here are denoted collectively as ${\cal B}_{4,F}$ while they are separated into reducible 
and irreducible Boxes and there are three possible categories of the form
\bea
\begin{tikzpicture}[scale=0.7]
\draw [] (1,0)--(3,-1.5);
\draw [] (1,0)--(3,1.5);
\draw [] (0,0) circle [radius=1];
\draw [] (-3,-1.5)--(-1,0);
\draw [] (-3,1.5)--(-1,0);
\end{tikzpicture}
\, , \hskip .3 cm
\begin{tikzpicture}[scale=0.7]
\draw [] (0.9,-0.5)--(3,-1.5);
\draw [] (0.9,0.5)--(3,1.5);
\draw [] (0,0) circle [radius=1];
\draw [] (-3,-1.5)--(-1,0);
\draw [] (-3,1.5)--(-1,0);
\end{tikzpicture}
\, , \hskip .3 cm
\begin{tikzpicture}[scale=0.7]
\draw [] (0.9,-0.5)--(3,-1.5);
\draw [] (0.9,0.5)--(3,1.5);
\draw [] (-0.9,0.5)--(-3,1.5);
\draw [] (-0.9,-0.5)--(-3,-1.5);
\draw [] (0,0) circle [radius=1];
\end{tikzpicture}
\nonumber
\eea
corresponding to \textit{C-Boxes} (or \textit{Candies}), \textit{T-Boxes} and \textit{S-Boxes} respectively, using the conventions in \cite{IrgesFotis1}.
The above diagrams will contribute as quantum corrections to three processes regarding the $4 A^3_\m$ scattering, the $\bar \phi$-$\phi$-$A^3_\m$-$A^3_\m$ vertex and the four-scalar interaction.
Here we are interested in the latter case even though all the above corrections include divergencies. Actually these divergencies will be absorbed, through the renormalization procedure, from the counter-terms of the lower dimensional vertices.
In that sense the calculation starts with the set of $C$-Boxes which includes the diagrams ${\cal B}^{C,1}_{4,\phi}$, ${\cal B}^{C,2}_{4,\phi}$ and ${\cal B}^{C,3}_{4,\phi}$.
All of them come in two channels, $s$ and $t$ with 
\be
s=(p_1+p_2)^2 \,\,\,\, {\rm and} \,\,\,\, t=(p_1+p_3)^2 \, ,
\ee
while their symmetry factor is $S_{{\cal B}^{C,1}_{4,\phi}}=S_{{\cal B}^{C,2}_{4,\phi}}=S_{{\cal B}^{C,3}_{4,\phi}}=1$.
The reason for this is that the external legs are particle-antiparticle pairs so they cannot be interchanged.
So the $s$-channel of the first Candy-diagram is 
\vskip .5cm
\begin{center}
\begin{tikzpicture}[scale=0.7]
\draw [dashed] (0,0)--(1.8,0);
\draw [photon] (0,0.9) circle [radius=0.9];
\draw [dashed] (-1.8,0)--(0,0);
\node at (-1.7,-0.4) {$p_1$};
\node at (1.7,-0.4) {$p_3$};
\node at (0,1.4) {$k$};
\draw [dashed] (-1,-1)--(0,0);
\draw [dashed] (1,-1)--(0,0);
\draw [<-, very thick] (1.4,0)--(1.2,0);
\draw [->, very thick] (-1.4,0)--(-1.2,0);
\draw [<-, very thick]  (-0.75,-0.7)--(-0.55,-0.5);
\draw [<-, very thick]  (0.55,-0.5)--(0.75,-0.7);
\node at (-0.8,-1.4) {$p_2$};
\node at (0.8,-1.4) {$p_4$};
\node at (4,0) {$=  \hskip .1 cm i {\cal B}^{C,1s}_{4,\phi} $};
\end{tikzpicture}
\end{center}
with momentum conservation condition $p_1+p_2+p_3+p_4=0$.
This is a four-point Tadpole coming from a dim-6 operator and its explicit expression reads
\bea\label{Bc1s4fi}
i {\cal B}^{C,1s}_{4,\phi} &=& 2 i g_4^2 \frac{c^{(6)}_1 }{\L^2}S_{{\cal B}^{C,1}_{4,\phi}} \int{ \frac{d^4 k}{(2 \pi)^4}}  \frac{-i \eta^{\m\n}}{ k^2} \eta_{\m\n} \Rightarrow \nonumber\\
{\cal B}^{C,1s}_{4,\phi} &=& 2 d g_4^2 \frac{c^{(6)}_1}{\L^2}  \int \frac{d^4 k }{(2 \pi)^4 i }  \frac{1}{ k^2}   \nonumber\\
 &=&  8 g_4^2 \frac{c^{(6)}_1}{\L^2} A_0(0) \, ,
\eea
while adding the $t$-channel we get
\be\label{Bc14fi}
 {\cal B}^{C,1}_{4,\phi} =  {\cal B}^{C,1s}_{4,\phi}  +  {\cal B}^{C,1t}_{4,\phi} = 16 g_4^2 \frac{c^{(6)}_1}{\L^2} A_0(0) \, ,
\ee
since the diagrams are channel-independent. 
Next comes the $s$-channel of the second Candy-diagram given by 
\vskip .5cm
\begin{center}
\begin{tikzpicture}[scale=0.7]
\draw [dashed] (0.9,0)--(2.5,1.5);
\draw [dashed] (0.9,0)--(2.5,-1.5);
\draw [dashed] (0,0) circle [radius=0.9];
\draw [dashed] (-2.5,1.5)--(-0.9,0);
\draw [dashed] (-2.5,-1.5)--(-0.9,0);
\node at (2.7,1.1) {$p_3$};
\node at (2.7,-1.1) {$p_4$};
\node at (-2.5,1.1) {$p_1$};
\node at (-2.5,-0.9) {$p_2$};
\node at (0,1.4) {$k+P_1$};
\node at (0,-1.4) {$k$};
\draw [<-, very thick] (-0.1,0.9)--(0.1,0.9);
\draw [->, very thick] (-0.1,-0.9)--(0.1,-0.9);
\draw [<-, very thick ] (-1.65,0.65)--(-1.85,0.85);
\draw [->, very thick ] (-1.65,-0.65)--(-1.85,-0.85);
\draw [->, very thick ] (1.65,0.65)--(1.85,0.85);
\draw [<-, very thick ] (1.65,-0.65)--(1.85,-0.85);
\node at (5,0) {$=\, \,i {\cal B}^{C,2s}_{4,\phi} $ \, ,};
\end{tikzpicture}
\end{center}
with $p_1+p_2+p_3+p_4=0$ and $P_1=p_1+p_2$.
Its evaluation yields
\bea\label{Bc2s4fi}
i {\cal B}^{C,2s}_{4,\phi} &=& - \Bigl( c^{(6)}_1 \frac{p^2} {\L^2}  \Bigr)^2 S_{{\cal B}^{C,2}_{4,\phi}} \int{ \frac{d^4 k}{(2 \pi)^4}}  \frac{i }{ k^2} \frac{i }{ (k+P_1)^2} \Rightarrow \nonumber\\
{\cal B}^{C,2s}_{4,\phi} &=&   \Bigl( c^{(6)}_1 \frac{p^2} {\L^2}  \Bigr)^2 \int \frac{d^4 k }{(2 \pi)^4 i }  \frac{1}{ k^2 (k+P_1)^2}   \nonumber\\
 &=&   \Bigl( c^{(6)}_1 \frac{p^2} {\L^2}  \Bigr)^2  B_0(k,k+\sqrt{s}) 
\eea
and collectively we get
\be\label{Bc24fi}
 {\cal B}^{C,2}_{4,\phi} =  {\cal B}^{C,2s}_{4,\phi}  +  {\cal B}^{C,2t}_{4,\phi} =   \Bigl( c^{(6)}_1 \frac{p^2} {\L^2}  \Bigr)^2  B_0(k,k+\sqrt{s}) +  \Bigl( c^{(6)}_1 \frac{p^2} {\L^2}  \Bigr)^2  B_0(k,k+\sqrt{t}) \, ,
\ee
where now the diagrams, at least to their finite part, depend on the channels.
Moreover, notice that even though the factor in front of $B_0$'s seems that scales like $1/\L^4$ it will not be neglected.
The reason is that it corrects a vertex proportional to $ c^{(6)}_1 p^2/\L^2 $ and not just $c^{(6)}_1$, so throwing it away will cost us in completeness.
The last contribution of this category, which originates from the $\bar \phi$-$\phi$-$A^3_\m$-$A^3_\m$ vertex, is
\vskip .5cm
\begin{center}
\begin{tikzpicture}[scale=0.7]
\draw [dashed] (0.9,0)--(2.5,1.5);
\draw [dashed] (0.9,0)--(2.5,-1.5);
\draw [dashed] (-2.5,1.5)--(-0.9,0);
\draw [dashed] (-2.5,-1.5)--(-0.9,0);
\node at (2.7,1.1) {$p_3$};
\node at (2.7,-1.1) {$p_4$};
\node at (-2.5,1.1) {$p_1$};
\node at (-2.5,-0.9) {$p_2$};
\draw [photon] (0,0) circle [radius=0.9];
\node at (0,1.4) {$k+P_1$};
\node at (0,-1.4) {$k$};
\node at (-1,0.7) {$\m$};
\node at (-1,-0.6) {$\n$};
\node at (1,0.7) {$\r$};
\node at (1,-0.6) {$\s$};
\draw [<-, very thick ] (-1.65,0.65)--(-1.85,0.85);
\draw [->, very thick ] (-1.65,-0.65)--(-1.85,-0.85);
\draw [->, very thick ] (1.65,0.65)--(1.85,0.85);
\draw [<-, very thick ] (1.65,-0.65)--(1.85,-0.85);
\node at (5,0) {$=\, \,i {\cal B}^{C,3s}_{4,\phi} $ \, .};
\end{tikzpicture}
\end{center}
Again momentum conservation forces that $p_1+p_2+p_3+p_4=0$ while the $s$-channel of the above diagram gives
\bea\label{Bc3s4fi}
i {\cal B}^{C,3s}_{4,\phi} &=& - 4 g_4^4 S_{{\cal B}^{C,3}_{4,\phi}} \int{ \frac{d^4 k}{(2 \pi)^4}}  \frac{-i \eta^{\m\r} Q_{\m\n}(p,k) }{ k^2}  \frac{-i \eta^{\n\s} Q_{\r\s}(p,k) }{ (k+P_1)^2} \Rightarrow \nonumber\\
{\cal B}^{C,3s}_{4,\phi} &=& 4 g_4^4  \int \frac{d^4 k }{(2 \pi)^4 i }   \left[  \eta^{\m\n} + \frac{ \eta^{\m\n} p_1\cdot p_2 + k^\m(p_2 - p_1)^\n } {2 \L^2}  +  \frac{2 ( \eta^{\m\n} k^2- k^\m k^\n )} { \L^2}   \right] \frac{1}{ k^2(k+P_1)^2} \times \nonumber\\
 && \left[  \eta_{\m\n} + \frac{ \eta_{\m\n} p_3\cdot p_4 + k_\m(p_3 - p_4)_\n } {2 \L^2}  +  \frac{2 ( \eta_{\m\n} k^2- k_\m k_\n )} { \L^2}   \right]  \nonumber\\
 &=& 4 g_4^4  \left[  4 ( 1+ \frac{ p_1\cdot p_2 + p_3\cdot p_4 } {2 \L^2} ) B_0(k,k+\sqrt{s})  + ( p_1^2 + p_4^2 - p_3^2 - p_2^2 )\frac{ B_0(k,k+\sqrt{s})  } {4 \L^2} + \frac{  12 } {\L^2}  A_0(0)    \right]  \, . \nonumber\\
\eea
The corresponding calculation for the $t$-channel is obtained from the above relation with the exchange $2 \leftrightarrow 3$ and $s \to t $, so collectively we get
\bea\label{Bc34fi}
 {\cal B}^{C,3}_{4,\phi} &=&  {\cal B}^{C,3s}_{4,\phi} +  {\cal B}^{C,3t}_{4,\phi} \nonumber\\
 &=& 4 g_4^4 \Bigl[  4 ( 1+ \frac{ p_1\cdot p_2 + p_3\cdot p_4  } {2 \L^2} ) B_0(k,k+\sqrt{s}) + 4 ( 1+ \frac{  p_1\cdot p_3 + p_2\cdot p_4 } {2 \L^2} ) B_0(k,k+\sqrt{t}) \nonumber\\
 & +&  ( p_1^2 + p_4^2 - p_3^2 - p_2^2 )\frac{ B_0(k,k+\sqrt{s})  } {4 \L^2} + ( p_1^2 + p_4^2 - p_3^2 - p_2^2 )\frac{ B_0(k,k+\sqrt{t})  } {4 \L^2} +   \frac{ 24 } {\L^2}  A_0(0)    \Bigr] \, . \nonumber\\
\eea
Adding \eq{Bc14fi}, \eq{Bc24fi} and \eq{Bc34fi} we end up with the full contribution of the $C$-Boxes to the 4-point scalar vertex
\bea\label{Bc4fi}
 {\cal B}^{C}_{4,\phi} &=& 16 g_4^2 \left(  c^{(6)}_1 +  6 g_4^2  \right) \frac{A_0(0) }{\L^2} +  \Bigl( \frac{c^{(6)}_1 p^2}{\L^2} \Bigr)^2 \Bigl( B_0(k,k+\sqrt{s}) + B_0(k,k+\sqrt{t})  \Bigr)  \nonumber\\
&+& 4 g_4^4 \Bigl[  4 ( 1+ \frac{ p_1\cdot p_2 + p_3\cdot p_4  } {2 \L^2} ) B_0(k,k+\sqrt{s}) + 4 ( 1+ \frac{  p_1\cdot p_3 + p_2\cdot p_4 } {2 \L^2} ) B_0(k,k+\sqrt{t}) \nonumber\\
 & +&  ( p_1^2 + p_4^2 - p_3^2 - p_2^2 )\frac{ B_0(k,k+\sqrt{s})  } {4 \L^2} + ( p_1^2 + p_4^2 - p_3^2 - p_2^2 )\frac{ B_0(k,k+\sqrt{t})  } {4 \L^2}  \Bigr] \, . 
\eea

The next contribution to the four-scalar vertex refers to the $T$-Boxes and the associated set contains the channels $s$ and $t$ while for each channel there are two possible topologies.
Notice that the $T$-Boxes are determined by two pairs of two linear combinations of the external momenta, 
$(P_1,P_2)$ and $(P_A,P_B)$. A useful choice for these pairs for the channels $T_{1,\cdots,4}$ is the following
\bea\label{t1234}
T_1 &:& (p_1+p_2, p_1 +p_2 + p_3)\, , \hskip .5 cm (p_1 + p_2 +2 p_3, 
0) \nonumber\\
T_2 &:& (p_1+p_3, p_1 +p_3 + p_4)\, , \hskip .5 cm (p_1 + p_3 +2 p_4,0) \nonumber\\
T_3 &:& (p_2, p_1 +p_2)\, , \hskip .5 cm (2 p_2 ,p_1 + p_2) \nonumber\\
T_4 &:& (p_1, p_1 + p_3)\, , \hskip .5 cm (2p_1,p_1 +p_3 )\, ,
\eea
and the counting of the internal momenta starts always from the photon-field with assigned loop-momentum $k$ and it is clockwise. 
The symmetry factor for this topology is $S_{{\cal B}^{T_{1,\cdots,4}}_{4,\phi}} = 1$.
Here it is enough to evaluate just one channel since the summation over $T_{1,\cdots,4}$ gives the full contribution of the $T$-Boxes.
Then the first $s$-channel is
\vskip .5cm
\begin{center}
\begin{tikzpicture}[scale=0.7]
\draw [photon] (-1,0) -- (1,1);
\draw [photon] (-1,0) -- (1,-1);
\draw [dashed] (1,1) -- (1,-1);
\draw [dashed] (-2,1) -- (-1,0);
\draw [dashed] (-2,-1) -- (-1,0);
\node at (-1,0.5) {$\m$};
\node at (-1,-0.5) {$\n$};
\node at (1,1.5) {$\r$};
\node at (1,-1.5) {$\s$};
\draw [dashed] (1,1) -- (2.5,1);
\draw [dashed] (1,-1) -- (2.5,-1);
\node at (1.7,1.5) {$p_3$};
\node at (1.7,-1.5) {$p_4$};
\node at (-1.9,1.5) {$p_1$};
\node at (-1.9,-1.50) {$p_2$};
\node at (-0.2,1.3) {$k+P_1$ };
\node at (0,-0.9) {$k$};
\node at (2.1,0) {$k+P_2$};
\draw [<-, very thick ] (-1.65,0.65)--(-1.85,0.85);
\draw [->, very thick ] (-1.65,-0.65)--(-1.85,-0.85);
\draw [<-, very thick] (1.7,1) -- (1.5,1);
\draw [<-, very thick] (1.5,-1) -- (1.7,-1);
\draw [->, very thick] (1,-0.1) -- (1,0.1);
\node at (6.5,0) {$=\,\, i {\cal B}^{T,1s}_{4,\phi}  $ };
\end{tikzpicture}
\end{center}
and its explicit form is given by
\bea
i {\cal B}^{T,1s}_{4,\phi}  &=& (-i) 2 g_4^4  \int{ \frac{d^4 k}{(2 \pi)^4}}  \frac{-i \eta^{\m\r} Q_{\r}(P_j,P_l,k) }{ k^2 } \frac{-i \eta^{\n\s} Q_{\s}(P_j,P_l,k) }{ (k+P_1)^2 }  \frac{i Q_{\m\n}(P_j,P_l,k) }{ (k+P_2)^2 }   \Rightarrow \nonumber\\
{\cal B}^{T,1s}_{4,\phi}  &=& - 2 g_4^4  \int \frac{d^4 k }{(2 \pi)^4 i }  \frac{ Q^{\m}(P_j,P_l,k) Q^{\n}(P_j,P_l,k) Q_{\m\n}(P_j,P_l,k) }{ k^2 (k+P_1)^2 (k+P_2)^2 } \nonumber\\
 &=& - 2 g_4^4  \int \frac{d^4 k }{(2 \pi)^4 i }  \left[  (k + P_A)^\m +  \frac{ (k + P_A )^\m(k + P_1)^2 } { \L^2}  - \frac{ (k + P_A)\cdot (k + P_1) \, (k + P_1)^\m } { \L^2} \right] \times  \nonumber\\
 &&\left[  (k + P_B)^\n +  \frac{ (k + P_B )^\n k^2 } { \L^2}  - \frac{ (k + P_B)\cdot k \, k^\n } { \L^2} \right] \frac{ 1 }{ k^2 (k+P_1)^2 (k+P_2)^2 }  \times \nonumber\\
&& \left[  \eta_{\m\n} + \frac{ \eta_{\m\n} p_2 \cdot p_1 + k_{\m} ( p_2 - p_1 )_\n  } {2 \L^2}  +  \frac{2 ( \eta_{\m\n} k^2- k_\m k_\n )} { \L^2} \right]  \nonumber
\eea
with $j= 1,2$ and $l=A,B$.
Expanding the brackets, reducing the integrals and keeping terms up to ${\cal O}(1/\L^2)$, the above expression becomes
\bea
{\cal B}^{T,1s}_{4,\phi} &=& - 2 g_4^4  \Biggl [  (1 + \frac{p_1 \cdot p_2}{2\L^2}) B_0(k,k+P_2-P_1) - \frac{P_A \cdot P_2}{2\L^2} B_0(k,k+P_2)  +  \Bigl ( - P_1\cdot P_2 + \frac{ P_A \cdot (P_2 - P_1) } { 2}  \nonumber\\ 
&+& \frac{  P_A \cdot(p_2-p_1) } { 8} - \frac{  (P_2+ P_1) \cdot(p_2-p_1) } { 4}   \Bigr ) \frac{B_0(k,k+P_2-P_1)}{\L^2}  \nonumber\\
&+& 3 P_B \cdot \Bigl (  P_A  -  \frac{ P_1 } { 4}    \Bigr ) \frac{B_0(k,k+P_2-P_1)}{\L^2}  + [ {\cal B}^{T,1s}_{4,\phi} ]_f \Biggr ] \, ,
\eea
where $[ {\cal B}^{T,1s}_{4,\phi} ]_f$ includes all the reduced and finite integrals of the $C_0$ and $C_\m$ form.
Then the complete contribution of the $T$-Boxes is given by
\be\label{Bt4fi}
{\cal B}^{T}_{4,\phi} = \sum_{(P_A,P_B)} \sum_{(P_1,P_2)} {\cal B}^{T,1s}_{4,\phi} ( P_1,P_2,P_A,P_B ) \, .
\ee
Note that there are three terms which do not include exclusively the capital momenta, nevertheless, when we consider the $t$-channels the replacement $p_1 \cdot p_2 \to p_1 \cdot p_3$ and $(p_2-p_1) \to (p_3-p_1)$ should take place.

The last set of 1-loop diagrams, correcting the 4-scalar vertex, regards the $S$-Boxes.
Recall that there is only one channel here which however is given in two different topologies since there are two possible ways to arrange the propagators inside the loop.
In that sense the diagrams are determined by seven linear combinations of the external momenta, $P_1$, $P_2$, $P_3$, $P_A$, $P_B$, $P_C$ and $P_D$ which are defined in the following.
Therefore the diagram of the first topology is given by
\vskip .5cm
\begin{center}
\begin{tikzpicture}[scale=0.7]
\draw [dashed] (-1,1) -- (1,1);
\draw [photon] (1,1) -- (1,-1);
\draw [dashed] (1,-1) -- (-1,-1);
\draw [photon] (-1,-1) -- (-1,1);
\draw [dashed] (-2,2) -- (-1,1);
\draw [dashed] (-2,-2) -- (-1,-1);
\draw [dashed] (2,2) -- (1,1);
\draw [dashed] (2,-2) -- (1,-1);
\draw [->, very thick] (-0.2,1) -- (0,1);
\draw [<-, very thick] (-0.2,-1) -- (0,-1);
%
\node at (-1.4,0.9) {$\m$};
%
\node at (-1.4,-0.9) {$\n$};
%
\node at (1.4,0.9) {$\r$};
%
\node at (1.4,-0.9) {$\s$};
%
\node at (-2.3,2.2) {$p_1$};
\node at (-2.3,-2.2) {$p_2$};
\node at (2.3,2.2) {$p_3$};
\node at (2.3,-2.2) {$p_4$};
%
\node at (0,1.5) {$k+P_1$ };
\node at (-1.3,0) {$k $};
\node at (0,-1.4) {$k +P_3$};
\node at (2,0) {$k+ P_2$};
\draw [<-, very thick] (-1.5,1.5) -- (-1.7,1.65);
\draw [->, very thick] (1.5,1.5) -- (1.7,1.65);
\draw [->, very thick] (-1.5,-1.5) -- (-1.7,-1.65);
\draw [<-, very thick] (1.5,-1.5) -- (1.7,-1.65);\nonumber
\node at (5,0) {$=\,\, i {\cal B}^{S,1}_{4,\phi} \, ,$};
\end{tikzpicture}
\end{center}
with symmetry factor $S_{{\cal B}^{S,1}_{4,\phi}} = 1$ and momentum conservation $p_1+p_2+p_3+p_4=0$.
Starting the counting of the loop momenta from the gauge propagator a useful choice for the $P$'s reads
\bea
&&P_1 = p_1, \,\,\,\, P_2 =  p_1 + p_3 \,\,\,\, {\rm and} \,\,\,\, P_3 = p_1 + p_3 + p_4 \nonumber\\
&&P_A = 2 p_1, \,\,\,\, P_B = 0, \,\,\,\, P_C = p_1+p_3=P_2  \,\,\,\, {\rm and} \,\,\,\, P_D =  p_1+p_3+2p_4 \, ,
\eea
then its explicit form reads
\bea
i {\cal B}^{S,1}_{4,\phi} &=&  g_4^4  \int{ \frac{d^4 k}{(2 \pi)^4}}  \frac{-i \eta^{\m\n} Q_{\m}(P_j,P_l,k) Q_{\n}(P_j,P_l,k) }{ k^2 } \frac{i  }{ (k+P_1)^2 }  \frac{-i \eta^{\r\s} Q_{\r}(P_j,P_l,k) Q_{\s}(P_j,P_l,k) }{ (k+P_2)^2 } \frac{i }{ (k+P_3)^2 }    \Rightarrow \nonumber\\
{\cal B}^{S,1}_{4,\phi} &=& g_4^4  \int \frac{d^4 k }{(2 \pi)^4 i }  \frac{ \eta^{\m\n} Q_{\m}(P_j,P_l,k) Q_{\n}(P_j,P_l,k) \eta^{\r\s} Q_{\r}(P_j,P_l,k) Q_{\s}(P_j,P_l,k) }{ k^2 (k+P_1)^2 (k+P_2)^2 (k+P_3)^2 } \nonumber\\
 &=&  g_4^4  \int \frac{d^4 k }{(2 \pi)^4 i }  \eta^{\m\n} \left[ (k + P_A)_\m   + \frac{(k + P_A )_\m k^2 - (k + P_A)\cdot k \, k_\m } { \L^2}  \right] \frac{ k_\n }{ k^2 (k+P_1)^2 (k+P_2)^2 (k+P_3)^2 }  \times \nonumber\\
 && \eta^{\r\s}  \left[ (k + P_C)_\r   + \frac{(k + P_C )_\r (k + P_2 )^2 - (k + P_C)\cdot (k + P_2 ) \, (k + P_2 )_\r } { \L^2}  \right] \times \nonumber\\
 && \left[ (k + P_D)_\s   + \frac{(k + P_D )_\s (k + P_2 )^2 - (k + P_D)\cdot (k + P_2 ) \, (k + P_2 )_\s } { \L^2}   \right]  \nonumber\\
 &=&  g_4^4  \int \frac{d^4 k }{(2 \pi)^4 i }  \frac{ k\cdot (k+P_A) \Bigl [ (k+P_2)\cdot (k+P_D)  \Bigr] }{ k^2 (k+P_1)^2 (k+P_2)^2 (k+P_3)^2 } \, , \nonumber
\eea
where here $j= 1,2,3$ and $l=A,B,C,D$.
Notice that for this specific type of diagrams the contribution of ${\cal O}(1/\L^2)$ terms is canceled and the result is the usual one in SQED.
So performing the products and reducing the integrals we get, similarly with the Appendix of \cite{IrgesFotis2}, that
\bea
 {\cal B}^{S,1}_{4,\phi} &=& g_4^4 \Bigl [ B_0(k,k+P_3-P_2) - 2P_1^\m C_\m(k,k+P_2-P_1,k+P_3-P_1) \nonumber\\
 &+& \Bigl (  P_D + P_2 + P_A  \Bigr )^\m C_\m(k+P_1,k+P_2,k+P_3) + P_1^2 C_0(k,k+P_2-P_1,k+P_3-P_1) \nonumber\\
 &+& \Bigl (  \eta^{\m\n} P_2\cdot P_D  + P_A^\m ( P_2 + P_D )^\n    \Bigr ) D_{\m\n}(k,k+P_1,k+P_2,k+P_3) \nonumber\\
 &+& P_2\cdot P_D \, P_A^\m D_{\m}(k,k+P_1,k+P_2,k+P_3)  \Bigr] \, .
\eea
Now the diagram of the second topology is
\vskip .5cm
\begin{center}
\begin{tikzpicture}[scale=0.7]
\draw [photon] (-1,1) -- (1,1);
\draw [dashed] (1,1) -- (1,-1);
\draw [photon] (1,-1) -- (-1,-1);
\draw [dashed] (-1,-1) -- (-1,1);
\draw [dashed] (-2,2) -- (-1,1);
\draw [dashed] (-2,-2) -- (-1,-1);
\draw [dashed] (2,2) -- (1,1);
\draw [dashed] (2,-2) -- (1,-1);
\draw [<-, very thick] (-1,0) -- (-1,0.2);
\draw [->, very thick] (1,0) -- (1,0.2);
%
\node at (-1.4,0.9) {$\m$};
%
\node at (-1.4,-0.9) {$\n$};
%
\node at (1.4,0.9) {$\r$};
%
\node at (1.4,-0.9) {$\s$};
%
\node at (-2.3,2.2) {$p_1$};
\node at (-2.3,-2.2) {$p_2$};
\node at (2.3,2.2) {$p_3$};
\node at (2.3,-2.2) {$p_4$};
%
\node at (0,1.5) {$k$ };
\node at (-2,0) {$k+P_3 $};
\node at (0,-1.5) {$k +P_2$};
\node at (2.1,0) {$k+ P_1$};
\draw [<-, very thick] (-1.5,1.5) -- (-1.7,1.65);
\draw [->, very thick] (1.5,1.5) -- (1.7,1.65);
\draw [->, very thick] (-1.5,-1.5) -- (-1.7,-1.65);
\draw [<-, very thick] (1.5,-1.5) -- (1.7,-1.65);\nonumber
\node at (5,0) {$=\,\, i {\cal B}^{S,2}_{4,\phi} \, ,$};
\end{tikzpicture}
\end{center}
with symmetry factor $S_{{\cal B}^{S,2}_{4,\phi}} = 1$ and momentum conservation $p_1+p_2+p_3+p_4=0$.
Following our loop-momentum counting a useful choice would be
\bea
&&P_1 = p_3, \,\,\,\, P_2 =  p_3 + p_4 \,\,\,\, {\rm and} \,\,\,\, P_3 = p_2 + p_3 + p_4 \nonumber\\
&&P_A = 0, \,\,\,\, P_B = 2 p_2+p_3+p_4, \,\,\,\, P_C = 2p_3=2 P_1  \,\,\,\, {\rm and} \,\,\,\, P_D =  p_3+p_4 = P_2 
\eea
and hence the above diagram gives directly
\bea
 {\cal B}^{S,2}_{4,\phi} &=& g_4^4 \Bigl [ B_0(k,k+P_3-P_2) - 2P_1^\m C_\m(k,k+P_2-P_1,k+P_3-P_1) \nonumber\\
 &+& \Bigl (  P_C + P_2 + 2P_1  \Bigr )^\m C_\m(k+P_1,k+P_2,k+P_3) + P_1^2 C_0(k,k+P_2-P_1,k+P_3-P_1) \nonumber\\
 &+& \Bigl (  \eta^{\m\n} P_2\cdot P_C  + 2P_1^\m ( P_2 + P_C )^\n    \Bigr ) D_{\m\n}(k,k+P_1,k+P_2,k+P_3) \nonumber\\
 &+& 2 P_2\cdot P_C \, P_1^\m D_{\m}(k,k+P_1,k+P_2,k+P_3) \Bigl ] \, .
\eea
Finally, the complete $S$-Box contribution is given by
\be\label{Bs4fi}
{\cal B}^{S}_{4,\phi}  =  {\cal B}^{S,1}_{4,\phi} (P_j,P_{j'})  +  {\cal B}^{S,2}_{4,\phi}(P_j,P_{j'}) \, ,
\ee
with $j=1,2,3$ and $j'=A,B,C,D$, while the full 1-loop correction to the $(\bar \phi \phi)^2$ vertex yields
\be\label{B4fi}
{\cal B}_{4,\phi} = {\cal B}^{C}_{4,\phi} + {\cal B}^{T}_{4,\phi} + {\cal B}^{S}_{4,\phi} \, .
\ee 
All the needed information to start the renormalization procedure for the boundary action, $S^{\rm b-h}$, is now in our hands and 
is contained in \eq{Mfi}, \eq{MAm} and \eq{B4fi}.
In what follows we use these equations to extract the counter-terms and the $\b$-functions of the associated couplings.

\subsection{Renormalization and $\b$-functions}\label{Renorm.bf}

Regarding the Boundary-Hybrid action given in \eq{SBHrd}, there is only one step left towards the calculation of the $\b$-functions and of the RG flows.
This step refers to the renormalization procedure, according to which the divergencies cancel after inserting appropriate counter-terms in the action.
An important point is the following:
Recall from \sect{QuCo} that every 1-loop diagram that we faced was reduced to scaleless integrals due to the absence of explicit mass terms in the classical Lagrangian.
Since our scheme is DR, these integrals in general vanish, however here we distinguish three cases following the Appendices of \cite{IrgesFotis2, FotisLetter2}.
According to those the scalar integrals of $A_0$-type are forced to vanish in DR while, for the $B_0$-type we have the freedom to separate the UV from the IR divergence.
So on our road to the $\b$-functions $A_0$'s will be absent and $B_0$'s will contribute a $2/\ve$ divergent term.
The third case refers to the finite integrals which, when scaleless, take a $0/0$ form and they can be replaced by arbitrary constants.

We now proceed with the renormalization of the couplings $g_{4,0}$ and $c^{(6)}_{1,0}$ as well as of the fields $A^3_{\m,0}$ and $\phi_0$.  
The subscript $0$ indicates bare quantities.
Regarding the gauge coupling recall that it contains $g_0$ and the anisotropy factor $\g_0$ which in principle both get renormalized.
Then their contribution will be hidden in the counter-term/$\b$-function of $g_{4,0}$.
Moreover, to simplify expressions, the alternative dimensionless in $d$-dimensions
coupling
\be\label{a40}
\a_{4,0} \equiv  \frac{1}{(4 \pi)^2} \m^{d-4} g_{4,0}^2
\ee 
can be considered. This and its 5d analogue will be used extensively in the following.
The counter-term for the gauge coupling is given by 
\bea\label{g40c}
g_{4,0} &=& g_4 + \d g_4 \Rightarrow \nonumber\\
g_{4,0} &=& g_4 ( 1 + \frac{\d g_4}{g_4} ) = ( 1+ \d_{g_4} ) g_4 = Z_{g_4} g_4 
\eea
while for the scalar self-coupling by
\bea\label{c610c}
c^{(6)}_{1,0} &=&c^{(6)}_1 + \d c^{(6)}_1  \nonumber\\
 &=& c^{(6)}_1 ( 1 + \frac{\d c^{(6)}_1}{c^{(6)}_1} ) = ( 1+ \d_{c^{(6)}_1} ) c^{(6)}_1 = Z_{c^{(6)}_1} c^{(6)}_1 \, .
\eea
In $d$-dimensions the scale independence relations of the bare couplings
\bea 
\m \frac{d g_{4,0} }{d \m } &=& \m \frac{d ( 4 \pi \m^{ \frac{4-d}{2} } \sqrt{\a_{4,0}} ) }{d \m } = 0 \label{dg4dm} \\  
\m \frac{d c^{(6)}_{1,0} }{d \m } &=& 0  \label{dc6dm}
\eea
generate the $\b$-function equations.
Similarly, for the anomalous dimensions of the fields we define
\bea
\phi_0  &=& \sqrt{Z_\phi} \phi = \sqrt{1 + \d_\phi} \phi \label{fi0fi} \\
A^3_{\m,0}  &=& \sqrt{Z_A} A^3_\m = \sqrt{1 + \d_A} A^3_\m \, . \label{Am0Am} 
\eea
The next step is to apply the above relations to our Lagrangian given by
\bea\label{Lb-h}
{\cal L}^{\rm b-h}_0 &=& - \frac{1}{4} F^3_{\m\n,0} F_0^{3,\m\n} + \frac{1}{2\xi} A^3_{\m,0} \pa^\m \pa_\n A_0^{3,\n} - \bar \phi_0 \Box \phi_0  - \frac{c_{1,0}^{(6)}}{4\L^2}  (\bar \phi_0 \phi_0) \bar \phi_0 \Box \phi_0 - \bar c_0^3 \Box c_0^3  \nonumber\\
&+& i g_{4,0}  \Bigl \{  \eta_{\m\r} - \frac{  \eta_{\m\r} \Box - \pa_\m \pa_\r  }{\L^2} \Bigr \} A^{3,\r}_0 \Bigl(  \bar \phi_0 \pa^\m \phi_0 - \phi_0 \pa^\m \bar \phi_0 \Bigr )  +  g^2_{4,0} (A^3_{\m,0})^2 \bar \phi_0 \phi_0   \nonumber\\
&+& \frac{  g^2_{4,0} }{2 \L^2}  \Bigl ( A^3_{\m,0} A^3_{\r,0} \pa^\r \bar \phi_0  \pa^\m \phi_0 + A^3_{\m,0} \pa^\r A^3_{\r,0} \pa^\m ( \bar \phi_0  \phi_0) \Bigr ) -  2 g^2_{4,0} \frac{ A^{3,\m}_0 ( \eta_{\m\r} \Box - \pa_\m \pa_\r ) A^{3,\r}_0   }{\L^2} \bar \phi_0 \phi_0 \nonumber\\
&+&  i \frac{ g_{4,0} \, c_{1,0}^{(6)} }{4 \L^2}A^3_{\m,0} \bar \phi_0 \phi_0 \Bigl(  \bar \phi_0 \pa^\m \phi_0 - \phi_0 \pa^\m \bar \phi_0 \Bigr ) + \frac{g^2_{4,0} \, c_{1,0}^{(6)} }{4 \L^2} (A^3_{\m,0})^2 ( \bar \phi_0 \phi_0 )^2  + {\cal L}^{\rm b-h}_l
\eea
where $ {\cal L}^{\rm b-h}_l$ corresponds to the 1-loop corrections.
Then substituting \eq{g40c}, \eq{c610c}, \eq{fi0fi} and \eq{Am0Am} into \eq{Lb-h} we obtain
\bea
{\cal L}^{\rm b-h}_0  = {\cal L}^{\rm b-h} + {\cal L}^{\rm b-h}_{\rm count.} +  {\cal L}^{\rm b-h}_{\rm correc.} \equiv {\cal L}^{\rm b-h} + {\cal L}^{\rm b-h}_{\rm 1-loop}  \nonumber
\eea 
with $ {\cal L}^{\rm b-h}$ the renormalized Lagrangian
\bea\label{Lb-hr}
{\cal L}^{\rm b-h} &=& - \frac{1}{4} F^3_{\m\n} F^{3,\m\n} + \frac{1}{2} A^3_{\m} \pa^\m \pa_\n A^{3,\n} - \bar \phi \Box \phi  - \frac{c_1^{(6)}}{4\L^2}  (\bar \phi \phi) \bar \phi \Box \phi - \bar c^3 \Box c^3  \nonumber\\
&+& i g_4  \Bigl \{  \eta_{\m\r} - \frac{  \eta_{\m\r} \Box - \pa_\m \pa_\r  }{\L^2} \Bigr \} A^{3,\r} \Bigl(  \bar \phi \pa^\m \phi - \phi \pa^\m \bar \phi \Bigr )  +  g^2_4 (A^3_{\m})^2 \bar \phi \phi   \nonumber\\
&+& \frac{  g^2_4 }{2 \L^2}  \Bigl ( A^3_{\m} A^3_{\r} \pa^\r \bar \phi  \pa^\m \phi + A^3_{\m} \pa^\r A^3_{\r} \pa^\m ( \bar \phi  \phi) \Bigr ) -  2 g^2_4 \frac{ A^{3,\m} ( \eta_{\m\r} \Box - \pa_\m \pa_\r ) A^{3,\r}  }{\L^2} \bar \phi \phi \nonumber\\
&+&  i \frac{ g_4 \, c_1^{(6)} }{4 \L^2}A^3_{\m} \bar \phi \phi \Bigl(  \bar \phi \pa^\m \phi - \phi \pa^\m \bar \phi \Bigr ) + \frac{g^2_4 \, c_1^{(6)} }{4 \L^2} (A^3_{\m})^2 ( \bar \phi \phi )^2 
\eea
and ${\cal L}^{\rm b-h}_{\rm 1-loop}$ the finite 1-loop Lagrangian which in momentum space becomes
\bea\label{L1-loop}
{\cal L}^{\rm b-h}_{\rm 1-loop} &=& \frac{1}{2} \Bigl \{ - \d_A \eta^{\m\n} q^2 + {\cal M}^{\m\n}_A  \Bigr \} A^3_\m A^3_\n + \Bigl \{ \d_\phi p_1^2 + {\cal M}_{\phi}  \Bigr \} \phi \bar \phi + \Bigl \{ ( \d c_1^{(6)}  + 2 c_1^{(6)} \d_\phi ) \frac{p^2}{\L^2} - {\cal B}_{4,\phi} \Bigr \} \frac{ (\bar \phi \phi)^2}{4} \nonumber\\
&+& \Bigl \{g_4 \d_3 \left( (p_1+p_2)_\m +  \frac{(p_1+p_2)_\m q^2 - (p_1+p_2)\cdot q \, q_\m }{\L^2} \right) + {\cal K}_{\bar \phi A \phi,\m}  \Bigr \} A^3_\m \bar \phi \phi \nonumber\\
&+& \Bigl \{ \d_4 g_4^2 \left( \eta^{\m\n} + \frac{ \eta^{\m\n} p_1 \cdot p_2 + q^\m_1 ( p_2 - p_1 )^\n  } {2 \L^2}  +  \frac{2 ( \eta^{\m\n} q_1^2- q^\m_1 q^{\n}_1 )} { \L^2} \right) +   {\cal B}^{\m\n}_{A \bar \phi A \phi}  \Bigr \} A_\m^3 A_\n^3 \phi \bar \phi \nonumber\\
&+& \Bigl \{ \d_5 \frac{ g_4 \, c_1^{(6)} }{ \L^2} (p_1+p_2)^\m +  {\cal B}^\m_{5, A (\bar \phi \phi )^2 }  \Bigr \} \frac{A_\m^3 ( \phi \bar \phi )^2}{4} +  \Bigl \{ \d_6 \frac{ g_4^2 \, c_1^{(6)} }{ \L^2} \eta^{\m\n} +  {\cal B}^{\m\n}_{6,A A ( \bar \phi \phi )^2 }  \Bigr \} \frac{A_\m^3 A_\n^3 (\phi \bar \phi)^2}{4} \, . \nonumber\\
\eea
Note that there is no counter-term for the Faddeev-Popov ghosts since they are decoupled from the theory. 

For the gauge-scalar three- and four-point vertices the following relations hold:
\bea\label{d3}
Z_3 &=& Z_{g_4}  Z_\phi \sqrt{Z_{A}}  \Rightarrow \nonumber\\
1 + \d_3 &=& ( 1 + \d_{g_4} ) ( 1 + \d_\phi ) ( 1 + \frac{1}{2} \d_{A} ) \Rightarrow \nonumber\\
\d_3 &=& \d_{g_4} + \d_\phi + \frac{1}{2} \d_{A} \, ,
\eea
and
\bea\label{d4}
Z_4 &=& Z_{g_4}^2  Z_\phi Z_{A}  \Rightarrow \nonumber\\
1 + \d_4 &=& ( 1 + 2 \d_{g_4} ) ( 1 + \d_\phi ) ( 1 + \d_{A} ) \Rightarrow \nonumber\\
\d_4 &=& 2 \d_{g_4} + \d_\phi + \d_{A} 
\eea
respectively, while for the five- and six-point vertices we get:
\bea\label{d5}
Z_5 &=& Z_{g_4} Z_{c_1^{(6)}} Z^2_\phi \sqrt{Z_{A}}  \Rightarrow \nonumber\\
1 + \d_5 &=& ( 1 + \d_{g_4} )  ( 1 + \d_{c_1^{(6)}}) ( 1 + 2 \d_\phi ) ( 1 + \frac{1}{2} \d_{A} ) \Rightarrow \nonumber\\
\d_5 &=& \d_{g_4}  + \d_{c_1^{(6)}} + 2\d_\phi + \frac{1}{2} \d_{A} \, ,
\eea
and
\bea\label{d6}
Z_6 &=& Z^2_{g_4} Z_{c_1^{(6)}} Z^2_\phi Z_{A}  \Rightarrow \nonumber\\
1 + \d_6 &=& ( 1 + 2 \d_{g_4} )  ( 1 + \d_{c_1^{(6)}}) ( 1 + 2 \d_\phi ) ( 1 +  \d_{A} ) \Rightarrow \nonumber\\
\d_6 &=& 2 \d_{g_4}  + \d_{c_1^{(6)}} + 2\d_\phi + \d_{A} 
\eea
respectively.
The Feynman rules for the counter-terms deriving from \eq{L1-loop} are
\begin{itemize}
\item Gauge boson 2-point function
\begin{center}
\begin{tikzpicture}
\draw[photon] (-3,0)--(-0.5,0) ;
\draw [thick] [fill=black] (-1.7,0) circle [radius=0.1];
\node at (2,0) {$=\displaystyle
 - i \d_A \eta_{\m\n} q^2$};
\end{tikzpicture}
\end{center}
\item Scalar 2-point function
\begin{center}
\begin{tikzpicture}
\draw[dashed] (-1,0)--(1.5,0);
\draw [<-, very thick] (1.1,0)--(0.9,0);
\draw [->, very thick] (-0.6,0)--(-0.4,0);
\draw [thick] [fill=black] (0.25,0) circle [radius=0.1];
\node at (4,0) {$=\displaystyle i \d_\phi p_1^2 $};
\end{tikzpicture}
\end{center}
\item The $A_\m$-$\phi$-$\bar \phi$ vertex counter-term
\begin{center}
\begin{tikzpicture}[scale=0.7]
\draw [dashed] (-2.5,1.5)--(-1,0);
\draw [dashed] (-2.5,-1.5)--(-1,0);
\draw[photon] (-1,0)--(1,0);
\draw [thick] [fill=black] (-1,0) circle [radius=0.1];
\draw [<-, very thick] (-1.5,0.55) -- (-1.7,0.75);
\draw [->, very thick] (-1.5,-0.55) -- (-1.7,-0.75);
\node at (4,0) {$=  \displaystyle i g_4 \d_3 Q_\m(p,q) $};
\end{tikzpicture}
\end{center}
\item The $(\bar \phi \phi)^2$ vertex counter-term
\begin{center}
\begin{tikzpicture}[scale=0.7]
\draw [dashed] (0,0)--(1.5,1.4);
\draw [dashed] (0,0)--(1.5,-1.4);
\draw [dashed] (-1.5,1.4)--(0,0);
\draw [dashed] (-1.5,-1.4)--(0,0);
\draw [thick] [fill=black] (0,0) circle [radius=0.1];
\draw [->, very thick]  (0.7,0.7)--(0.8,0.8);
\draw [<-, very thick]  (0.7,-0.7)--(0.8,-0.8);
\draw [->, very thick] (-0.9,-0.9)--(-0.7,-0.7);
\draw [<-, very thick] (-0.7,0.7)--(-0.5,0.5);
\node at (4.6,0) {$= \hskip .1 cm \displaystyle  i ( \d c_1^{(6)} + 2 c_1^{(6)} \d_\phi ) \frac{p^2} {\L^2} \, .$};
\end{tikzpicture}
\end{center}
\item The $A_\m$-$A_\n$-$\phi$-$\bar \phi$ vertex counter-term
\begin{center}
\begin{tikzpicture}[scale=0.7]
\draw [dashed,thick] (0,0)--(1.5,1.4);
\draw [dashed,thick] (0,0)--(1.5,-1.4);
\draw [photon] (-1.5,1.4)--(0,0);
\draw [photon] (-1.5,-1.4)--(0,0);
\draw [thick] [fill=black] (0,0) circle [radius=0.1];
\draw [<-, very thick] (0.65,0.55) -- (0.85,0.75);
\draw [->, very thick] (0.65,-0.55) -- (0.85,-0.75);
\node at (4,0) {$=\displaystyle 2 i \d_4 g_4^2 Q_{\m\n}(p,q) $};
\end{tikzpicture}
\end{center}
\item Four-scalars one-photon vertex counter-term 
\begin{center}
\begin{tikzpicture}[scale=0.7]
\draw [dashed] (0,0)--(1.5,1.4);
\draw [dashed] (0,0)--(1.5,-1.4);
\draw [dashed] (-1.5,1.4)--(0,0);
\draw [dashed] (-1.5,-1.4)--(0,0);
\draw [photon] (0.02,1.4)--(0.02,0);
\draw [thick] [fill=black] (0,0) circle [radius=0.1];
\draw [->, very thick]  (0.7,0.7)--(0.8,0.8);
\draw [<-, very thick]  (0.7,-0.7)--(0.8,-0.8);
\draw [->, very thick] (-0.9,-0.9)--(-0.7,-0.7);
\draw [<-, very thick] (-0.7,0.7)--(-0.5,0.5);
\node at (4.6,0) {$= \hskip .1 cm \displaystyle i \d_5 g_4 \, c_1^{(6)} \frac{  (p_1 + p_2)_\m }{\L^2}  \, .$};
\end{tikzpicture}
\end{center}
\item Four-scalars two-photons vertex counter-term
\begin{center}
\begin{tikzpicture}[scale=0.7]
\draw [dashed] (0,0)--(1.5,1.4);
\draw [dashed] (0,0)--(1.5,-1.4);
\draw [dashed] (-1.5,1.4)--(0,0);
\draw [dashed] (-1.5,-1.4)--(0,0);
\draw [photon] (0.02,-1.4)--(0.02,0);
\draw [photon] (0.02,1.4)--(0.02,0);
\draw [->, very thick]  (0.7,0.7)--(0.8,0.8);
\draw [<-, very thick]  (0.7,-0.7)--(0.8,-0.8);
\draw [->, very thick] (-0.9,-0.9)--(-0.7,-0.7);
\draw [<-, very thick] (-0.7,0.7)--(-0.5,0.5);
\draw [thick] [fill=black] (0,0) circle [radius=0.1];
\node at (4.2,0) {$= \hskip .1 cm \displaystyle  2 i \d_6 \, g^2_4 \frac{ c_1^{(6)} }{\L^2} \eta_{\m\n} \, .$};
\end{tikzpicture}
\end{center}
\end{itemize}
The renormalization conditions needed to make the theory finite at 1-loop are in order. 
For the gauge boson propagator, diagrammatically, we have that  
\be
\begin{tikzpicture} [scale=0.9]
\draw [photon,thick] (-2.3,0)--(-1.2,0);
\draw [thick] [fill=gray] (-0.5,0) circle [radius=0.8];
\draw [photon,thick] (0.3,0)--(1.3,0);
\node at (2,0) {$+$};
\draw [photon,thick] (2.5,0)--(3.6,0);
\draw [thick] [fill=black] (3.7,0) circle [radius=0.1];
\draw [photon,thick] (3.8,0)--(4.9,0);
\node at (6,0) {$=\, \, 0 \, .$};
\end{tikzpicture}
\nonumber
\ee
This implies that the contracted gauge propagator satisfies
\bea\label{RCAm}
- \frac{1}{3} \left(\eta^{\m\n} -\frac{q^\m q^\n}{q^2}   \right) ( -  \d_A \eta_{\m\n} q^2) + {\cal M}_A(q^2) = 0 \, . 
\eea
The second condition demands that 
\be
\begin{tikzpicture} [scale=0.9]
\draw [dashed] (-2.3,0)--(-1.2,0);
\draw [thick] [fill=gray] (-0.5,0) circle [radius=0.8];
\draw [dashed] (0.3,0)--(1.3,0);
\draw [<-, very thick] (1,0)--(0.8,0);
\draw [->, very thick] (-1.8,0)--(-1.6,0);
\node at (2,0) {$+$};
\draw [<-, very thick] (3.3,0)--(3.1,0);
\draw [->, very thick] (4.2,0)--(4.4,0);
\draw [dashed] (2.5,0)--(3.6,0);
\draw [thick] [fill=black] (3.7,0) circle [radius=0.1];
\draw [dashed] (3.8,0)--(4.9,0);
\node at (5.8,0) {$=\, \, 0 $};
\end{tikzpicture}
\nonumber
\ee
which, as equation, reads
\bea\label{RCfi}
\d_\phi p_1^2 + {\cal M}_{\phi}(p^2_1) = 0 \, .
\eea 
Finally, the last condition refers to the four-scalar vertex and demands that 
\be
\begin{tikzpicture} [scale=0.9]
\draw [dashed] (0,0)--(1.5,1.4);
\draw [dashed] (0,0)--(1.5,-1.4);
\draw [dashed] (-1.5,1.4)--(0,0);
\draw [dashed] (-1.5,-1.4)--(0,0);
\draw [->, very thick]  (0.7,0.7)--(0.8,0.8);
\draw [<-, very thick]  (0.7,-0.7)--(0.8,-0.8);
\draw [->, very thick] (-0.9,-0.9)--(-0.7,-0.7);
\draw [<-, very thick] (-0.7,0.7)--(-0.5,0.5);
\draw [thick] [fill=gray] (0.0,0) circle [radius=0.6];
\node at (1.9,0) {$+$};
\draw [dashed] (3.5,0)--(4.8,1.2);
\draw [dashed] (3.5,0)--(4.8,-1.2);
\draw [dashed] (2.2,1.2)--(3.5,0);
\draw [dashed] (2.2,-1.2)--(3.5,0);
\draw [->, very thick]  (4.25,0.7)--(4.35,0.8);
\draw [<-, very thick]  (4.25,-0.7)--(4.35,-0.8);
\draw [->, very thick] (3,0.5)--(2.8,0.7);
\draw [<-, very thick] (3,-0.5)--(2.8,-0.7);
\draw [thick] [fill=black] (3.5,0) circle [radius=0.1];
\node at (6,0) {$=\,\,  {0}$};
\end{tikzpicture}
\nonumber
\ee
or
\bea\label{RC4fi}
( \d c_1^{(6)}  + 2 c_1^{(6)} \d_\phi ) \frac{p^2}{\L^2} - {\cal B}_{4,\phi}(p^2)  = 0\, . 
\eea

Now for the determination of the counterterms we need to evaluate the 1-loop diagrams, obtained in \sect{QuCo}, which here is done with the help of dimensional regularization.
Let us start with the condition of the vacuum polarization which includes $ {\cal M}_A$ whose complete contribution is given in \eq{MAm}.
In DR this becomes 
\be\label{DRMA}
{\cal M}_A = \frac{1}{16 \pi^2} \Bigl [ \frac{2 q^2 g_4^2}{3} \frac{1}{\ve} + \frac{4 q^4 g_4^2 }{3\L^2} \frac{1}{\ve}  \Bigr ] + [ {\cal M}_A]_f
\ee
and as a consequence \eq{RCAm} gives
\be\label{cdA}
\d_A = \frac{1}{16 \pi^2} \Bigl [ - \frac{2 g_4^2}{3} \frac{1}{\ve} - \frac{4 q^2 g_4^2}{3\L^2} \frac{1}{\ve} \Bigr ] \, .
\ee
Next consider the scalar propagator whose 1-loop correction is given in \eq{Mfi} and in DR reads 
\be
{\cal M}_\phi = \frac{1}{16 \pi^2} \Bigl [ - 4 p_1^2 g_4^2 ( 1 - \frac{ p_1^2 }{\L^2} )\frac{1}{\ve} +   \Bigr ] +  [{\cal M}^{\phi A_\m}_\phi ]_f \, . 
\ee
Substituting this back to the condition \eq{RCfi} we get
\be\label{cdfi}
\d_\phi = \frac{1}{16 \pi^2} \Bigl [  4 g_4^2 \frac{1}{\ve} -  \frac{ 4  g_4^2 p_1^2 }{\L^2} \frac{1}{\ve}\Bigr ] \, .
\ee
As we have already mentioned, due to gauge invariance the relation $\d_\phi = \d_3$ holds.
So using \eq{d3} we can extract the counterterm of the gauge coupling as
\bea\label{cdg4}
\d_{g_4} &=& -\frac{1}{2} \d A \Rightarrow \nonumber\\
\frac{\d{g_4}}{g_4} &=& \frac{1}{16 \pi^2} \Bigl [ \frac{ g_4^2}{3} \frac{1}{\ve} + \frac{2 q^2 g_4^2}{3\L^2} \frac{1}{\ve} \Bigr ] \Rightarrow \nonumber\\
\d{g_4} &=&  \frac{1}{16 \pi^2} \Bigl [ \frac{ g_4^3}{3} \frac{1}{\ve} + \frac{2 q^2 g_4^3}{3\L^2} \frac{1}{\ve} \Bigr ] \, .
\eea
Finally for the last condition we need $ {\cal B}_{4,\phi}$ which is given in \eq{B4fi} and in DR gives
\be\label{B4phip}
{\cal B}_{4,\phi} =   \frac{1}{16 \pi^2} \Bigl [  \frac{4 (c^{(6)}_1\, p^2)^2}{\L^4\ve} + \frac{52 g_4^4}{\ve} + \frac{2 g_4^4 \{ f_C(p_i) + f_T(p_i) + f_S(p_i) \} }{\L^2\ve} \Bigr ] +  [{\cal B}_{4,\phi}]_f \, ,
\ee
where $i=1,2,3,4$ and $f_C$, $f_T$ and $f_S$ are functions of the external momenta coming from the $C$-, $T$- and $S$-Boxes respectively.
Substituting the above in the condition of \eq{RC4fi} the counter-term of $c_1^{(6)}$, for $p^2 \equiv p_1^2$, yields
\be\label{cdc61}
\d c_1^{(6)} =  \frac{1}{16 \pi^2} \Bigl [  4 (c^{(6)}_1)^2 \frac{p_1^2}{\L^2}  - 8 c^{(6)}_1 g_4^2 + 52 g_4^4 \frac{\L^2}{p_1^2} + \frac{8 c^{(6)}_1 g_4^2 p_1^2}{\L^2} + \frac{2 g_4^4 \{ f_C(p_i) + f_T(p_i) + f_S(p_i) \} }{p_1^2} \Bigr ] \frac{1}{\ve} \, .
\ee
Note that $\d_A$, $\d_\phi$ and $\d g_4$ reduce to the values of the usual SQED (see \cite{IrgesFotis2}) when the HDO are decoupled.
As the next step should be to determine the $\b$-function of the couplings, it is natural to choose an off-shell momentum scheme, $p^2\ne0$, since the classical Lagrangian lacks of explicit mass terms.
This is actually necessary otherwise the calculation of $c^{(6)}_1$'s $\b$-function is ambiguous.
At this stage, it may seem that our model has two dimensionless couplings, $c^{(6)}_1$ and $g_4$ and three dimensionful scales $\m$, $\L$ and $v$ 
in case a vev develops. 
On the other hand, the action \eq{SBHrd} is a product of the orbifold lattice which, for the boundary, 
inherits the model only with $g_4$, $\m$ and $v$ so there is an apparent mismatch of the independent parameters.
Nevertheless this is not true because $\L$ and $v$ depend on the regularization scale,
as a consequence the only independent scale is $\m$.
In addition, as we will see in the next section, the minimization condition of the scalar potential will induce 
a relation between $c^{(6)}_1$ and $g_4$, leaving us with the correct number of parameters.

The off-shell regularization scheme exploited here is choosing $p_i^2 = q_i^2 = \L^2$, which together with the momentum and channel conservation relations
\be
p_1+p_2+p_3+p_4=0 \,\,\,\, {\rm and} \,\,\,\, \{s=4\L^2,t=0\} \nonumber
\ee
fix
\be
p_1 \cdot p_2 = p_3 \cdot p_4 =  \L^2,  \,\,\,\, {\rm and} \,\,\,\,  p_1 \cdot p_3 =  p_2 \cdot p_4 = p_1 \cdot p_4 =  p_2 \cdot p_3 = -\L^2 \, . \nonumber
\ee
Using the above scheme to simplify our counter-terms, we obtain from \eq{cdfi}, \eq{cdg4} and \eq{cdc61} 
\bea
\d_\phi &=& 0 \label{dfiv} \\
\d{g_4} &=&  \frac{1}{16 \pi^2} \Bigl [  \frac{g_4^3}{\ve}   \Bigr ] \, , \label{dg4v}\\
\d c_1^{(6)} &=&  \frac{1}{16 \pi^2} \Bigl [  4 (c^{(6)}_1)^2 + 34 g_4^4  \Bigr ] \frac{1}{\ve} \label{dc61v}
\eea
respectively.
Now we are in position to determine the $\b$-functions of the two couplings.
For this we work in $d$-dimensions so we need to know the dimensionality of the couplings when $d\ne4$.
Keep in mind that the classical dimensions of the gauge and scalar fields, determined from the corresponding kinetic terms, are $d_{A^3_\m} = \frac{d-2}{2}$ and $d_\phi = \frac{d-2}{2}$ respectively.
Starting with the gauge coupling, its associated operators are $A^3_\m \bar \phi \pa \phi $ and $(A^3_\m)^2 \bar \phi \, \phi $ and a dimensional analysis of the latter gives
\bea
2d_{g_4} + 2d_{A^3_\m} + 2 d_\phi &=& d \Rightarrow \nonumber\\
2d_{g_4} + d-2 + d -2 &=& d  \Rightarrow \nonumber\\
d_{g_4}  &=& \frac{4-d}{2}
\eea
in accordance with \eq{dg4dm}.
For the scalar self-coupling there is only one associated operator, $(\bar \phi \phi)^2 \bar\phi \Box \phi$ suppressed by $1/\L^2$, indicating that
 \bea
d_{c^{(6)}_1} + 2 + 4 d_\phi -2 &=& d \Rightarrow \nonumber\\
d_{c^{(6)}_1} + 2(d-2) &=& d  \Rightarrow \nonumber\\
d_{c^{(6)}_1}  &=& 4-d \, .
\eea
These classical dimensions and since \eq{a40} and \eq{dg4v} determine $\d \a_4 = 2 \a_4^2/\ve$, 
together with \eq{dg4dm} yield the $\b$-function of the gauge coupling
\bea
\b_{\a_4} = - d_{\a_4} \, \a_4 + 2 \a^2_4  &{\rm or}&  \b_{g_4 \m^{\frac{-\ve}{2}}} = - d_{g_4} \, g_4 \m^{\frac{-\ve}{2}} + \frac{g^3_4 \m^{\frac{-3\ve}{2}}}{16\pi^2} \Rightarrow \nonumber\\
\b_{\a_4} = - \ve \a_4 + 2 \a^2_4  &{\rm or}&  \b_{g_4 \m^{\frac{-\ve}{2}}} = - \frac{\ve}{2} g_4 \m^{\frac{-\ve}{2}} + \frac{g^3_4 \m^{\frac{-3\ve}{2}}}{16\pi^2}\, . \nonumber
\eea
For $c^{(6)}_1$, using \eq{dc6dm}, we obtain the $\b$-function
\bea
\b_{c^{(6)}_1} &=&  -d_{c^{(6)}_1} \, c^{(6)}_1 + \frac{1}{16 \pi^2} \Bigl [  4 (c^{(6)}_1)^2 + 34 g_4^4  \Bigr ] \Rightarrow \nonumber\\
\b_{c^{(6)}_1} &=&  -\ve c^{(6)}_1 + \frac{1}{16 \pi^2} \Bigl [  4 (c^{(6)}_1)^2 + 34 g_4^4  \Bigr ] \,  .\nonumber
\eea
For completeness let us just present now the corresponding $\b$-function of the bulk gauge coupling.
There is no need for extra calculations here since the bulk lattice action, given in \eq{SBf}, after considering the naive continuum limit 
in Minkowski space, yields the Lagrangian
\bea\label{LBc}
{\cal L}^B &=& - \frac{1}{4} F^A_{\m\n} F^{A,\m\n} + \frac{1}{16 \L^2} ( D^\m F^A_{\m\n} ) ( D_\m F^{A,\m\n} ) - \frac{ g_5 }{24 \L^2} f_{ABC} F^A_{\m\n} F^B_{\n\r} F^C_{\r\m}    \nonumber\\
&+& ( \overline{ D_\m \Phi^A}) ( D^\m \Phi^A ) - \frac{1}{4 \L^2} ( \overline{ D^2 \Phi^A}) ( D^2 \Phi^A )   \Biggr] \, .
\eea
The above is a 5d version of the Lee-Wick gauge model \cite{Grinstein,Casarin} from where we can extract the bulk $\b$-functions
by generalizing the results to $d=5$.
Now the Lee-Wick Lagrangian is given in \cite{Grinstein} and it is
\bea\label{LWL}
{\cal L} &=& - \frac{1}{4} F^A_{\m\n} F^{A,\m\n} + \frac{1}{2 m^2} ( D^\m F^A_{\m\n} ) ( D_\m F^{A,\m\n} ) + \frac{ k_F \, g }{2 m^2} f_{ABC} F^A_{\m\n} F^B_{\n\r} F^C_{\r\m}  + ( D_\m \phi^A)^* ( D^\m \phi^A )   \nonumber\\
&-& \frac{1}{m^2} \Bigl \{ \d_1( D^2 \phi^A)^* ( D^2 \phi^A ) + i g \d_2 \, (\phi^A)^* ( D_\m F^{\m\n} )D_\n \phi^A  + g^2 \d_3 \, (\phi^B)^* F^A_{\m\n} F^{A,\m\n} \phi^B \Bigr \}   \Biggr] 
\eea
and the $d=4$ $\b$-function that derives from it, is
\be\label{bgLW}
\b(g) =  -\frac{g^3}{16\pi^2} \Biggl [ \left( \frac{43}{6} - 18 k_F + \frac{9}{2}k_F^2 \right) C_2 
- \left( \frac{\d_1 + 6 \d_3}{3\d_1} \right) n_\phi  \Biggr] \, .
\ee
$C_2$ is the Casimir operator in the adjoint representation, which for $SU(2)$ is $C_2=2$.
$n_\phi$ is the number of scalar fields and $k_F$, $\d_1$ and $\d_3$ are couplings multiplying the HDO.
Back to our model, the only coupling in \eq{LBc} is $g_5$, a consequence of the lattice origin of the action, 
so there should be only one independent $\b$-function which should be extracted from \eq{bgLW}.
A direct comparison of \eq{LBc} and \eq{LWL} can be made by setting $\d_2=\d_3 = 0$ 
in the latter and defining $16 \L^2 = 2 m^2$, $k_F = -2/3$ and $\d_1 = 2$ in the former, to obtain
\bea
{\cal L}^B &=& - \frac{1}{4} F^A_{\m\n} F^{A,\m\n} + \frac{1}{2 m^2} ( D^\m F^A_{\m\n} ) ( D_\m F^{A,\m\n} ) + \frac{ k_F \, g_5 }{2 m^2} f_{ABC} F^A_{\m\n} F^B_{\n\r} F^C_{\r\m}    \nonumber\\
&+& ( \overline{ D_\m \Phi^A}) ( D^\m \Phi^A ) - \frac{\d_1}{ m^2} ( \overline{ D^2 \Phi^A}) ( D^2 \Phi^A ) \, .
\eea 
The 1-loop part of the $\b$-function of our bulk action, $\b^1_{g_5}$, can be then obtained by substituting the above parameters together with $n_\phi=1$ in $\b(g)$. This yields
\be\label{b1g5}
\b^1_{g_5} =  - \Bigl [ \frac{125}{6} \Bigr] \frac{g_5^3}{16\pi^2}  \, .
\ee
Finally, considering $(F^A_{\m\n})^3/\L^2$ as the associated operator of the coupling $g_5$, dimensional analysis indicates that
\bea\label{dg5}
d_{g_5} + 3 d_{F_{\m\n}} - 2 &=& d \Rightarrow \nonumber\\
d_{g_5} + 3 \frac{d}{2} - 2 &=& d \Rightarrow \nonumber\\
d_{g_5} &=& \frac{4-d}{2}
\eea
in $d$-dimensions, with $d_{F_{\m\n}} = d/2$.
Hence including the tree level part, the complete 1-loop $\b$-function of the gauge coupling becomes
\be
\b_{g_5 \m^{\frac{-\ve}{2}}} = -\frac{\ve}{2} g_5 \m^{\frac{-\ve}{2}} - \frac{125}{6}  \frac{g_5^3 \m^{\frac{-3\ve}{2}}}{16\pi^2}  \,\,\,\, {\rm or} \,\,\,\,  \b_{\a_5} = - \ve \a_5 - \frac{125}{6 N} \a_5^2 \nonumber
\ee
where we have formed the dimensionless combination $g_5 \m^{-\ve/2}$ and following our steps on the boundary, defined the dimensionless in $d$-dimensions
gauge coupling
\be\label{a5g5}
\a_5 = \frac{2 N \, g_5^2 }{16 \pi^2} \m^{-\ve} \, . 
\ee
Note that in 4-dimensions $ \b_{\a_5}$ is such that the coupling is asymptotically free, a well known characteristic of non-Abelian gauge theories.
The contribution of the HDO affects only the numerical factor in front of the 1-loop $\b$-function.

We collect the $\b$-functions for the boundary and bulk couplings in $d=4-\ve$ dimensions and for $N = 2$:
\bea
\b_{\a_4} &=& - \ve \a_4 + 2 \a^2_4 \,\,\,\, {\rm or} \,\,\,\, \b_{g_4 \m^{\frac{-\ve}{2}}} = - \frac{\ve}{2} g_4 \m^{\frac{-\ve}{2}} + \frac{g^3_4 \m^{\frac{-3\ve}{2}}}{16\pi^2} \label{betag4} \\
\b_{c^{(6)}_1} &=&  -\ve c^{(6)}_1 + \frac{1}{16 \pi^2} \Bigl [  4 (c^{(6)}_1)^2 + 34 g_4^4  \Bigr ] \label{betac61} \\
\b_{\a_5} &=& - \ve \a_5 - \frac{125}{12} \a_5^2  \,\,\,\, {\rm or} \,\,\,\, \b_{g_5 \m^{\frac{-\ve}{2}}} = -\frac{\ve}{2} g_5 \m^{\frac{-\ve}{2}} - \frac{125}{6}  \frac{g_5^3 \m^{\frac{-3\ve}{2}}}{16\pi^2} \, .\label{bg5a5}
\eea

\section{The Higgs phase}\label{aBHm}

The action in \eq{SBHrd} represents a version of massless SQED, enhanced by dimension-6 operators.
If instead of the dimension-6 operators a scalar, quartic polynomial self interaction term was present, it would be just the classic Coleman-Weinberg model \cite{Coleman}.
It is natural then to ask if and in what ways the boundary theory of \sect{Renorm.bf}  is different from it.
We therefore perform next such a comparison, as we analyze the Higgs phase.
A short review, including all the main results of the CW model relevant to our discussion, is presented in Appendix \ref{CWm}.
In the second part of this section we also consider the case where pure polynomial terms are added to our model.
This would be of course inconsistent with our construction but it could be easily realized in less restricted models.

\subsection{The scalar potential and a comparison to the Coleman-Weinberg model}\label{cCWm}

This comparison is going to be very useful because not only will it show the differences between the two models but 
we can exploit at the same time their similarities.
Let us work in momentum space and take $d=4$, following Appendix \ref{CWm}.
First we have to construct the improved effective potential that corresponds to \eq{Lb-h}.
Next comes its minimization through which we will see if there exists a non-trivial minimum and whether it imposes a relation between the couplings,
as it does in \eq{le4}.
Then we determine the scalar and gauge field masses and from those the scalar-to-gauge mass ratio.

\begin{figure}[!htbp]
\centering
\includegraphics[width=9cm]{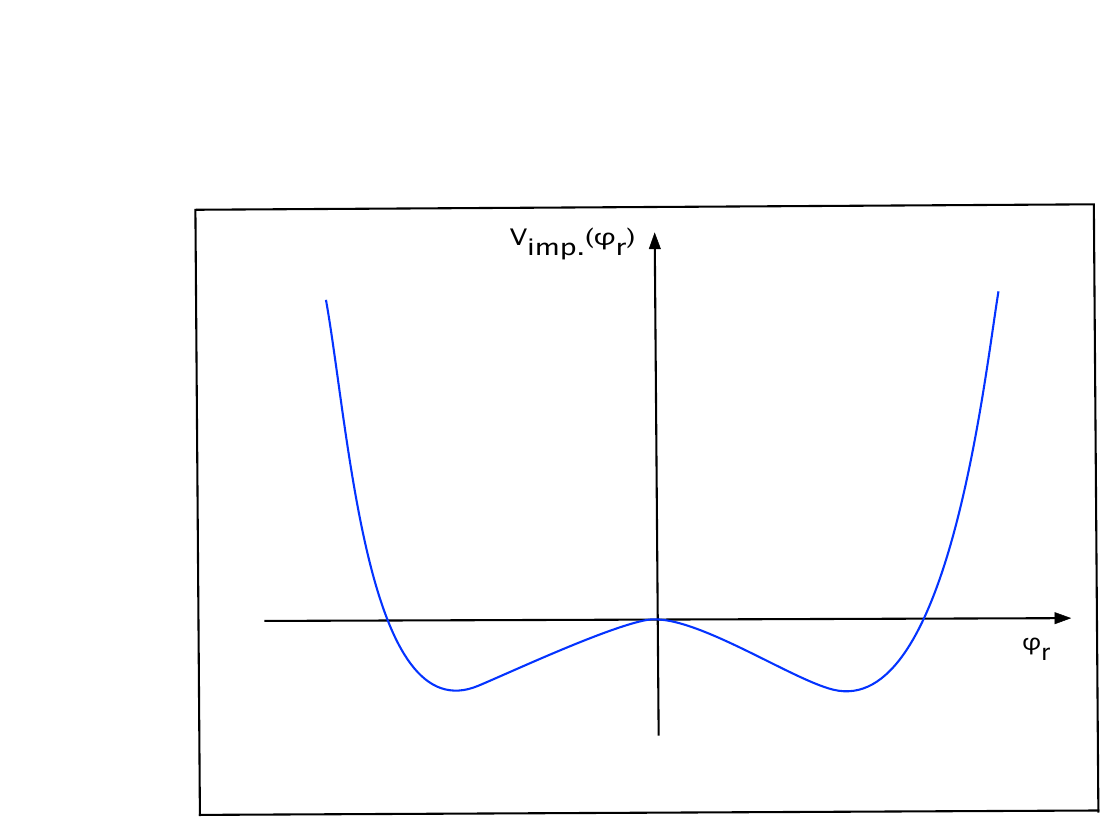}
\caption{\small The potential in \eq{Vimpf}.  \label{mexicanhat}}
\end{figure}

The starting point is the renormalized but yet unimproved potential which, up to total derivatives, is 
\bea\label{VR}
V(\bar \phi, \phi) &=& - \frac{c_1^{(6)} p^2 }{4\L^2}  (\bar \phi \phi)^2 \equiv - \frac{c_1^{(6)} p_1 \cdot p_3 }{4\L^2}  (\bar \phi \phi)^2 \Rightarrow \nonumber \\
V(\bar \phi, \phi) &=& \frac{c_1^{(6)}}{4}  (\bar \phi \phi)^2 \, ,
\eea
where the choice $p^2 \equiv p_1 \cdot p_3 =  - \L^2$ comes from the off-shell scheme introduced in the previous section.
Note that we used this choice so as to get rid of the unphysical overall minus sign which makes momentum space uncomfortable. 
One way to see the effect of the improvement is to notice that $\phi$ could take the place of the renormalization scale, 
a fact that can be seen through the definition of the effective mass.
In particular this is given as the second derivative of $V$
\be
\frac{\pa^2 V(\bar \phi, \phi)}{\pa \bar \phi \pa\phi } =  m_{\rm eff}^2(\m) =  c_1^{(6)} \bar \phi \phi \, ,
\ee
with $\bar \phi, \phi$ the only running parameters in the r.h.s.
This information can be inherited in the potential if instead of looking at the running of $\m$ we focus on the running of $\phi$, using the identifications
\be\label{offshell}
\m^2 \to m^2_{\rm eff}(\m) = c_1^{(6)} \bar \phi \phi(\m) \,\,\,\, {\rm and} \,\,\,\, m^2_{\rm eff}(\m_R) =  c_1^{(6)} v^2
\ee
with $\m_R$ an arbitrary renormalization scale and
$v$ the vacuum expectation value (vev) of the scalar field.
To make the argument clear take the fourth derivative of the full 1-loop bare potential of \eq{Lb-h} which, in momentum space and in the same scheme as \eq{VR}, is
\bea\label{V4}
V^{(4)} &\equiv&  \frac{\pa^2 }{\pa \bar \phi^2} \frac{\pa^2 }{ \pa\phi^2 } V_0(\bar \phi, \phi) =  c_{1}^{(6)}  + \d c_1^{(6)}  + 2 c_1^{(6)} \d_\phi + {\cal B}_{4,\phi} \nonumber\\
 &\equiv&   c_{1}^{(6)} + \d c_1^{(6)} + {\cal B}_{4,\phi} \, ,
\eea
where \eq{dfiv} was used.
Recall that in our case the HDO made $\d_\phi$ vanish, which is not the case in the usual CW model (see Appendix \ref{CWm}).
Essentially $V^{(4)}$ gives a reformulation of the renormalization condition in \eq{RC4fi}.
Since the above relation is about a physical quantity, connected to the Green functions,
it should be finite, forcing us to absorb the divergent part of $ {\cal B}_{4,\phi}$ in the remnant counterterm.
Here comes the crucial part now, since in the regularization scheme of \eq{offshell} the contribution of the scaleless $B_0$ integrals can be rewritten as
\be
B_0 \equiv \frac{2}{\ve} + \ln \frac{ \m^2 }{ \m^2_R} = \frac{2}{\ve} + \ln \frac{ m^2_{\rm eff}(\m) }{ m^2_{\rm eff}(\m_R) } = \frac{2}{\ve} + \ln \frac{ \bar \phi \phi }{ v^2 } \, .
\ee  
Substituting this into \eq{V4} (using \eq{B4phip}) and canceling the divergent part we get
\be\label{V4b}
V^{(4)} =  c_{1}^{(6)} + \frac{1}{32 \pi^2} \Bigl [  4 (c^{(6)}_1)^2 + 34 g_4^4  \Bigr ] t \equiv  
c_{1}^{(6)}(t) \,\,\,\, {\rm and} \,\,\,\, \d c_1^{(6)} = \frac{- 1}{16 \pi^2} \Bigl [  4 (c^{(6)}_1)^2 + 34 g_4^4  \Bigr ] \frac{1}{\ve}\, ,
\ee
with $t =  \ln ( \bar \phi \phi / v^2) $ and the condition $V^{(4)} =  c_{1}^{(6)}$, for $t=0$, at work.
As a side comment note that when the Callan-Symanzik operator hits on $V^{(4)}$, one obtains
\bea
\left[- 2 \frac{\pa}{\pa t} + \b^1_{c_1^{(6)}} \frac{\pa}{\pa c_1^{(6)}} + \b^1_{g_4} \frac{\pa}{\pa g_4} - 2 \g_\phi \right] V^{(4)} &=& 0 \Rightarrow \nonumber\\
\frac{\pa c_{1}^{(6)}(t)  }{\pa t} &=& \frac{1}{2}\b^1_{c_1^{(6)}} \Rightarrow \nonumber\\
c_{1}^{(6)}(t) &=& c_{1}^{(6)} + \b^1_{c_1^{(6)}} \frac{t}{2} \, ,
\eea
where $\b^1_{c_1^{(6)}}(\b^1_{g_4})$ is the 1-loop part of the $\b$-function of $c_1^{(6)}( {\rm of} \,g_4)$ while the anomalous dimension of $\phi$, given by $\g_\phi$, vanishes.
Then comparing the above to \eq{V4b} fixes $\b^1_{c_1^{(6)}}$ to
\be\label{bc61imp}
\b^1_{c_1^{(6)}} =  \frac{1}{16 \pi^2} \Bigl [  4 (c^{(6)}_1)^2 + 34 g_4^4  \Bigr ] \, ,
\ee
matching the loop part of \eq{betac61}.
Notice that in the CW model, the corresponding Callan-Symanzik equation gives for the quartic coupling the $\b$-function of \eq{blCW}.
There, a cross term between the quartic and gauge coupling exists due to the appearance of the anomalous dimension \eq{gCW}.
Here the presence of the HDO leads to $\g_\phi=0$, hence the cross term is missing in $\b^1_{c_1^{(6)}} $.
The improved 1-loop effective potential is now easily obtained by integrating with respect to the scalar field the renormalized $V^{(4)}$, to finally arrive in momentum space at
\be\label{Vimpm}
V_{\rm imp.}(\bar \phi, \phi) =  \frac{c_1^{(6)}}{4}  (\bar \phi \phi)^2 + \Bigl \{ 2 (c_1^{(6)})^2 + 17 g^4_4 \Bigr \}  \frac{(\bar \phi \phi)^2}{64 \pi^2} \Bigl( \ln\frac{\bar \phi \phi}{v^2} - 3 \Bigr) \, .
\ee
The corresponding effective potential from the CW analysis is given in \eq{VCW1} and it is apparent that taken at face value,
our potential in the chosen renormalization scheme is essentially a CW potential with the HDO coupling $c_1^{(6)}$ playing the role of the quartic coupling $\l$.
There are differences though that are quite important with the most obvious ones hiding in the numerical factors. 
To see an example of the effect of the different numerical factors, we proceed with the minimization of the potential.
Being of the CW type, the potential is expected to be of a no-scale nature, yielding a constraint between couplings rather than determining a vev.
Following \cite{Coleman} and defining $\bar \phi \phi = [ ( A^1_5)^2 +( A^2_5)^2]/2 \equiv \phi_r^2$, we first rewrite the potential as
\be\label{Vimpm2}
V_{\rm imp.}(\phi_r) =  \frac{c_1^{(6)}}{4}   \phi_r^4 + \Bigl \{ 2 (c_1^{(6)})^2 + 17 g^4_4 \Bigr \}  \frac{ \phi_r^4}{64 \pi^2} \Bigl( \ln\frac{ \phi_r^2}{v^2} - 3 \Bigr) \, 
\ee
and then find its minimum:
\bea\label{mincond.}
\frac{\pa V_{\rm imp.}(\phi_r)}{\pa \phi_r} \Bigl |_{\phi_r=v} &=& 
\frac{-( 10 (c_1^{(6)})^2 + 85 g_4^4 - 32 \pi^2 c_1^{(6)} ) v^3 }{32\pi^2} =0 \label{mincond.1} \Rightarrow \\
c_1^{(6)} &=& \frac{85}{32 \pi^2}  g_4^4 \, ,\label{mincond.2}
\eea
where $(c_1^{(6)})^2$ was neglected with respect to $c_1^{(6)}$ since the latter is approximately 32 times bigger than the former at the above relation.
Substituting \eq{mincond.2} into \eq{Vimpm} we end up with 
\be\label{Vimpf}
V_{\rm imp.}(\phi_r) =  \frac{17 g^4_4 \phi_r^4}{128 \pi^2} \Bigl( 2 \ln\frac{\phi_r^2}{v^2} - 1 \Bigr)  + {\cal O}(g^8_4)\, ,
\ee
which justifies our choice to neglect $(c_1^{(6)})^2$.
Notice that both \eq{mincond.2} and \eq{Vimpf} have analogues in the CW analysis, with the corresponding results given by \eq{le4} and \eq{VCW2} respectively.
Now, \eq{mincond.} indicates a non-trivial minimum at $\langle \phi_r \rangle = v$, as
the shape of the potential is of the standard Mexican hat form, see \fig{mexicanhat}.
\FloatBarrier
%
The vev triggers the spontaneous breaking of the gauge symmetry, allowing us to perform in the potential the shift
$\phi_r \to h + v$ with $h$ the physical scalar.
Then,
\be\label{mfr}
 m^2_h  \equiv \frac{\pa^2 V(h)}{\pa h^2 } \Bigl |_{h=0} =  \frac{210}{8 \pi^2} g_4^4 v^2 \, .
\ee
Notice the difference with respect to CW numerical factor in the scalar mass. It arises due to the different overall numerical factor in \eq{Vimpf}
that affects the magnitude of the curvature of the potential at the minimum and originates from the higher derivative operator with coupling $c_1^{(6)}$.
Since the gauge symmetry is broken, we expect the gauge field to develop a mass.
To leading order in $\L$ only the operator $(A^3_{\m})^2 \bar \phi \phi$ contributes after the shift and from that we obtain a gauge boson mass
\be\label{mA3}
m^2_{A^3_{\m}} = g_4^2 v^2 \equiv m^2_Z \, ,
\ee
the same as in the CW model.
Therefore the above two expressions for the masses determine, at tree level, the scalar-to-gauge-field mass ratio
\bea\label{mratio}
\frac{ m^2_h}{m^2_Z } \equiv \r^2_{\rm bh} &=& \frac{210}{8 \pi^2} g_4^2  \Rightarrow \nonumber\\
\r_{\rm bh} &=&  \sqrt{\frac{210}{8 \pi^2} \, } g_4 \simeq 1.64 \, g_4 \, .
\eea
In the last line we computed the numerical factor for later convenience.
The corresponding CW analysis results in a scalar and gauge mass given in \eq{SmCW} and \eq{VmCW} respectively, detemining the mass ratio in \eq{mSmV}:
\be\label{rmCWR}
\r_{\rm CW} = \sqrt{\frac{3}{8 \pi^2}} e  \simeq 0.19 \, e \, .
\ee
Let us discuss some numerics.
To begin, the appearance of the anisotropy is rather important because it allows $\r_{\rm bh}$ to reach its Standard Model value $1.38$
for reasonable values of $g$ (recall, $g_4=g\sqrt{\g}$).
The analogous observation from the non-perturbative point of view in \cite{IrgesK6} was that close to the Higgs-Hybrid phase transition and for $\g\simeq 0.50$,
a $\r_{\rm bh}\simeq 1.40$ can be measured. Away from the phase transition or in the absence of anisotropy, the mass ratio is far from this value.
As an example consider the triple point in \fig{npPD} which is reached for $\g \simeq 0.79$  \cite{IrgesK5}.
Then, for our model, \eq{mratio} gives $\r_{\rm bh} \simeq 1.45 \, g$, 
which for $g \simeq 0.95$ reproduces the SM value, while in the CW model we would need much larger values of $e$ to reach the same result.
Another way to see the difference is to suppose that $g$ and $e$ are of the same order. 
Then for $\g \simeq 0.79$ and $g\simeq e$, we get $\r_{\rm bh} \simeq 7.60 \, \r_{\rm CW}$.
Regarding the dependence of $\r_{\rm bh}$ on $g_4$ in \eq{mratio},
forcing \eq{mratio} and \eq{rmCWR} to give the SM value, fixes $g_{4} \simeq 0.84$ and $e \simeq 7.20$ respectively, 
indicating that the former can be consistent with perturbation theory, not so much the latter.
Interestingly, this value for $g_4$, when substituted back into \eq{mincond.2}, gives $c^{(6)}_1 \simeq 0.13$ close to the SM value for the Higgs self-coupling.
Recall that $c^{(6)}_1$ plays the role of the scalar quartic coupling here so this is a non-trivial coincidence.
On the other hand, setting $e \simeq 7.20$ into \eq{le4} we find $\l \simeq 1123.20$. 

The above numerical discussion did not take into account constraints that originate from the non-perturbative dynamics. 
We will insert later information of this sort in the discussion
and look again at the numbers but first we have to understand better the running of the various parameters involved.
So, let us now look at the scale dependence of the couplings.
This can be done for the two models by solving the equations
\be\label{a4aeRG}
\m \frac{d \, g_4(\m)}{d \m} = \b_{g_4} \hskip 1cm {\rm and} \hskip 1cm \m \frac{d \, e(\m)}{d \m} = \b_e
\ee
respectively, choosing as IR boundary conditions $m_R,M_R$, $g_4(m_R) \equiv g_{4,R}$ (or $g(m_R) \g(m_R) \equiv g_R \, \g_R$) and $e(M_R) \equiv e_R $.
In order to locate the UV limit of the running in our case, recall first that the masses depend on $g_4$ and $v$ only.
We will choose from now on a scheme where the vev is kept frozen at a value $v=v_\ast$,
in which case the running of the masses is determined by that of $g_4$ only.
Both evolutions implied by \eq{a4aeRG} are not interrupted, in principle, but from a Landau pole at some extremely large UV scale.
A crucial difference with respect to the CW case is that in our model the running must be halted in the UV by the quantum phase transition, 
way before the Landau pole is hit.
Let us call this scale $\m_\ast$ and see in the following section whether it can be identified more precisely.

\begin{figure}[!htbp]
\centering
\includegraphics[width=9cm]{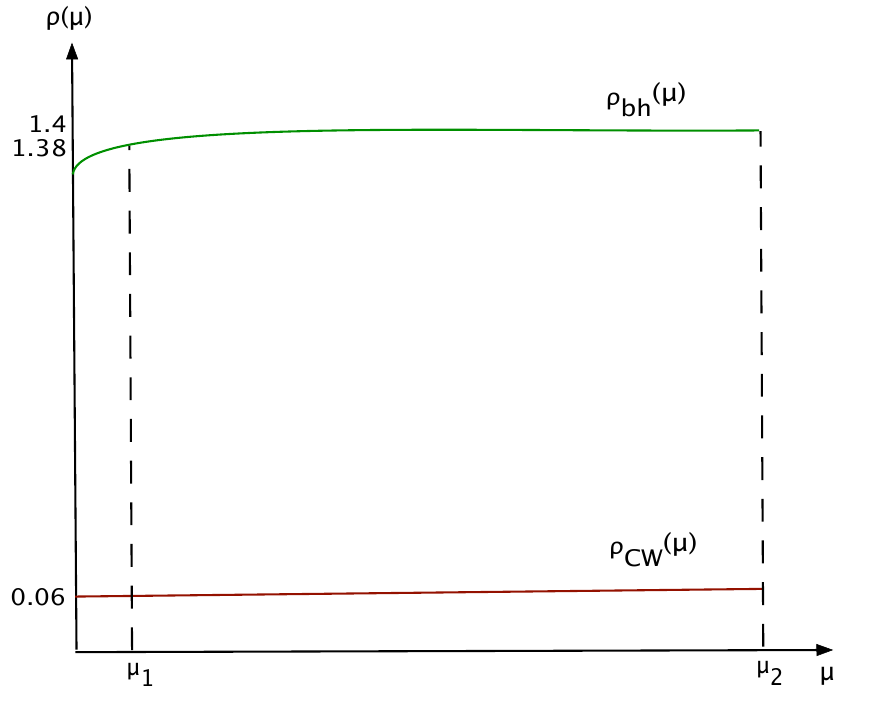}
\caption{\small The running of the $\r$-parameter with respect to the energy scale $\m$ for the Boundary-Hybrid model (green line) and the Coleman-Weinberg model (red line).
As $\m$ increases, $\r_{\rm bh}(\m)$ increases and reaches the Standard Model value at $\m_1 
$. 
At that scale it is approximately 23 times bigger than $\r_{\rm CW}(\m_1)=0.06$.  \label{rbhrCW}}
\end{figure}

Solving the RG equations of \eq{a4aeRG} gives
\be\label{a4mu}
g_4(\m) = \frac{g_{4,R}}{\sqrt{ 1- \frac{g^2_{4,R}}{16 \pi^2} \ln\frac{\m^2}{m_R^2} } } \hskip 1cm {\rm or} \hskip 1cm \a_4(\m) = \frac{\a_{4,R}}{1- \a_{4,R} \ln\frac{\m^2}{m_R^2} } 
\ee
for our boundary coupling and 
\be\label{aemu}
 e(\m) = \frac{e_R }{\sqrt{ 1- \frac{e_R^2}{48 \pi^2} \ln\frac{\m^2}{M_R^2} } }
\ee
for the Coleman-Weinberg case, both evaluated for $d=4$.
On \fig{rbhrCW} we compare the two evolutions for the case of $M_R=m_R=91.1$ GeV, $g_{4,R} = 0.83$ and $e_R=0.31$.
\FloatBarrier
%
The CW evolution appears as a straight line just because the running of $g_4$ is 3 times faster than that of $e$.
The running on the figure starts at $m_R=91.1$ GeV and ends at $10^5 $ GeV which is an appropriate range to illustrate clearly the difference between the two models.
The lower values of the $\r$-parameters are $\r_{\rm bh}(m_R) \simeq 1.36$ and $\r_{\rm CW}(m_R) \simeq 0.06$.
The running leads to a small increase from the starting values due to the logarithmic running,
but the increase is relatively more substantial for $\r_{\rm bh}$.
Of course this also means that the boundary coupling $g_4$ could reach its Landau pole much faster
compared to the CW coupling $e$. Numerically,
the Landau pole for the $g_4$ gauge coupling is located at
\be\label{a4LP}
\m_{4,{\rm L}} = e^{\frac{ 8\pi^2 }{ g^2_{4,R}}} m_R \approx 5.3 \times 10^{51} \, {\rm GeV}
\ee
and for the CW coupling at
\be\label{aeLP}
\m_{e,{\rm L}} = e^{\frac{ 24 \pi^2 }{ e^2_R}} M_R \approx 2.6 \times 10^{1072} \, {\rm GeV} \, .
\ee
There is of course a chance for $\r_{\rm CW}(\m)$ to also reach the SM value if $e(\m)$ becomes large enough.
This can be however realized only when $\m$ approaches $\m_{e,{\rm L}} $ where perturbation theory breaks down anyway.
%
\begin{figure}[!htbp]
\centering
\includegraphics[width=8cm]{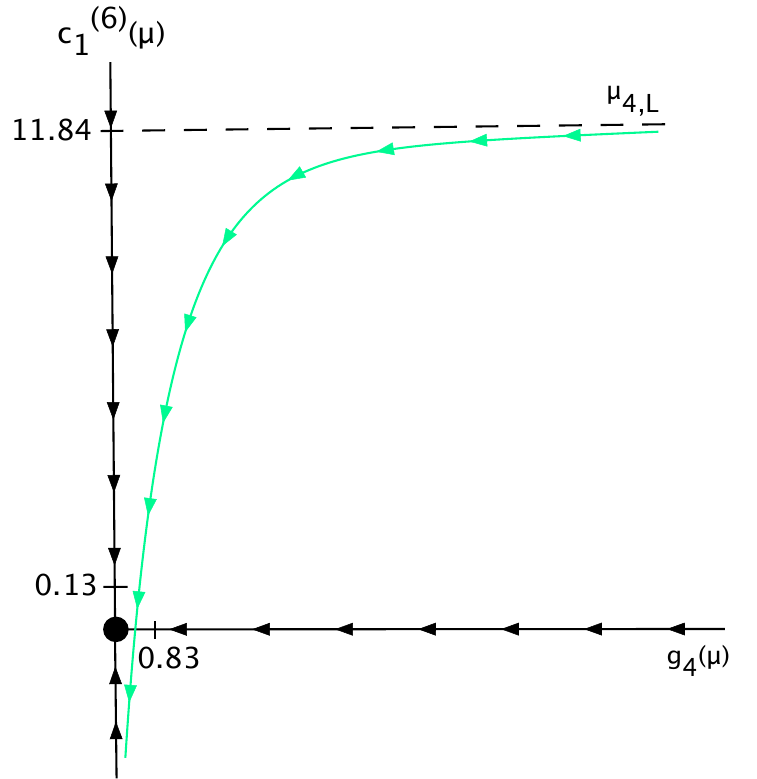}
\caption{\small The simultaneous running of the HDO and gauge couplings $c^{(6)}_1(\m)$ and $g_4(\m)$.
The arrows point towards the IR, so both couplings increase in the UV.
As $\m \to 0$ the Gaussian fixed point, $\bullet$, is approached and then $g_4 \to 0$ while the HDO coupling reaches $-\infty$.
At the starting point of the running, $\m=m_R$, the couplings read $(c^{(6)}_{1,R},g_{4,R}) = (0.13, 0.83)$
while in the UV, where $g_4$ reaches the Landau pole $\m_{4,{\rm L}}$, yield $(c^{(6)}_1(\m_{4,{\rm L}}),g_4(\m_{4,{\rm L}})) = (11.84, \infty)$. \label{c61g4}}
\end{figure}
%

Before we get into the non-trivial effects of the phase transition, let us construct the RG flow diagram 
on the $c_1^{(6)}-g_4$ plane, neglecting its presence.
For that purpose we need the running of \eq{a4mu} while we evaluate $c^{(6)}_1(\m)$ solving \eq{dc6dm} 
for the $\b$-function of \eq{betac61} (neglecting the $(c^{(6)}_1)^2$ contribution
).
Then the RG equation reads
\be
\m \frac{d c^{(6)}_1(\m) }{d \m} = \b_{c^{(6)}_1} = -\ve c^{(6)}_1 + 34 g_4^4
\ee 
which in $d=4$ gives
\be
c^{(6)}_1(\m) = c^{(6)}_{1,R} + \frac{17 \, g_4^4 }{16\pi^2} \ln \frac{\m^2}{m^2_R} \, .
\ee
The combined evolution of the couplings is seen in \fig{c61g4}. 

Below we summarize the similarities and differences between our boundary effective action (before taking into account non-perturbative dynamics) and the CW model: 
\begin{itemize}
\item At the classical level, the CW Lagrangian contains a polynomial $\phi^4$ term, a marginal in $d=4$ operator, as the only contribution to the scalar potential.
Here, due to the origin of the boundary effective action, there are only derivative terms.
After field redefinitions, a dimension-6 derivative operator plays the role of the scalar potential but it also contributes to the vertices.

\item In the CW model a mass counter-term is introduced from the start, despite the absence of a classical mass term.
This breaks scale invariance already at the classical level.
In our case we do not need such a counter-term.
This is crucial if we want to assign the responsibility for the simultaneous breaking of scale and gauge symmetries to the HDO.

\item Renormalization yields $\b_\l$ and $\b_{c^{(6)}_1}$ for the CW and our case respectively.
The former contains a cross term between the quartic and gauge coupling which is absent from the latter.
The reason is that the effect of the HDO, for the chosen renormalization scheme, leads to the vanishing of the anomalous dimension of the scalar field.
This anomalous dimension is non-zero in the CW case.

\item After renormalization, both 1-loop effective potentials indicate the existence of a non-trivial minimum.
Around the minimum, they differ only by a multiplicative constant. This constant affects crucially though
the scalar-to-gauge boson mass ratio. The operators that appear in the effective action, 
determine the speed of the gauge coupling running: due to the HDO the $g_4$ coupling and mass ratio run 3 times faster than the corresponding quantities in the CW model.

\item The running of the coupling in the CW model does not stop until the Landau pole. 
In our case the running of $g_4$ is similar but is expected to be stopped by the phase transition.

\end{itemize} 

\subsection{What if polynomial potential terms where allowed?}\label{bfifi3}

In contrast to the CW Lagrangian, where the potential is represented by the usual marginal operator $(\bar \phi \phi)^2$, 
here the only potential-like term 
\be
\frac{O^{(6)}_1}{\L^2} = \frac{ \bar \phi \phi \Box \bar \phi \phi}{\L^2} \nonumber
\ee
is a 6-dimensional derivative operator.
In a general $U(1)$ gauge theory coupled to a complex scalar we could have other dimension-6 operators, for example 
\be\label{Ob6}
\left(\frac{O_1^{(6)}}{\L^2}, \frac{O_2^{(6)}}{\L^2}, \frac{O_3^{(6)}}{\L^2}\right) = \left( \frac{\bar \phi \phi \Box \bar \phi \phi}{ \L^2} , \frac{\bar \phi \Box^2 \phi}{\L^2}, \frac{(\bar \phi \phi)^3}{\L^2} \right) \, .
\ee
Note that we have used the usual box derivative since this analysis is supposed to be done after expanding the covariant derivatives. 
However, as it is demonstrated in \cite{FotisLetter1, FotisLetter2}, after an appropriate field redefinition only one of them stays independent.
In other words a Lagrangian, enhanced by the operators in \eq{Ob6} is in fact equivalent to a Lagrangian that contains
only the polynomial term $O_3^{(6)}$. Let us call such a basis, the W-basis.
The same is true if we insert also dim-8 operators and an extended W-basis can be constructed. It will contain only
\be\label{Wb68}
\left(\frac{O_3^{(6)}}{\L^2}, \frac{O_4^{(8)}}{\L^4}\right) = \left( \frac{(\bar \phi \phi)^3}{\L^2}, \frac{(\bar \phi \phi)^4}{\L^4} \right) \, ,
\ee
as HDO, which actually play the role of a non-trivial scalar potential with associated couplings $c^{(6)}_3$ and $c^{(8)}_4$.
Then we would need these operators to behave like the marginal operator of the CW model.
However, if only $O_3^{(6)}$ or only $O_4^{(8)}$ is present then there can be no CW mechanism in progress 
since neither of them contributes as a mass term through radiative corrections.
To be more specific the only possible 1-loop diagram is
\be\label{O3d23}
\begin{tikzpicture} [scale=0.9]
\draw [dashed] (6,0)--(7.6,0);
\draw [dashed] (7.6,0.5) circle [radius=0.5];
\draw [dashed] (7.6,0)--(9.1,0);
\draw [dashed] (7.6,0)--(7,-1);
\draw [dashed] (7.6,0)--(8.2,-1);
\end{tikzpicture}
\nonumber
\ee
when only $O_3^{(6)}$ plays the role of the potential and
\be
\begin{tikzpicture} [scale=0.9]
\draw [dashed] (6,0)--(7.6,0);
\draw [dashed] (7.6,0.5) circle [radius=0.5];
\draw [dashed] (7.6,0)--(9.1,0);
\draw [dashed] (7.6,0)--(7,-1);
\draw [dashed] (7.6,0)--(8.2,-1);
\draw [dashed] (7.6,0)--(6,-0.7);
\draw [dashed] (7.6,0)--(9.2,-0.7);
\end{tikzpicture}
\nonumber
\ee
when the scalar potential is made only from $O_4^{(8)}$.
In the former the radiative correction gives rise to a quartic term while in the latter to a $(\bar \phi \phi)^3$ term, 
notwithstanding both cases correspond to scaleless tadpole-integrals and vanish in DR.
Another way to express this conclusion is to look at the $\b$-functions of $c^{(6)}_3$ and $c^{(8)}_4$.
These are evaluated, at 1-loop order, in \cite{FotisLetter1, FotisLetter2} and show that if $V \sim O_3^{(6)}$ or $V \sim O_4^{(8)}$, 
the associated couplings do not run. Therefore a CW analysis is meaningless.
On the other hand, when both operators in \eq{Wb68} appear in the potential, then a non-trivial scalar potential is constructed, of the form
\be
V^{(8)} \sim \frac{ c_3^{(6)}}{\L^2} (\bar \phi \phi)^3 + \frac{ c^{(8)}_4}{\L^4} (\bar \phi \phi)^4  \, ,\nonumber
\ee
whose phase diagram indeed possesses a branch with spontaneously broken internal symmetry.
Nevertheless, this effect is trivial since the running of $c^{(8)}_4$ is such that it sets the phase diagram unstable.
The above arguments are presented in detail in \cite{FotisLetter1, FotisLetter2}.
Thus, another important point is that even if the W-basis of \eq{Wb68} were allowed 
we would not manage to construct a non-trivial phase diagram with SSB at work.

\section{The effective action near the Higgs-Hybrid phase transition}\label{HDOPD}

The extra step we would like to take in this section is to build in the effective action certain non-perturbative features 
that have been observed via Monte Carlo simulations on the lattice.
We start with a comment on the nature of higher dimensional operators in the effective action.
We have seen above how they affect quantitatively the scalar mass and the $\b$-functions.
At a more qualitative level, one question is whether the HDO are of a classical or a quantum nature. 
In \cite{FotisLetter2} we argued that when the suppressing scale $\L$ is an internal scale,
they must have a quantum origin. In DR for example one can identify $\L$ with the regulating scale $\m$ or alternatively with a fixed scale
derived from $\m$, such as $\m_\ast$.
We have already used this fact throughout. In addition, in \cite{FotisLetter2} it was demonstrated that the (1-loop) quantization of a Lagrangean with
vertices deriving from the HDO is equivalent to the quantization of a Lagrangean without HDO but taking into account all possible operator insertions.
These two arguments allow us to characterize the HDO as quantum corrections which is not an unimportant detail
since then, in the presence of a Higgs mechanism, scale and internal symmetry break spontaneously and simultaneously by quantum effects.

In order to be able to build in the boundary effective action the effects of the presence of the phase transition, we now make two assumptions. 
The first assumption states that dimensional reduction occurs in the vicinity of the Higgs-Hybrid phase transition.
Since all dimensions are infinite, the dimensional reduction must develop due to localization. 
This was in fact observed numerically in \cite{IrgesK6} where it was demonstrated that near the Higgs-Hybrid phase transition in the Higgs phase
and in the entire Hybrid phase, the lattice becomes layered in the fifth dimension. 
This means in particular that the $U(1)$ gauge-scalar effective action of the boundary slice in the Higgs phase contains a 4d gauge coupling (identified as $g_4$) 
and the dynamics of a bulk slice in the Hybrid phase is to a good approximation a 4d $SU(2)$ Yang-Mills theory, associated with a 4d coupling $g_s$.
The previous construction of the boundary effective action in the Higgs phase was of course motivated by and is consistent with this non-perturbative fact.
We will also exploit the consequences of the dimensional reduction in the bulk of the Hybrid phase reflected by $g_s$ in the following but before that, we need
one more assumption.
Therefore our second assumption is that the bulk-driven transition is reached by $g_4$ and $g_s$ in the UV.
Concrete non-perturbative evidence for the validity of this assumption we have in the Higgs phase where the masses of the gauge and scalar fields
decrease in units of the lattice spacing $a_4$ as the phase transition is approached \cite{IrgesK6}. Regarding the behaviour of $a_4$ as the phase transition is approached
from the side of the Hybrid phase we do not have concrete numerical evidence but we can motivate this assumption by imagining an RG flow in the Hybrid phase
that starts from the phase transition that separates the 5d Confined and the Hybrid phase and approaches the Hybrid-Higgs phase transition.
The evolution must be of an asymptotically free type (i.e. the evolution of $g_s$), monotonically flowing from the IR to the UV, where it hits the 
Hybrid-Higgs phase transition. 

These two facts constrain the dynamics in a radical way. The most notable constraint is that then, RG flows that emanate from a given point on the 
Higgs-Hybrid phase transition and extend in the two phases, are necessarily correlated \cite{Corfu2}. We will make these statements concrete below but
first we have to clarify a couple of technical points. The first concerns the fact that the Higgs-Hybrid phase transition is bulk driven, which means that
it is of a five-dimensional nature, independent of the boundary conditions. Its location on the phase diagram was 
determined in \cite{IrgesFotis2} using the $\ve$-expansion and corresponds to the blue line of \fig{npPD}.
Here we repeat the construction to some extent but also generalize it in several ways.
One generalization is related to the renormalization of the anisotropy factor $\g$.
In \cite{IrgesFotis2} it was held constant, a fact that was connected to the identification of the phase transition in the $\ve$-expansion
as a WF fixed point. This is a simplification as the phase transition is really of first order and it was consistent in \cite{IrgesFotis2} only because
the effective action was obtained via a LO expansion in the lattice spacing and no HDO were present.
Here, the effective action is computed to NLO with dimension-6 HDO induced in it and the anisotropy is free to run due to quantum corrections.
The HDO must be suppressed by a scale which defines a cut-off. This cut-off must be an internal to the system scale and is necessary 
to define an effective action without a continuum limit, such as one that is appropriate near a first order phase transition.
Practically this means that in the present analysis the location of the phase transition will be identified by the matching of the RG flows
of the effective 4d couplings $g_4$ and $g_s$ rather than as a 5d WF fixed point. 
This is consistent with the non-perturbative picture because the first method defines a finite cut-off, while the second yields a continuum limit by construction.
To put it in simple words, in the effective action a second order phase transition is seen as a 5d WF fixed point while a first order phase transition
is seen as the point where 4d RG flows meet. 
Of course, these two methods should differ by a small amount in the UV parametrized by the effect of the HDO.
The matching of the 4d RG flows, made possible by our assumptions above, encodes therefore in an indirect way the 5d nature of the phase transition.
The other technical point we need to settle concerns the connection between lattice and continuum parameters mentioned in the Introduction, expressed as \cite{IrgesFotis2}
\be
\m = \frac{F(\b_4,\b_5)}{a_4}\, .
\ee
We have already used this relation in the construction of the continuum effective action from the lattice action, where 
the non-trivial lattice coupling dependence of $F$ was responsible for the appearance of general couplings in front of the HDO.
Now we try to make one more step in the characterization of $F$, using the
fact that here the anisotropy runs.
Let us distinguish $F$ in the Higgs and Hybrid phases by denoting it as $F_4 \equiv F(\b_4,\b_5)$ and $F_s \equiv F(\b_{4,s},\b_{5,s})$ respectively.
Then 
\be\label{F4}
F_4 = a_4(\m) \, \m 
\ee
for the former and 
\be\label{F5}
F_s = a_{4,s}(\m) \, \m 
\ee
for the latter and the difference between $F_4$ and $F_s$ originates from the way that the lattice spacing connects to $\m$ in the two phases.
The observation here is that since the phase transition is a line of UV points, the lattice spacing necessarily sweeps through the same values 
on the two sides near a point on the phase transition, that is $a_4=a_{4,s}$, thus it is a regularization choice to take $F_4(\m)=F_s(\m)$. 
On the phase transition where RG flows from the two sides meet, they assume of course the common value $F_4(\m_\ast)=F_s(\m_\ast)\equiv F_\ast$.
In \cite{IrgesFotis2} it was demonstrated that the choice $F_\ast={\rm const.}$ reproduces qualitatively well the Higgs-Hybrid phase transition 
in the limit that it is a line of WF fixed points, therefore here we continue using this approximation. 

Moving one step further we can express these relations through the lattice spacing of the extra dimension inserting the anisotropy.
Using the tree level relation $\g=a_4/a_5$ and then promoting to running parameters, we obtain
\be\label{F4Fs}
F_4 = a_5(\m) \, \m \, \g(\m) \,\,\,\, {\rm and} \,\,\,\, F_s = a_{5,s}(\m)\, \m \, \g_s(\m)
\ee
in the Higgs and Hybrid phase respectively, with $\g$ and $\g_s$ representing the anisotropy in each phase. Then, using the above approximation, we have that
\be\label{ggsFFs}
\frac{\g(\m) }{\g_s(\m)} \equiv \frac{a_{5,s}(\m)}{a_5(\m)} \, .
\ee

\subsection{Matching RG flows in the Higgs and Hybrid phases}\label{MHHP}

In order to match RG flows, we need to evaluate $g_s(\m)$, the coupling in the Hybrid phase.
According to \cite{IrgesK4,IrgesK5} localization holds approximately near the Hybrid-Higgs phase transition on both sides (see \fig{npPD}), 
while as we move deeper in the Hybrid phase the 5d space becomes more and more layered. In the limit $\b_5=0$, the 5d space is exactly (and trivially) layered.
This means that inside the entire Hybrid phase we see approximate 4d slices with $SU(2)$ gauge group in the bulk and a $U(1)$ 
theory on the boundary in either a Coulomb or a Confined phase, with the slices almost decoupled from each other.
It is sufficient for our discussion to focus on one of the bulk slices. 
This means that we know exactly how to construct the RG flow of its gauge coupling, especially towards the UV where it becomes asymptotically free.
The only point that needs care is the fact that in the bulk too, we have HDO in the effective action.
Fortunately, according to the formalism developed at the end of \sect{Renorm.bf} the corresponding $\b$-function 
is that of the usual 4d Lee-Wick gauge model, \cite{Grinstein}, given in \eq{bgLW} which for $\d_2=\d_3 = 0$, $k_F = -2/3$ and $\d_1 = 2$ becomes
\be
\b_{\a_s} = - \frac{125}{12} \a_s^2  \,\,\,\, {\rm or} \,\,\,\, \b_{g_s} =  - \frac{125}{6} \frac{g_s^3 }{16\pi^2}
\ee
with $g_s (\a_s = g_s^2/16 \pi^2)$ the 4d dimensionless $SU(2)$ coupling.
Its running is given by
\bea\label{gsmu1}
g_s(\m) &=& \frac{g_{s,R}}{\sqrt{ 1+ \frac{125 g^2_{s,R}}{96 \pi^2} \ln\frac{\m^2}{m^2_R} } } \hskip 1cm
{\rm or} 
\hskip 1cm \a_s(\m) = \frac{\a_{s,R}}{1+ \frac{125 \a_{s,R}}{6} \ln\frac{\m^2}{m^2_R} } \Rightarrow \nonumber\\
g_s(\m) &=&  \frac{c_s}{\sqrt{\ln\frac{\m}{\L_s} } } \hskip 1cm {\rm or} \hskip 1cm \a_s(\m) =  \frac{c'_s}{\ln\frac{\m}{\L_s}  }
\eea
with 
\be\label{LQmR}
\L_s = e^{-\frac{c_s^2}{ g^2_{s,R}} } m_R = e^{-\frac{c'_s}{ \a_{s,R} }} m_R 
\ee
and $c_s = \sqrt{48 \pi^2/125}$ and $c'_s = 3/125$.
If we had a general LW gauge model, \eq{gsmu1} would suggest that the coupling has the usual asymptotically free behaviour
reaching zero in the continuum limit.
However here, due to the Higgs-Hybrid transition and the assumption that approaching it from either side drives the system towards the UV,
the running in the Hybrid phase should be related to that of the Higgs phase.
A matching of all physical observables at a generic point along RG flows is expected to be extremely hard but
the matching of gauge couplings only, should be possible at the scale $\m_\ast$.
There, the running of $g_s(\m)$ stops and it never reaches the continuum limit and the model inherits a finite cut-off.
This is rather unusual as it defines a 4d Yang-Mills theory with a finite UV cut-off. One should keep in mind of course that
this phase is not physical from the Higgs phase boundary point of view, it just regulates the Higgs phase, where the interesting physics takes place.

To be more specific, consider the auxiliary running couplings in the Higgs and Hybrid phases, \eq{a4mu} and \eq{gsmu1} respectively and invert
both with respect to the regulating scale.
Then the former gives that
\be\label{mua4}
\m = \exp \Bigl [ \frac{\a_4(\m) - \a_{4,R} }{2 \a_4(\m) \, \a_{4,R} } \Bigr] m_R\, ,
\ee 
while in the Hybrid phase that
\be\label{muas}
\m =e^{\frac{c'_s}{\a_s(\m)}} \L_s \, .
\ee 
As both move towards the UV trying to reach the phase transition at a common point where $\m=\m_\ast$, they must necessarily assume common values.
Hence, equating \eq{mua4} with \eq{muas} we get 
\be\label{a4ma4Ras}
\frac{\a_4(\m) - \a_{4,R} }{2 \a_4(\m) \, \a_{4,R} } - \frac{c'_s}{\a_s(\m)} = \ln \frac{\L_s}{m_R}\, ,
\ee
where $m_R$ is not independent from $\L_s$ due to \eq{LQmR}.
Using this, we obtain the relation 
\be\label{a4masm}
\frac{\a_4(\m) - \a_{4,R} }{2 \a_4(\m) \, \a_{4,R} }  = c'_s \frac{\a_{s,R} - \a_s(\m)}{\a_s(\m)\a_{s,R}}
\ee
which makes the relation between the running of the couplings in the two sides of the phase transition explicit.
Here comes the crucial step, since we are interested in finding the scale $\m_\ast$ where the RG flows meet and the couplings coincide.
This results in
\bea
\frac{\a_4(\m_\ast) - \a_{4,R} }{2 \a_4(\m_\ast) \, \a_{4,R} } &=& c'_s \frac{\a_{s,R} - \a_s(\m_\ast)}{\a_s(\m_\ast)\a_{s,R}} \Rightarrow \nonumber\\
\frac{\a_\ast - \a_{4,R} }{2 \a_\ast \, \a_{4,R} } &=& c'_s \frac{\a_{s,R} - \a_\ast}{\a_\ast \a_{s,R}} \Rightarrow \nonumber\\
\a_\ast &=& \frac{ \a_{4,R} \, \a_{s,R} (1 + 2 c'_s)}{\a_{s,R} + 2 c'_s \a_{4,R} } \, ,\label{aast}
\eea
with $\a_4(\m_\ast) = \a_s(\m_\ast) = \a_\ast$, thereby the cut-off implied by \eq{muas} being equal to
\bea\label{muast}
\m_\ast &=& e^{\frac{c'_s}{1 + 2 c'_s} [ \frac{1}{\a_{4,R}} + \frac{2 c'_s}{\a_{s,R}} ] } \L_s \, .
\eea
The scale where the phase transition occurs as well as $\a_\ast$ in \eq{aast} depend, apart from the input scale $\L_s$, 
on the arbitrary reference values $\a_{4,R}$ and $\a_{s,R}$.
Similarly, $m_R$ is fixed as soon as $\a_{s,R}$ is fixed from \eq{LQmR}.
We add the value of the scalar mass at the phase transition which will be needed later:
\be\label{mhast}
m_{h\ast} = \sqrt{\frac{210}{8\pi^2}} 16\pi^2 v_\ast \a_\ast\, .
\ee

How far is the cut-off $\m_\ast$ from a continuum limit, equivalently how far the first order phase transition implied by the finite
cut-off in \eq{muast} is from a second order phase transition?
For this, let us look at the way that the bulk coupling runs, taking into account the HDO.
The relevant RG equation, using the $\b$-function of \eq{bg5a5} for $\ve=-1$ this time, yields
\be\label{a5m}
\a_5(\m) = \frac{\a_{5,R}}{C \, \a_{5,R} + (1 - C \, \a_{5,R}) \frac{M_R}{\m}  } \, ,
\ee
with $C = 125/12$ and $M_R, \a_{5,R} \equiv \a_5(M_R)$ arbitrary parameters.
Following \cite{IrgesFotis2}, if we demand \eq{bg5a5} to vanish at 1-loop order there appears both a Gaussian and a Wilson-Fisher fixed point 
given by\footnote{We denote the WF fixed point by a 'star', which is (slightly) different from the RG flow matching point, denoted before by an 'asterisk'.}
\be\label{1C}
\a_{5\bu} = 0 \,\,\, {\rm and} \,\,\, \a_{5\star} = \frac{1}{C} = 0.096\, ,
\ee
where $F$, in \eq{ma4}, obtains a constant value\footnote{See the relative discussion in Sect. (4.2.1) of \cite{IrgesFotis2}.} which reads
\be\label{Fstar}
F(\b_{4\star}, \b_{5\star}) \equiv F_\star \simeq 1.51 ,
\ee
taking into account the HDO.
The scale where the WF fixed point is reached is given by 
\bea\label{m5star}
\m_{\star} &=& \frac{\a_{5\star}}{\a_{5,R}} \frac{(C \, \a_{5,R} -1 )}{(C \, \a_{5\star} -1 )} M_R \Rightarrow \nonumber\\
\m_{\star} &\to& \infty
\eea
that is, in the continuum limit. Clearly the difference $\m_\star-\m_\ast$ is not a good distance measure on the phase diagram because it is always infinite.
Instead, we can compute the value of $\a_{5\ast}=\a_5(\m_\ast)$ and compare it to $\a_{5\star}$. Clearly the 
difference $\a_{5\star}-\a_{5\ast}$ is now finite.
For example, with $M_R = m_R = 5.55$ GeV and $\a_{5,R} = \a_{s,R}=0.014 $ (these choices are justified in the following sections), we have $\a_{5\ast} = 0.083$, a value not far from \eq{1C}.
From this example we only keep that $\a_{5\ast} < \a_{5\star}$, showing that the first order phase transition is above the 
WF (blue) line on the phase diagram of \fig{npPD}. This means that moving from the side of the Higgs phase towards the phase transition,
one hits on the first order phase transition before the continuum limit, in the form of a WF fixed point, is met. This is consistent with our main assumption that the phase transition
is a UV point. The interesting fact is that for the other side of the phase transition, this behaviour does not imply the opposite, as one could naively conclude.
For this, we point to Figure 2 of \cite{IrgesFotis2}, where it was shown that the flow beyond the WF point can be characterized as a Landau branch,
where the system, as it moves towards the WF fixed point from above, sees a Landau pole, thus a finite cut-off, not a continuum limit.
Again, this is consistent with the assumption that the phase transition is seen as a UV point also from the other side.
\begin{figure}[!htbp]
\centering
\includegraphics[width=10cm]{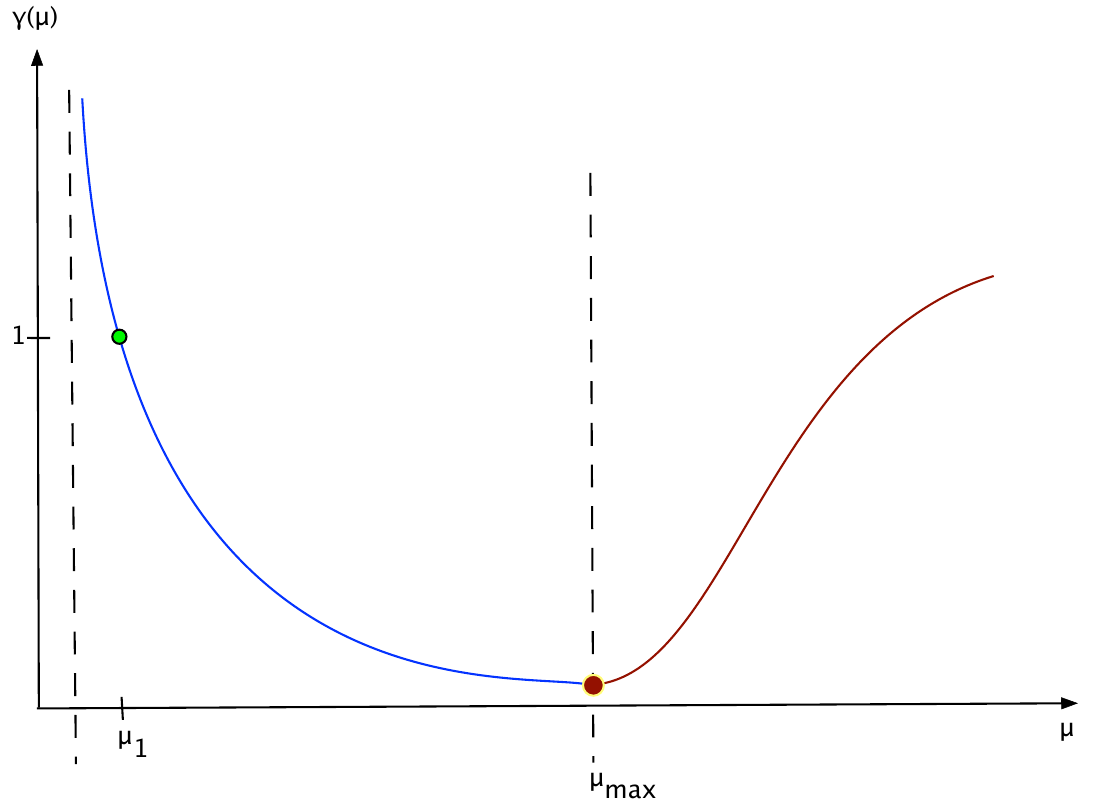}
\caption{\small The running of the anisotropy $\g(\m)$ as a function of the dynamical scale. 
The green bullet represent the scale $\m_1 $ where the anisotropy becomes $\g(\m_1) = 1$ while the red one shows $\g_{\rm min} \equiv \g(\m_{\rm max})$, the minimum value of the anisotropy.
The behaviour of $\g$ between the dashed lines (blue curve) resembles the bulk-driven phase transition of \fig{npPD}.
Passing towards $\m_{\rm max}$ the running of the anisotropy (red line) changes its behaviour and increases with increasing $\m$.
 \label{gammamu}}
\end{figure}
%

The last piece of technical information we are missing is the running of the anisotropy parameter $\g$ and $\g_s$ for the Higgs and Hybrid phase respectively.
Let us start with the former which in principle is fixed by already known results, since at the classical level $\g^2=\b_5/\b_4$ (equivalently use
the relations in \eq{defg4}) and the running of $\b_4$ and $\b_5$ are determined through the running of $\a_4$ and $\a_5$. 
All we need is to promote \eq{defg4} to running couplings, combined with \eq{F4}.
Then we have
\be
g^2_4(\m) = g^2_5(\m) \, \m \frac{\g(\m) }{F_4}
\ee
or
\bea
4 \frac{g^2_4(\m)}{16 \pi^2} &=& 4 \frac{g^2_5(\m) \, \m}{16 \pi^2}  \frac{\g(\m) }{F_4} \Rightarrow \nonumber\\
4 \a_4(\m) &=& \a_5(\m) \frac{\g(\m) }{F_4} \, . \nonumber
\label{gamu1}
\eea
Solving for $\g(\m)$ and following the discussion above \eq{ggsFFs} we get
\be
\g(\m) = 4 F_\ast \frac{\a_4(\m)}{\a_5(\m)} \, 
\ee
and all we have to do is to combine \eq{a4mu} and \eq{a5m} to obtain
\be\label{gamu2}
\g(\m) = 4 F_\ast \frac{\a_{4,R}}{\a_{5,R}}\frac{ C \, \a_{5,R} + (1- C \, \a_{5,R}) \frac{M_R}{\m} }{1 - 2 \a_{4,R} \ln \frac{\m}{m_R}} \, ,
\ee
with $F_\ast $ given by \eq{Fstar} without much loss of generality since the cut-off and WF lines are very close.
Computing the flow, we are lead to \fig{gammamu} which presents the running of the anisotropy parameter.
According to the figure, the running of $\g(\m)$, in the region between the dashed lines, resembles the line of the bulk-driven transition of \fig{npPD}.
The interesting point here is that there is a minimum value, $\g_{\rm min}$, of the anisotropy or a maximum scale, $\m_{\rm max}$, after which the running of $\g$ (red line in \fig{gammamu}) changes its behaviour.
\begin{figure}[!htbp]
\centering
\includegraphics[width=8cm]{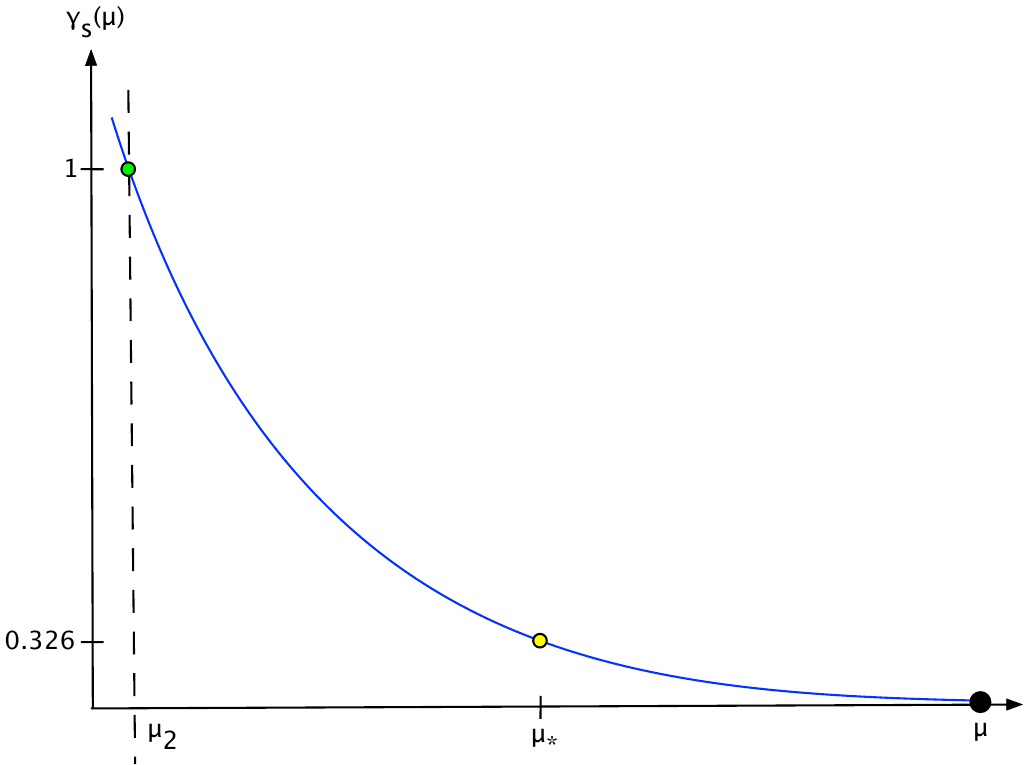}
\caption{\small 
The running of the anisotropy $\g_s(\m)$ as a function of the dynamical scale.
The green bullet represent the scale $\m_2 $ where the anisotropy becomes $\g_s(\m_2) = 1$ while the yellow one the scale $\m_\ast$ at which $\g_s(\m_\ast) \equiv \g_(\m_\ast)$.
The black bullet (Gaussian fixed point), shows the limit $\m \to \infty $ where $\g_s \to 0$, corresponding to the minimum value of the anisotropy.
The behaviour of $\g_s$ stays always similar to the bulk-driven phase transition of \fig{npPD}.
 \label{gsammamu}}
\end{figure}
%
In particular this maximum scale is
\be\label{mumax}
\m_{\rm max} = 32.5 \, W_L( 0.063 \, e^{\frac{0.4974}{\a_{4,R} } } )
\ee
whose value will be useful in the next section.
The function $W_L(x)$ is the product logarithm or the Lambert function.
The running of $\g$ is such that any RG flow that starts from the isotropic point $\g=1$ span over values $\g<1$ in the UV,
which is the direction of approach of the phase transition, when the change of $\b_4$ is mild along the RG flow.

Next we focus on the Hybrid phase and on $\g_s(\m)$.
For the associated running we follow an other path than the Higgs phase using the relation between lattice and continuum couplings.
The needed ingredients are $\b_4(\m)$ and $\b_5(\m)$ as functions of the running couplings $\a_4(\m)$ and $\a_s(\m)$, given by \eq{a4mu} and \eq{gsmu1} respectively.
Then using also that $\g = \b_5/\b_4$ and $\g_s = \b_{5,s}/\b_{4,s}$ we get
\be\label{b4b4s}
\b_4(\m) = \frac{1}{4 \pi^2 \a_4(\m)} \hskip 1cm {\rm and} \hskip 1cm \b_5(\m) = \g^2(\m) \b_4(\m)\, .
\ee
for the Higgs phase and
\be\label{b5b5s}
\b_{4,s}(\m) = \frac{1}{4 \pi^2 \a_s(\m)} \, \hskip 1.5cm \b_{5,s}(\m) = \g_s^2(\m) \b_{4,s}(\m)
\ee
for the Hybrid phase.
Since we do not know how to compute $\g_s(\m)$, we can exploit localization:
Since during the evolution along a 4d slice changes little the localization property, we can safely assume that
$\beta_{5,s}(\m)$ is constant along a flow (we move along $\b_5={\rm const.}$ lines) in \eq{b5b5s}.
Hence,
\bea
\b_{5,s}(\m) &=& \g_s^2(\m) \b_{4,s}(\m) \equiv \b_{5,s\ast} \Rightarrow \nonumber\\
\g_s(\m) &=& \sqrt{ \frac{\b_{5,s\ast} }{\b_{4,s}(\m)} } = 2 \pi \sqrt{ \b_{5,s\ast} \, \a_s(\m) } \label{gsmu}
\eea
where $\b_{5,s\ast}$ is the value of $\beta_{5,s}(\m)$ on the phase transition.
To be more quantitative let us choose $\a_{s,R} = 0.014$, $\a_{4,R} = 0.00435$ (these choices are justified in the following) and demand that $\g_s(\m_\ast) \equiv \g(\m_\ast)$.
This suggests that
\be\label{b5sast}
\b_{5,s\ast} \simeq 0.6
\ee
fitting to a good approximation the non-perturbative results \cite{IrgesK5,IrgesK6}.
Computing the flow, with these choices, we are lead to \fig{gsammamu} which presents the running 
of the anisotropy parameter whose form resembles the blue line of \fig{npPD}.
Note that even though $\g(\m_1) = \g_s(\m_2) = 1$ the associated scales do not admit the same value, so $\m_1 \ne \m_2$.

\subsection{Physics in the vicinity of the phase transition}\label{OTPT}

Now we are ready to return to the numerical discussion and ask the sharper question of whether the RG dynamics
allows realistic numbers to be generated. In order to facilitate the discussion we collect the relevant equations 
and simplify the notation, by defining $\a_{4,R}\equiv x$, $\a_{s,R}\equiv y$ and $c_s'\equiv c$. We have \eq{LQmR}, \eq{muas} and \eq{mhast} that 
read in this notation
\be\label{hierarchy}
m_R = \L_s e^{\frac{c}{y}}, \hskip .5 cm \m_\ast = \L_s e^{\frac{c}{\a_\ast}}, \hskip .5cm m_{h\ast} = \sqrt{\frac{210}{8\pi^2}} 16\pi^2 v_\ast\a_\ast
\ee
where $c=3/125$ and
\be\label{asxy}
\a_\ast = (1+2c)\frac{xy}{y+2cx}\, .
\ee
If we keep $\L_s$ fixed, the model is parametrized by the constants $x$, $y$ and $v_\ast$.
A necessary condition for the validity of the effective action is that these scales obey the hierarchy
\be\label{mRhast}
m_R < m_{h\ast} < \m_{\ast}\, .
\ee\label{effineq}
Beyond this constraint, we would like to see if we can generate in addition a Standard Model-like spectrum, that is
\be\label{SMspec}
m_{h\ast} \simeq 125\, {\rm GeV} \hskip .5cm {\rm and}\hskip .5 cm\r_{\rm bh} > 1
\ee
It turns out that the window of parameters that solve \eq{hierarchy} is small. 
If we focus on \eq{mRhast} the first restriction we get is $\a_{\ast} < y$ which combined with \eq{asxy} suggests the condition $x < y$.
Such a case is expected
to be obeyed between a confined and a deconfined coupling and then we get the right hierarchy only in the range $0< x < 0.01$ approximately, for
any reasonable value of $v_\ast$, which we will assume to be $v_\ast \sim O(100\, {\rm GeV})$.
If in addition we impose \eq{SMspec}, the solution becomes even more constrained.
There are four variables, $x, y, v_\ast$ and $\L_s$ which need care two of which can be fixed by a physical motivation.
In fact, we can set $g_{s,R}$ equal to the SM strong coupling $g (m_Z) \simeq 1.5$ which fixes $y \simeq 0.014$.
For $\L_s$ there are two interesting scenarios, $\L_s \equiv \L_{\rm QCD} \simeq 200$ MeV and $\L_s \equiv m_p \simeq 1000$ MeV, with the latter equal to the proton mass.
We give two examples: for $\L_s\simeq 200$ MeV, $x\simeq 0.00330$, $y\simeq 0.014$ and $v_\ast\simeq 142$ GeV, 
we get $m_R\simeq 1.11$ GeV, $m_{h\ast}\simeq 125$ GeV, $\m_\ast\simeq 223.3$ GeV and $\r_{\rm bh}\simeq 1.2$.
The value of $\r_{\rm bh}$ increases towards its SM value $\sim 1.38$ if we use our second choice for $\L_s$, which corresponds to $1000$ MeV. 
Then the above range for $x$ shifts by a bit and 
presents an alternative set of numbers: for $x\simeq 0.00435$, $y\simeq 0.014$ and $v_\ast \simeq 108.2$ GeV we obtain 
$m_R\simeq 5.55$ GeV, $m_{h\ast}\simeq 125.1$ GeV, $\m_\ast\simeq 209.1$ GeV and $\r_{\rm bh}\simeq 1.373$.
Note that the latter implies $m_{Z\ast} \simeq 91.1$ GeV for the gauge boson mass and $c^{(6)}_{1\ast} \simeq 0.12$ for the HDO coupling \eq{mincond.2}, 
therefore in this example the observables take quite Standard Model-like values.
Furthermore, the obtained $m_R$ and $y$ justify our discussion below \eq{m5star} which shows that the model reaches its first order phase transition before the continuum limit.
Finally, $x$ (or $\a_{4,R}$) for the above two cases when inserted in \eq{a4LP} gives $\m_{4,\rm L} \simeq 2 \times e^{60} $ GeV and $\m_{4,\rm L} \simeq 1 \times e^{50} $ GeV respectively.
This implies that $\m_\ast << \m_{4,\rm L}$ and hence the model remains consistent.

We are now ready to construct the perturbative phase diagram using the above numerical analysis and draw RG flow lines on it.
\begin{figure}[!htbp]
\centering
\includegraphics[width=11cm]{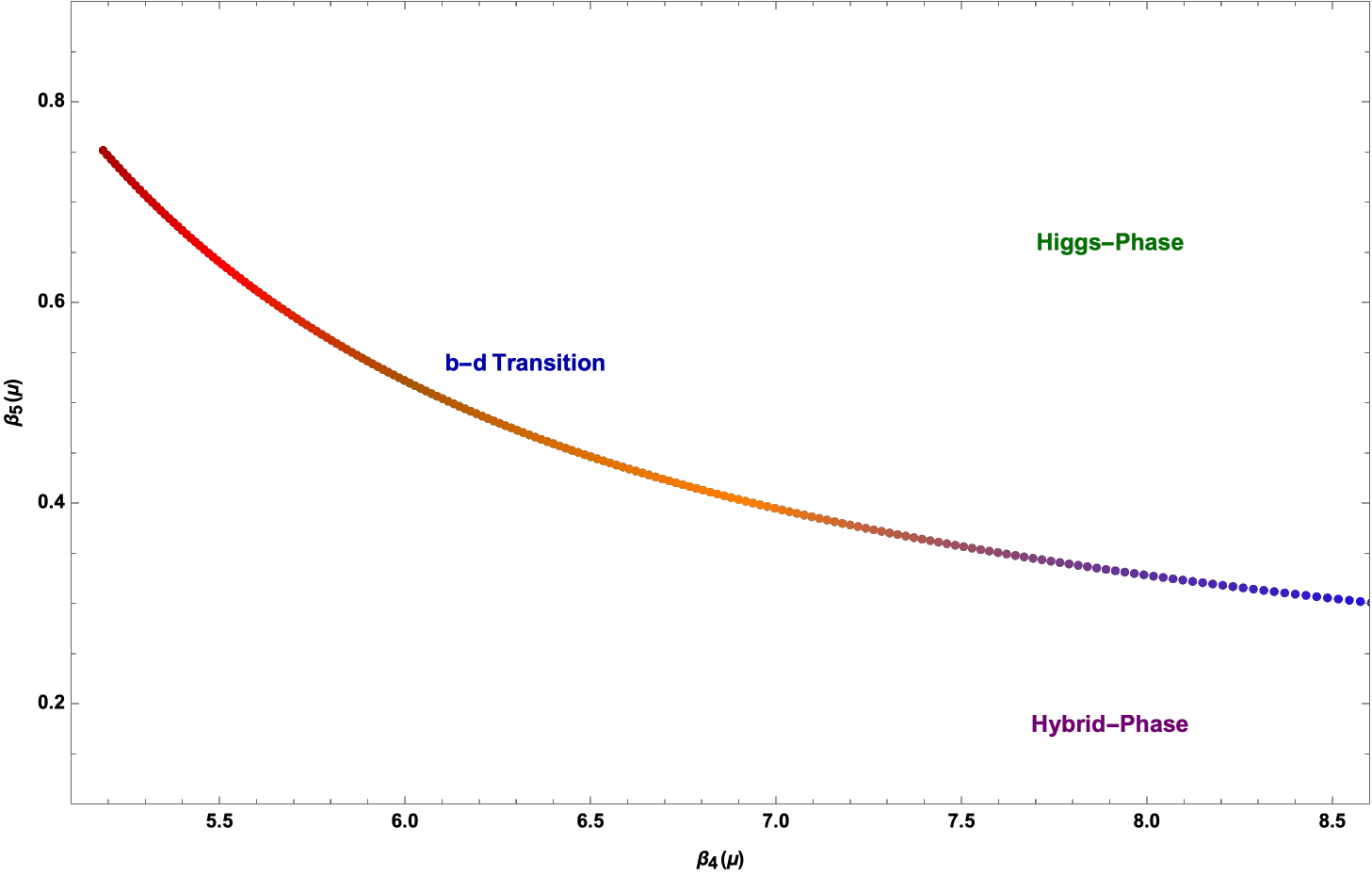}
\caption{\small 
The perturbative phase diagram.
The featuring phase transition is weak first order and bulk-driven (b-d Transition) separating the Higgs from the Hybrid phase.
The phase line is constructed by pairs of $(\b_{4\ast}, \b_{5\ast})$ obtained at the cut-off scale $\m_\ast$ by means of the matching procedure described in the text,
while its colour code reflects that the cut-off scale increases from the IR (left, redder dots) to the UV (right, bluer dots).
 \label{pPD}}
\end{figure}
%
Starting with the phase diagram what we expect is the existence of a weak, bulk-driven, first order phase transition which separates the Higgs from the Hybrid phase, deduced from the matching condition of the flows at the natural cut-off $\m_\ast$.
The (re)construction of the boundary-driven transition is out of the scope of the current work.

Our construction algorithm is the following:
We choose to present the phase diagram on the plane of running couplings $\b_4$ and $\b_5$, which is analogous to the lattice parametrization.
These are connected to the perturbative $\a_4(\m)$ and $\a_s(\m)$ through \eq{b4b4s} and \eq{b5b5s} respectively.
The next step is to express the couplings $\b_4(\m)$ and $\b_5(\m)$ as functions of the matching condition parameters in \eq{hierarchy} and \eq{asxy} that is in terms of $\L_s, y, v_\ast$ and $x$.
Among them the first two are already fixed in our examples and now we choose to keep also $v_\ast$ fixed and let $x$ to run freely.
Here we work with the set of parameters that gives the SM spectrum: $\L_s \simeq 1000$ MeV and $y \simeq 0.014$, which combined 
fix also $m_R \simeq 5.55$ GeV, as well as $v_\ast \simeq 108.2$ GeV.
Then the phase diagram is constructed by varying $x$ and after that collecting the produced pairs $( \b_4(\m_\ast), \b_5(\m_\ast) )$.
As a last comment, the numerical analysis below \eq{SMspec} gives us the upper bound of $x$, $x_{\rm max} \simeq 0.004736$,
for which \eq{hierarchy} admits the minimum cut-off $\m_{\ast,\rm min} \simeq 136.1$ GeV.
It would be nice to also have an upper bound.
Recall that our model stops generating a viable SM spectrum when $\r_{\rm bh} \simeq 1$ which corresponds to $x_{\rm min} \simeq 0.002706$ and $\m_{\ast,\rm max} \simeq 5123$ GeV.
Actually this is an allowed, almost forced upper bound for our example since \eq{mumax} gives for $x_{\rm min}$ 
the scale $ \m_{\rm max} \simeq 5751\, {\rm GeV} > \m_{\ast,\rm max} $, above which the running of $\g$ is given by the undesired red line of \fig{gammamu}.
With these in our mind the perturbative phase diagram of our model is depicted in \fig{pPD} showing indeed the existence of a first order phase transition.
This bulk-driven transition (b-d Transition) is the result of a series of finite cut-offs and 
even though it resembles that of \cite{IrgesFotis2} (blue line of \fig{npPD}) its location is slightly above the latter, 
so the RG flows will hit on it before reaching the WF line (see the discussion below \eq{m5star}).
\begin{figure}[!htbp]
\centering
\includegraphics[width=7.7cm]{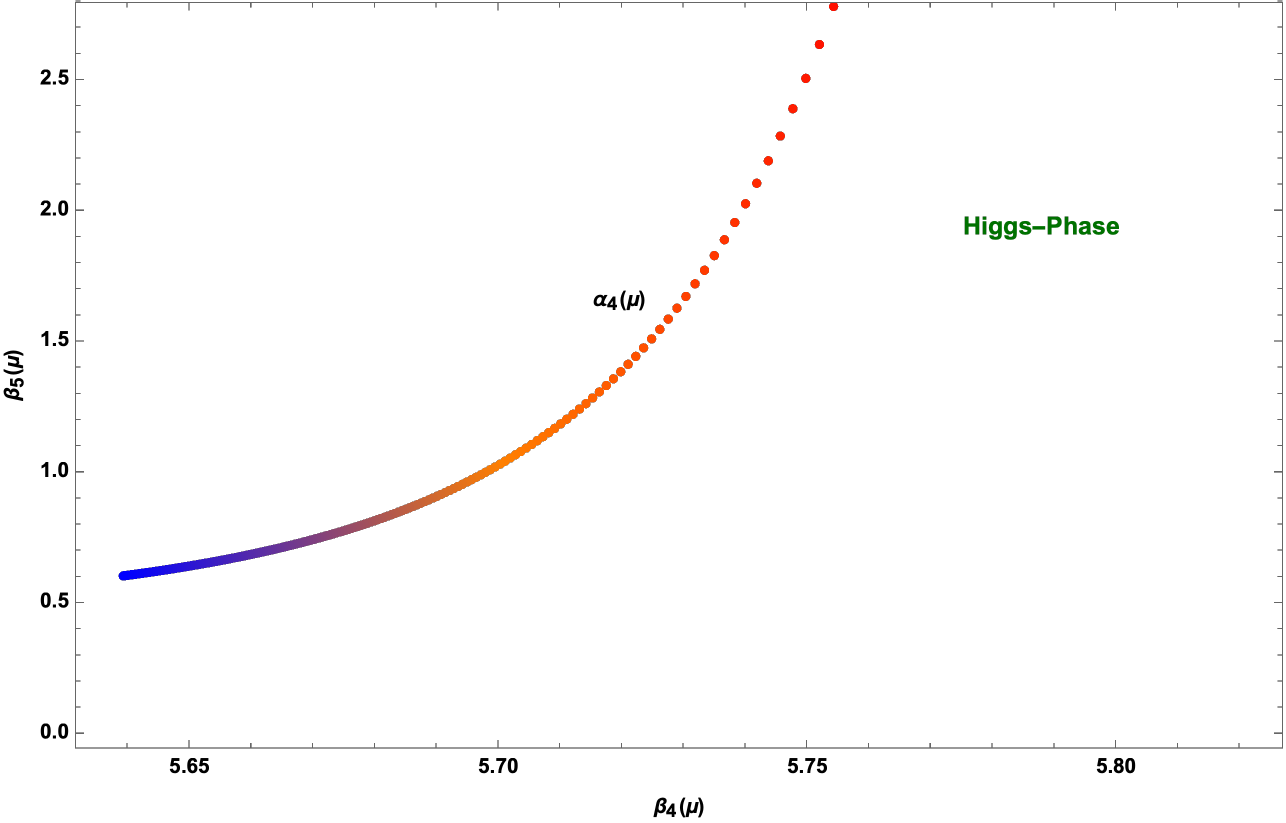}\hfill
\includegraphics[width=7.7cm]{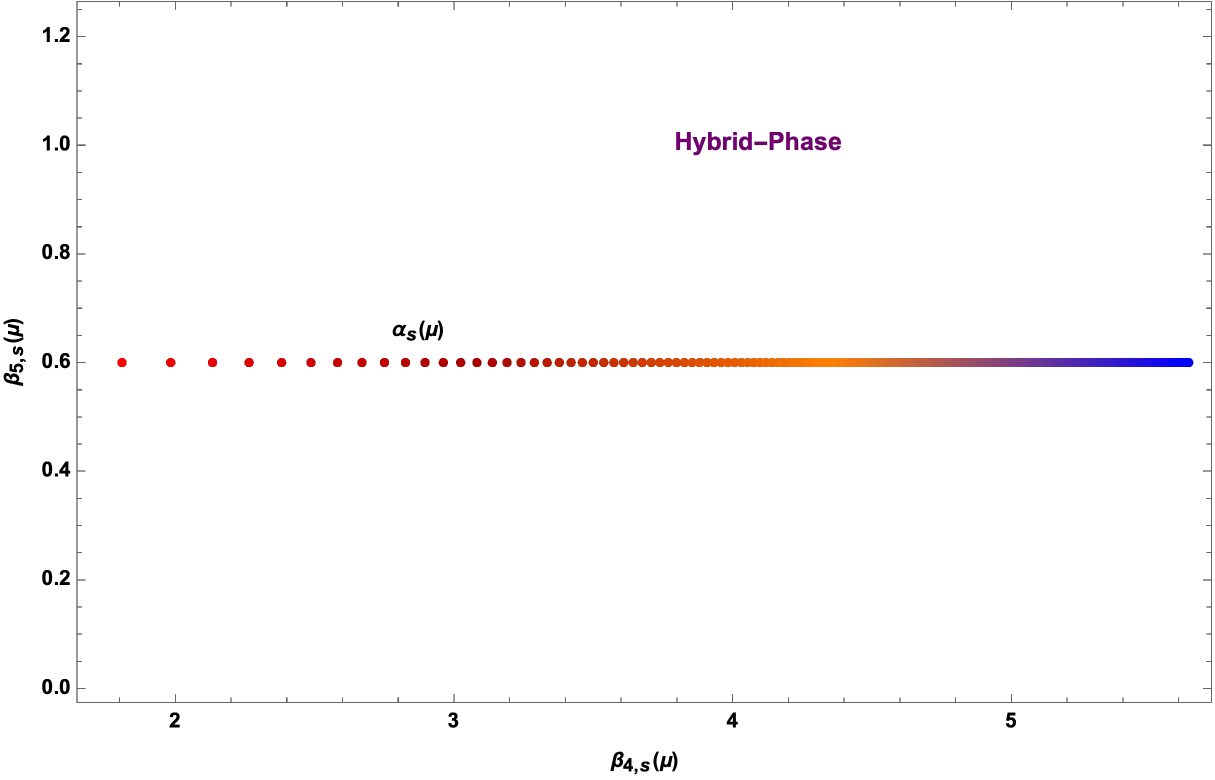}
\caption{\small The RG flow of the running coupling $\a_4(\m)$ in the Higgs phase (left) and $\a_s(\m)$ in the Hybrid phase (right) depicted on the $\b_4-\b_5$ and $\b_{4,s}- \b_{5,s}$ plane respectively.
For the latter the only running parameter is $\b_{4,s}(\m)$ while $\b_{5,s}(\m)$ is kept fixed at $0.6$.
The colouring of the curves represents the increase of the dynamical scale from the IR (redder dots) to the UV (bluer dots) reaching eventually the cut-off, $\m_\ast$.
The latter is the scale where both RG flows end hitting the same point on the line of phase transitions.
 \label{PD1}}
\end{figure}
%
\begin{figure}[!htbp]
\centering
\includegraphics[width=11cm]{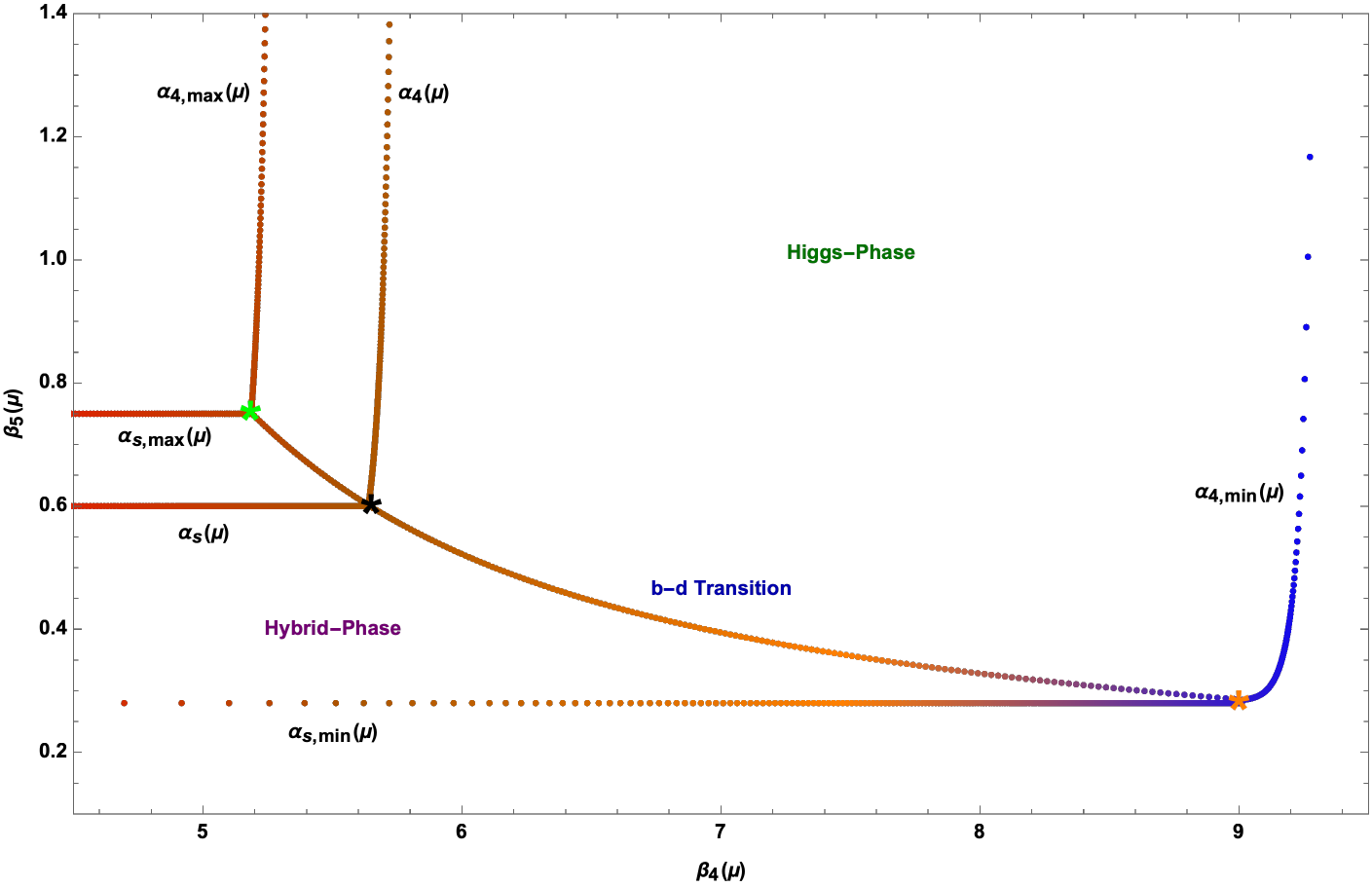}
\caption{\small The running of the couplings $\a_4(\m)$ and $\a_s(\m)$ (Higgs and Hybrid phase respectively) as the RG flow moves towards the phase transition in the UV.
When the cut-off scale is reached the RG flows hit the point $\ast$ of the phase transition.
Here we show three pairs of RG flows whose end points are located 
at $\m_{\ast, \rm min} \simeq 136.1$  GeV (green star), $\m_\ast \simeq 209.1$ GeV (black star) and $\m_{\ast, \rm max} \simeq 5123$ GeV (orange star) respectively.
The flow whose cut-off is $\m_\ast$ and whose flow ends on the black star generates the Standard Model spectrum.
 \label{RGPD}}
\end{figure}
%

For the construction of the RG flows we follow almost the same path with the previous case.
As before the necessary ingredients are the couplings $\a_4(\m)$ in \eq{a4mu} and $\a_s(\m)$ in \eq{gsmu1},
as well as their counterparts \eq{b4b4s} and \eq{b5b5s}. We choose to focus on the RG flow corresponding to the set of parameters
\be\label{fp}
(\L_s,v_\ast,x,y) \equiv (1000 \, {\rm MeV}, 108.2 \, {\rm GeV}, 0.00435, 0.014 )
\ee
which produced the SM-like spectrum.
The difference with respect to the phase diagram algorithm is that now these are kept fixed and the only parameter which runs freely is $\m$.
For the Higgs phase the RG flow is obtained by the simultaneous running of $\b_4(\m)$ and $\b_5(\m)$, \eq{b4b4s}.
In the Hybrid phase on the other hand the only running parameter is $\b_{4,s}(\m)$, according to the discussion below \eq{b5b5s}, 
since $\b_{5,s}$ is constant and given by \eq{b5sast} when the set of values in \eq{fp} is used.
Collectively these arguments result in the running couplings in \fig{PD1}.
Keep in mind that the RG flow in both phases, when the common cut-off scale is reached hits on the same point of the phase transition 
and the associated running ends, a consequence of the matching condition.
In the current scenario this scale is $\m_\ast \simeq 209.1$ GeV and corresponds to the point $( \b_{4\ast}, \b_{5\ast} ) = ( 5.64, 0.60)$.
Of course this is not the only possible choice since by changing $x$ we get RG flows which 
end on different points on the phase transition, corresponding to different cut-off scales.
For example for $x \equiv x_{\rm min} = 0.002706$ and $x \equiv x_{\rm max} = 0.004736$ 
we obtain $m_{h\ast}\simeq 78.3$ GeV, $\m_\ast\simeq 5123$ GeV, $\r_{\rm bh}\simeq 1$ 
and $m_{h\ast} \equiv \m_\ast \simeq 136.1$ GeV and $\r_{\rm bh}\simeq 1.43$ respectively.
The associated endpoints are $( \b_{4\ast}, \b_{5\ast} )_{\rm min} \equiv ( 9.08, 0.28)$ and 
$( \b_{4\ast}, \b_{5\ast} )_{\rm max} \equiv ( 5.2, 0.75)$, while neither case gives a SM-like spectrum.
Note that if we had changed, apart from $x$, also the other variables we could have come up with different RG flows which however would not refer to the phase diagram of \fig{pPD}.
This is the reason why in the above examples we varied only $x$ keeping $\L_s, y$ and $v_\ast$ fixed.
The simultaneous running of the above RG flows towards the first order phase transition is presented in \fig{RGPD}. 
%
\begin{figure}[!htbp]
\centering
\includegraphics[width=11cm]{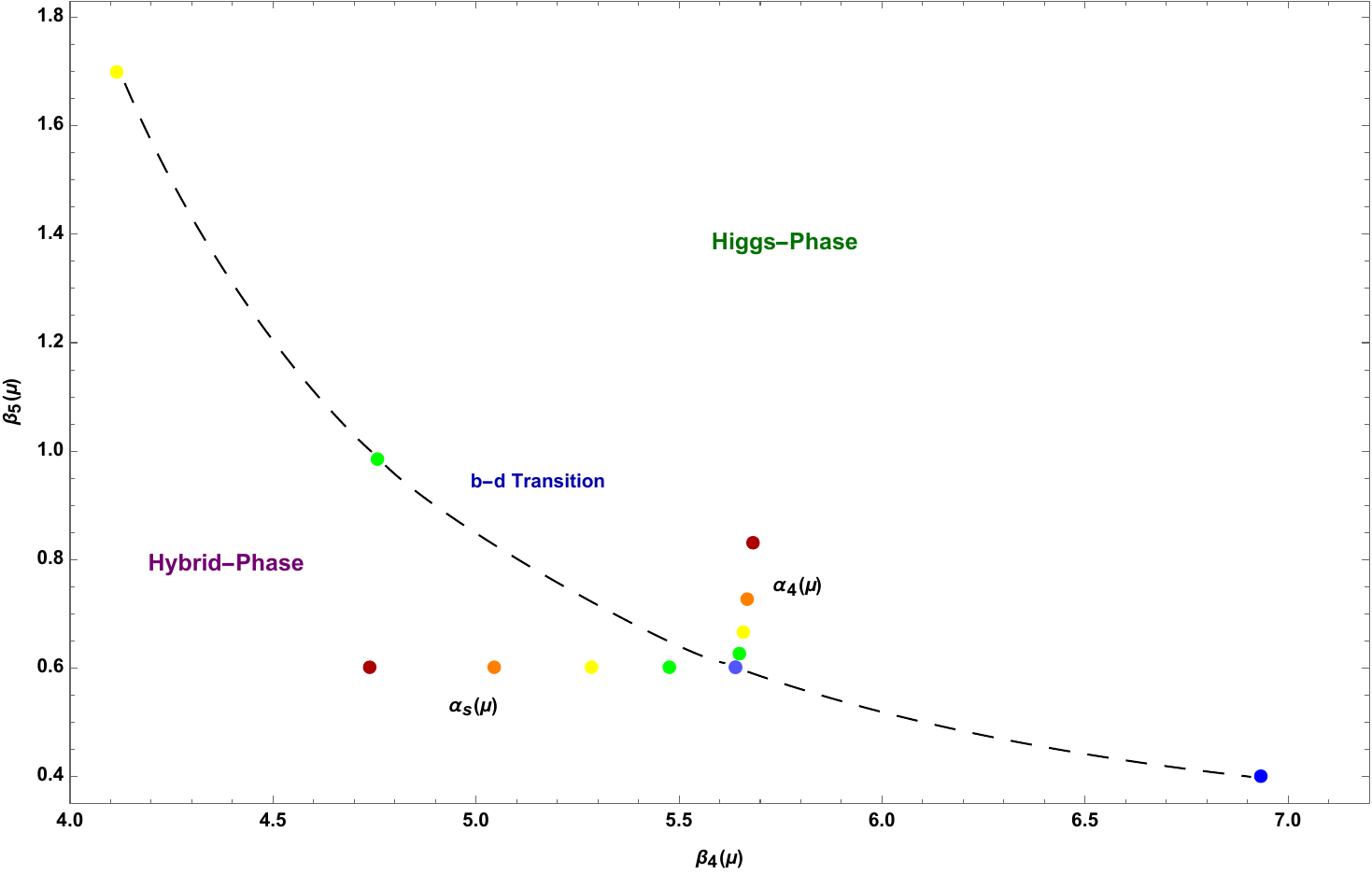}
\caption{\small Zoom around the phase transition for the SM-like example in \eq{fp}. The colour code convention regarding the scale value of each point follows light frequencies.
 \label{PDRGL}}
\end{figure}
%
In \fig{PDRGL} we zoom around the phase transition for the set of parameters in \eq{fp} where we show a few points of \fig{RGPD}.
In this plot, the colour code is strict for all points. The colour of each point represents the value of $\m$ that corresponds to it, with red
representing the IR, blue the UV and with intermediate frequency colours representing analogous intermediate scales. 
Points of same colours along the RG lines have the same $\m$, which illustrates explicitly the correlation 
of the RG flows in the two phases. The flows meet on the phase transition on the blue point where $\m_\ast=209.1$ GeV.

Up to now, we have not said much about the HDO which played the role of the potential in the boundary effective action.
Let us now comment on the character of this operator using the scaling dimension, $\Delta_{O^{(l)}}$, following the notation in \cite{FotisLetter2}.
In our starting Lagrangian $c_1^{(6)}$, before its connection to $g_4$, had the role of the quartic coupling of the potential so the operator of interest is
\be
O^{(6)}_1 = (\bar \phi \phi) \Box (\bar \phi \phi) \, . \nonumber
\ee
Its classical dimension is given by
\bea
d_{O^{(6)}_1} &=& 2 + 2d - 4 \nonumber\\
 &=& 2d - 2
\eea
and in $d=4$ it is $d_{O^{(6)}_1} = 6$, as expected.
We are interested in the way that the behaviour of this operator changes due to the quantum corrections.
Following standard terminology, the nature of an operator is decided by the quantity $\Delta_{O^{(l)}}-d$ according to:
\be\label{Oprule}
 \Delta_{O^{(l)}}-d :
\begin{cases}
< 0 \rightarrow {\rm relevant}\\
> 0 \rightarrow {\rm irrelevant}\\
=0 \rightarrow {\rm RG\, equation}
\end{cases}
\ee
and an appropriate definition for the scaling dimension $\Delta_{O^{(l)}}$ is
\be\label{sc.d.Ol}
\Delta_{O^{(l)}} = d_{O^{(l)}} + \g_{O^{(l)}}\, ,
\ee
with $ \g_{O^{(l)}}$ the anomalous dimension of the operator, defined as
\be\label{an.d.1}
\g_{O^{(l)}} = \frac{\partial \b^1_{c^{(l)}} (c^{(l)})}{\partial c^{(l)} } \Bigg |_{c^{(l)} \to c^{(l)}_\ast} \, ,
\ee
with $c^{(l)}$ the associated to $O^{(l)}$ coupling.
For example, when quantum corrections are turned off, $\g_{O^{(6)}_1}=0$ and then \eq{Oprule} in $d=4$ shows that
\be
\Delta_{O^{(6)}_1} - 4 = 2 > 0
\ee
exposing the irrelevant nature of $O^{(6)}_1$ near the Gaussian fixed point.
On the other hand, when quantum corrections are presented we solve \eq{an.d.1} for the coupling $c^{(6)}_1$ with the associated $\b$-function \eq{betac61}.
In $d=4$ this reads
\be
\g_{O^{(6)}_1} = \frac{c^{(6)}_1}{2 \pi^2} 
\ee
and on the matching point yields for $(\L_s,v_\ast,x,y) = (200 \, {\rm MeV}, 142 \, {\rm GeV}, 0.00330, 0.014 )$ and $(\L_s,v_\ast,x,y) = (1000 \, {\rm MeV}, 108.2 \, {\rm GeV}, 0.00435, 0.014 )$
\be
\g^\ast_{O^{(6)}_1} = 0.022 \,\,\, {\rm and} \,\,\, \g^\ast_{O^{(6)}_1} = 0.025 
\ee
respectively,
which still render $\Delta_{O^{(6)}_1} >0$ and the associated operator irrelevant.
Now this is a puzzle because the only operator in the potential is $O^{(6)}_1$ so somehow it should be able to drive the running in a relevant way.
Indeed notice that it is inserted in the Lagrangian in momentum space as $\frac{p^2}{\m_\ast^2} (\bar \phi \phi)^2$ since it is a derivative operator.
This means that its effects become less (more) important when $p^2$ decreases with respect to (increases towards) $\m^2_\ast$.
Therefore, as the coupling approaches its upper value, the operator itself
\be
\frac{O^{(6)}_1}{\m^2_\ast} \equiv \frac{p^2}{\m^2_\ast} (\bar \phi \phi)^2 \longrightarrow (\bar \phi \phi)^2\, ,
\ee
tending towards a usual quartic term.
Hence, the above renders $O^{(6)}_1$ more and more marginal (or less and less irrelevant) and by taking into account the running of $g_4$ it can be characterized as marginally relevant.

\subsection{Lines of Constant Physics}\label{ELCP}

In our final section we construct Lines of Constant Physics\footnote{An LCP should not be confused with an RG flow.
The latter originate from the usual RG equations which involve only the renormalized couplings while the former, 
in the spirit of Statistical Physics, is determined via the bare couplings. This is of course a characteristic of the lattice formulation.\cite{IrgesK4}} 
for the Boundary-Hybrid action, in the Higgs phase.
The necessary ingredients are two observables as functions of $\m$, which we can take $\r_{\rm bh}(\m)$ and $m_h(\m)$.
An LCP as typically constructed on the lattice is a line in the space of bare couplings along which these observables are constant,
so one would also need a relation between bare and renormalized quantities.
Our effective action however does not contain an explicit bare mass term so the relation between $m_{h,0}$ and $m_h$ is undefined and
the same is true also for the gauge boson mass.
\begin{figure}[!htbp]
\centering
\includegraphics[width=11cm]{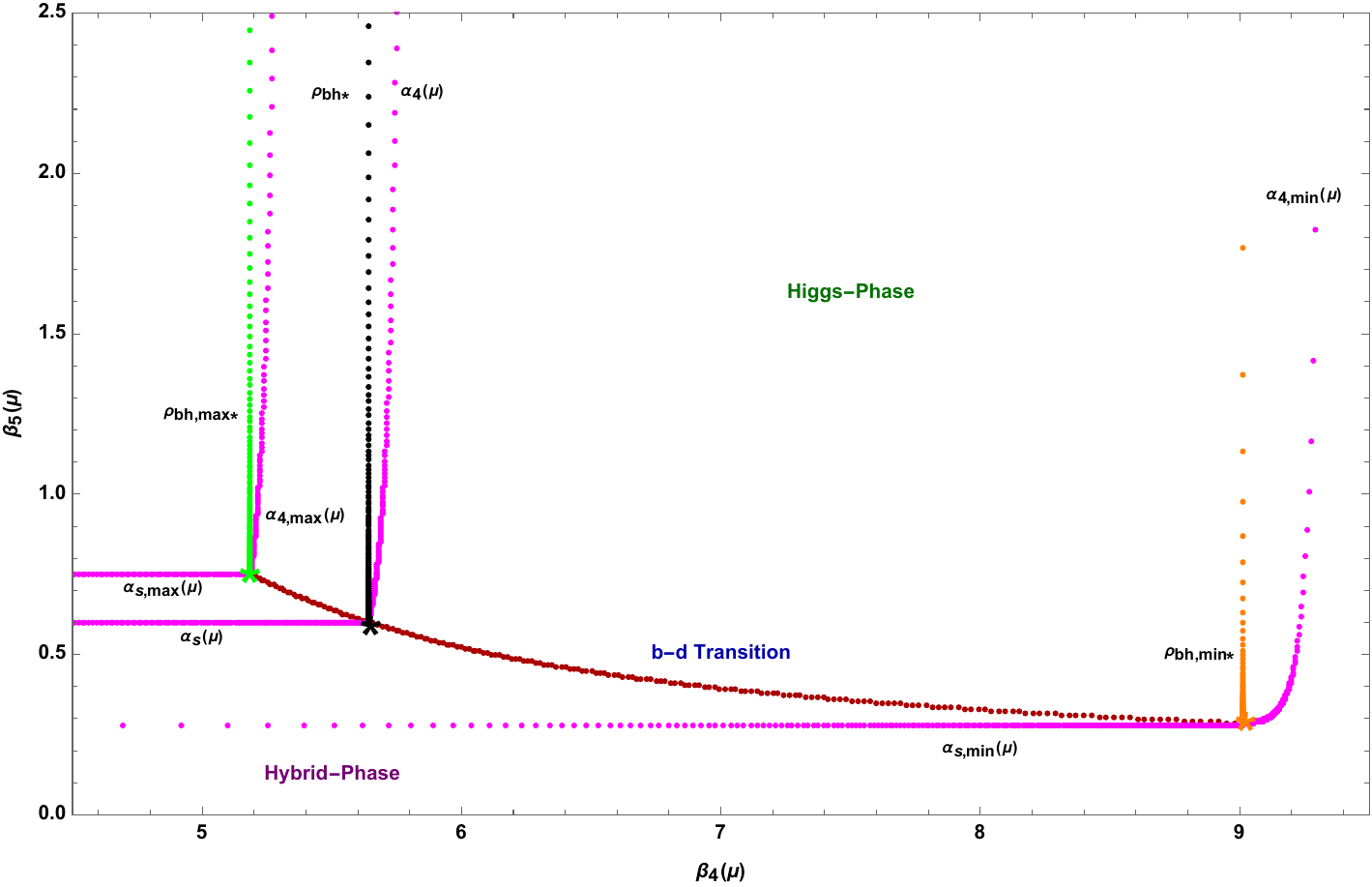}
\caption{\small The Lines of Constant Physics (green, black and orange line) for the Boundary-Hybrid model in the Higgs phase.
Along these lines the mass ratio and the scalar mass are kept fixed. 
However, only the curve of $\r_{\rm bh\ast}$ (black line) corresponds to the Standard Model spectrum.
The other two LCP refer to the cases $ \r_{\rm bh, min\ast} < \r_{\rm bh\ast}$ (orange line) and $ \r_{\rm bh, max\ast} > \r_{\rm bh\ast}$ (green line).
The RG flows of $\a_4(\m)$ and $\a_s(\m)$ (magenta lines) for these three cases have also been drawn for a comparison with the LCP.
 \label{RGPDLCP}}
\end{figure}
%
We must come up with a way of drawing the LCP on our phase diagram whose axes are the running and not the bare couplings.
A strategy can be the following:
Consider the observables mentioned above and recall from \sect{cCWm} that in our scheme the vev stays frozen at $v_\ast = {\rm const.}$
This means that when $\m$ changes then both $\r_{\rm bh}$ and $m_h$ change only as functions of $\a_4$.
As already explained, for the example of \eq{fp} we get $m_{h\ast}\simeq 125.1$ GeV and $\r_{\rm bh_\ast} \simeq 1.373$ 
on the phase transition and we now construct our LCP for this case.
In order to be able to do it, we need a relation between the observables and the 
lattice couplings which here is given by the first part of \eq{b4b4s} and which evaluates on $\m_\ast$ to
\be\label{b4srs}
\b_{4\ast} = \frac{105}{\r^2_{\rm bh}(\m_\ast) \pi^2} \simeq 5.64 \, .
\ee
Bare in mind for later use that for $\r_{\rm bh, min\ast} <  \r_{\rm bh\ast}$ we get $\b_{4,\rm max\ast} > \b_{4\ast}$ while $\r_{\rm bh, max\ast} >  \r_{\rm bh\ast}$ gives $\b_{4,\rm min\ast} < \b_{4\ast}$.
Even though $\b_{4\ast}$ stays fixed along the LCP, this is not the case for $\b_5$ since the second part of \eq{b4b4s}, given the running of the anisotropy, indicates that
\be
\b_5(\m) = \g^2(\m) \b_{4\ast}\, .
\ee
As a consequence our LCP will correspond to vertical lines which start from $\ast$ and move into the Higgs phase, while $\r_{\rm bh}(\m_\ast)$ and $m_h(\m_\ast) $ are kept fixed.
The projection of the LCP on the phase diagram is depicted in \fig{RGPDLCP} by the black line.
The following comments are in order:
First observe that for both the LCP and the RG running there is a lower and an upper bound.
The two bounding scales ($m_R$ and $\m_\ast$ for both the Higgs and the Hybrid phase
in the IR and UV respectively) ensure that the RG flow is short hence only a small amount of fine tuning, for the Higgs mass, is needed.
In the same spirit, the LCP shows that the masses can stay fixed for a short scale region before a change of phase occurs, 
during which the masses are insensitive to changes of the dynamical scale. In this region a stable four-dimensional effective theory with a finite cut-off can be constructed.
For completeness here we have considered also the cases $x \equiv x_{\rm min}$ (or $\r_{\rm bh, min\ast} $) and $x \equiv x_{\rm max}$ (or $\r_{\rm bh, max\ast}$), analyzed in the previous section, whose RG flows were drawn on \fig{RGPD}.
On the phase transition the former gives that $\b_{4,\rm max \ast} \simeq 9.08 $ and the latter that $\b_{4,\rm min \ast} \simeq 5.2$ while the associated LCP correspond to the orange and green line of \fig{RGPDLCP} respectively.

How much tuning is necessary in order to render our model, \eq{SBHrd} enhanced by some non-perturbative information, 
a viable candidate for generating the Standard Model spectrum and at the same time give 
a resolution to the Higgs mass hierarchy problem?
This is an important issue to discuss and we should mention that here, there are two kinds of fine tuning at work.
The first type is the usual quantitative tuning associated with the cut-off, alas being realized in a rather mild way since the presence of the
phase transition as detected by perturbation theory via the matching of the RG flows, enforces systematically a low cut-off.
The second type of tuning is related to the freedom of choosing the initial data $\L_s,v_\ast,\a_{4,R}$ and $\a_{s,R}$
(the bulk parameters $\a_{5,R}$ and $M_R$ are fixed by the Hybrid phase characteristics since both have the same origin.
Hence, we can set $\a_{5,R} \equiv \a_{s,R}$ and $M_R = m_R$ implying that the associated RG flows have the same initial data without loss of generality), which amounts to
picking the RG flow that respects the scale hierarchy constraint \eq{mRhast} and generates a realistic spectrum, as in \eq{SMspec}.
Clearly the picking of a special RG flow involves an infinite amount of fine tuning, but this is the case for any effective theory with a Higgs mechanism.
The quantitative aspect of the choice of the initial parameters is the real fine tuning that we have to discuss in more detail.
In fact, in the previous numerical analysis we were varying only $\a_{4,R}$ while the other three parameters were kept fixed.
To get a better feel for the amount of fine tuning,
let us fix the dimensionfull parameters $\L_s$ and $v_\ast$, let $\a_{4,R}$ and $\a_{s,R}$ vary and see how much fine tuned is a
chosen RG flow line. To be concrete we take the SM-like example with $\L_s \simeq 1000 \, {\rm MeV}$ and $v_\ast \simeq 108.2 \, {\rm GeV}$ (see below \eq{SMspec} for the motivation). 
Then, we have the following cases:
\begin{itemize}
\item Case I: $ \a_{s,R} \ge {\cal O}(10^{-1})$.
For $\a_{4,R} \simeq 0.00435$ we get
$m_R\simeq 1.27$ GeV, $m_{h\ast}\simeq 126.7$ GeV, $\m_\ast\simeq 195.5$ GeV and $\r_{\rm bh}\simeq 1.38$.
This case resembles the SM spectrum, however at $\m_\ast$ \eq{a5m} gives $\a_{5\ast} = 0.096025$ or $\a_{5\star} - \a_{5\ast} < 0 $ 
which suggests that the first order phase transition is below the WF line.
This is forbidden since $\m_\star = \infty$ while $\m_\ast$ is finite.
\item Case II: $ \a_{s,R} \sim {\cal O}(10^{-2})$.
Here there are two sub-cases.
For $ 0.01 \le \a_{s,R} \le 0.024 $ and $\a_{4,R} \simeq 0.00435$ we get exactly the SM spectrum, $m_{h\ast}\simeq 125.1$ GeV and $\r_{\rm bh}\simeq 1.373$, with $m_R\simeq 5.55$ GeV and $\m_\ast\simeq 209.1$ GeV so both constraints, \eq{mRhast} and \eq{SMspec}, are full-filled.
For $ 0.026 \le \a_{s,R} \le 0.098 $ we get a viable spectrum, again only for $\a_{4,R} \simeq 0.00435$, which is similar but not identical with the SM one.
In both sub-cases the first order phase transition is above the WF line since $\a_{5\star} - \a_{5\ast} > 0$.
\item Case III: $ \a_{s,R} \le {\cal O}(10^{-3})$.
For $\a_{4,R} \simeq 0.0054$ we obtain again a viable SM spectrum, $m_{h\ast}\simeq 125.2$ GeV, $\m_\ast\simeq 208.5$ GeV and $\r_{\rm bh}\simeq 1.373$ with $\a_{5\star} - \a_{5\ast} > 0$.
Nevertheless \eq{SMspec} is full-filled, here we get $m_R\simeq 2.64 \times 10^{10}$ GeV which violates \eq{mRhast}.
Note that we can choose the unreasonable scale $\L_s \sim 10^{-10} \, {\rm GeV}$ to get the same, valid, spectrum however, the cut-off now reads $\m_\ast \sim 10^{-8}$ GeV rendering also this choice problematic.
\end{itemize}
In the realistic scenario, Case II, the amount of fine tuning corresponds
to forcing $\a_{s,R}$ to remain in the range $0.010 \le \a_{s,R} \le 0.098$, since $\a_{4,R} \simeq 0.00435$ stays essentially constant when 
\eq{mRhast} and \eq{SMspec} are satisfied.
For completeness let us briefly refer to the case where $\a_{4,R} \simeq 0.00435$ and $\a_{s,R} \simeq 0.014$ are kept fixed while $\L_s$ and $v_\ast$ are free to run.
In the same spirit with the above, \eq{mRhast} and \eq{SMspec} are full-filled when $\L_s$ remains in the range $0.6 \, {\rm GeV} \le \L_s \le 16 \, {\rm GeV}$ while $v_\ast$ admits the value $v_\ast \simeq 108.2 $ GeV.

In conclusion, if we imagine that $\L_s$ and $v_\ast$ are somehow known, then the fine tuning of an RG flow that respects 
the physical constraints \eq{mRhast} and \eq{SMspec}, is equal to or less than $O(10^2)$.
What is interesting in this way of quantifying the fine tuning is that the lattice may actually be capable of fixing these dimensionfull parameters.

\section{Conclusion}

We constructed the 1-loop effective action of an $SU(2)$ gauge theory in five dimensions with boundary conditions that leave a
$U(1)$-complex scalar theory on the boundary, located at the origin of a semi-infinite fifth dimension.
At an exclusively perturbative level, the boundary theory is a version of the Coleman-Weinberg model where the quartic term
is replaced by a dimension-6 derivative operator. A qualitatively similar to the CW model Higgs mechanism is at work but with different coefficients
in the scalar mass and the $\b$-functions that change things towards a more realistic direction.
If in addition we impose on the effective action non-perturbative features known from the lattice, the system becomes highly constrained.
The picture is that the model possesses a non-trivial phase diagram where the phases are separated by first order, quantum phase transitions located in the UV.
If we are interested to use the model as a cartoon of a possible origin of the Standard Model Higgs sector, then it turns out that
we have to sit on, or near the interface of the phase transition that separates the Higgs phase and a layered-type of phase
which in the orbifold model is called the Hybrid phase. There, dimensional reduction happens via localization in both phases
and the effective action must be constructed with a dynamically generated finite cut-off but also with RG flows that 
are correlated below and above the phase transition.

The connection between perturbative and non-perturbative effects is made possible by \eq{ma4} with the function $F(\b_4,\b_5)$
approximated by a constant near the phase transition. The numerical value of the constant can be chosen to leading order 
(i.e. for an action with up to dimension-4 operators) so that it reproduces
the line of phase transitions determined in lattice simulations, as shown in part I of this work.
The importance of this relation can be also seen here in part II by considering the truncation of the action that the lattice action generates.
In principle, the expansion in the lattice spacings generates, in addition to the dimension-4 operator, an infinite tower of higher dimensional operators, of which we have kept here only
the lowest order, dimension-6 contribution. Keeping the entire tower would just reproduce of course the lattice plaquette action, whose other non-perturbative features
we also know from lattice Monte Carlo simulations. To add to the phase transition, we mention the localization in its vicinity and the existence
of a Higgs phase with a definite mass spectrum. We have seen that the truncated version of the action reproduces consistently the latter feature,
which gives us confidence that it is a reasonable truncation, all the effects of the neglected higher dimensional ($>6$) operators being
packaged in the chosen value for the constant $F$. Indeed, the numerical value we use in part II is slightly different from the one 
in part I, due to the fact that here we have kept the next to leading order, dimension-6 operator in addition to the dimension-4 operator of part I.
The Monte Carlo analysis therefore ensures that keeping operators of even higher dimension, 
can not alter qualitatively the physics near the phase transition but it would most likely further shift the value of $F$ by a bit.
Suppose on the other hand that we place ourselves in a regime of the phase diagram where there is no localization.
Such a regime is for example the interior of the Higgs phase, far from the quantum phase transition.
Non-perturbatively this results in a nonsensical spectrum: any mass $m$ in units of 4d lattice spacing becomes $m a_4>1$,
implying that the interpretation of the spectrum as the spectrum of a four-dimensional theory is wrong. This regime is
genuinely five-dimensional and all the issues of the nonrenormalizability of higher dimensional gauge theories are present.
In the continuum effective approach this would be reflected by the impossibility to find a solution to the 
system of inequalities \eq{effineq}: one would not be able to construct a meaningful four-dimensional quantum theory with a spectrum
not dominated by the cut-off. The only issue left to discuss regarding the truncation concerns the choice of the renormalization scheme, $p^2=\m_\ast^2$.
Apparently, all higher derivative operators containing a factor of the form $(\Box/\m_\ast^2)^m$ in this scheme become equally important as our dimension-6 operator.
Again, the lattice comes to our rescue. Apart from the indirect argument related to $F$ presented above, we note that these operators appear
classicaly with ever decreasing coefficients since they originate from the expansion of the 'exponential' plaquette action. 
They will still contribute after renormalization, but they will not change the physics since we know the full non-perturbative picture from Monte Carlo simulations,
which is already successfully reproduced by the truncated to dimension-6 version.
The property that seems to be purely non-perturbative though, is the localization. It is a fact that has been established by simulations
and it is only indirectly confirmed here by the sensible four-dimensional boundary effective action that we have analyzed. Its detailed dynamical nature
is not understood up to date however, beyond lattice simulations. 
In short, localization in a specific regime of the phase diagram is a non-perturbative fact that must be built in our effective action by hand and its validity can be confirmed a posteriori 
by the consistency (internal and with respect to the lattice simulations) of the dimensionally reduced effective action.
It is clear from this discussion that the truncation and localization issues are correlated.
The closer we are to the phase transition, the more the boundary decouples from the bulk and a more truncated 4d effective action can reproduce the physics.
Exactly on the phase transition, its location can be reproduced by an action truncated to dimension-4. Moving a bit away from the phase transition into the 
interior of the Higgs phase, in order to reproduce the spectrum we need at least a dimension-6 operator. Moving further in the interior,
we need to add more and more operators, finally reproducing the 5d plaquette action and localization is lost.
These considerations extend also to the generalizations that couple the gauge fields to gravity.
In order to be consistent with observations, gravity must also be localized (at least) on the boundary. Clearly such a dynamical mechanism for gravity
is even less understood analytically than the localization of gauge fields but contrary to the latter, models coupled to gravity are not amenable to numerical lattice investigations at present.
Any model of this kind that involves gravity must therefore have localization built in as an assumption and the validity of this assumption must be checked 
by the internal consistency of the construction.

We summarize our main technical results:
\begin{itemize}
\item The $\b$-functions of the boundary couplings \eq{betag4}, \eq{betac61} and of the bulk coupling, \eq{bg5a5}.
\item The renormalized scalar potential of the boundary effective action near the 5d quantum phase transition, \eq{Vimpm},
generated exclusively by derivative operators. In addition, using the observation that quantizing the action with higher dimensional operators is equivalent in the spirit
of generalized effective field theory to quantizing it without them but including them instead in the effective theory as expectation values of quantum operators \cite{FotisLetter2}, 
we saw that the scalar potential of \eq{Vimpm} is responsible for the simultaneous, spontaneous breaking of scale and gauge symmetries.
A crucial fact that ensures this property is that the scale suppressing the higher dimensional (derivative) operator is the internal scale $\m_\ast$,
the finite value of the regulating scale generated by the first order quantum phase transition in the interior of the phase diagram, present due to its five-dimensional origin.
\item The boundary mass spectrum in the Higgs phase, given in \eq{mfr} and \eq{mA3}.
\item The correlated running of the respective 4d effective couplings in the Higgs and Hybrid phases near the phase transition, \eq{a4masm}.
The former is a boundary flow and the latter is a bulk flow, governed by common values of the regulating scale $\m$. 
They meet on the phase transition where $\m=\m_\ast$ is finite.
\item The RG flows in the Higgs and Hybrid phases shown in \fig{PD1}, \fig{RGPD}, the zoom around the phase transition in \fig{PDRGL} and the Lines of Constant Physics in \fig{RGPDLCP}.
\end{itemize}

An important question is if we have provided in the context of our construction with an alternative resolution to the Higgs mass hierarchy problem. 
The fine tuning involved is about one part in a hundred and it is of a different
type compared for example to the supersymmetric SM, being related to the choice of a "physical RG flow" on the phase diagram.
More specifically what we saw is that the dynamics do not allow a high cut-off for the effective action, which indeed does ameliorate the fine tuning problem.
On the other hand, in order to satisfy the constraint of a proper scale hierarchy, a considerable fine tuning of the free parameters 
(the initial data for the gauge coupling RG flows and the scalar vev) is
necessary and this becomes more enhanced when a realistic spectrum is asked for.
This essentially amounts to a fine tuning in the process of picking a physical RG flow line (equivalently of a Line of Constant Physics)
out of a continuum of RG flows on the phase diagram. Once such a physical RG flow is picked, there is very little fine tuning
that takes place along it.

The phenomenology of models of localization is expected to be quite different from the phenomenology
of models with a Kaluza-Klein spectrum. We have not performed a related analysis in the present paper,
instead we leave it for a future work.
On the other hand it should be possible to do several related measurements on the lattice.
For example, the numerical methods of \cite{Roman} could be applied to see to which extent the potential \eq{Vimpm} agrees with 
the full non-perturbative potential. A question that arises here is whether on the lattice, the scalar potential is a potential for
a fundamental field or a composite. Perturbation theory can not tell the difference nonetheless
the two approaches should agree on the shape of the potential.
It is also most likely possible to measure numerically the quantity $\L_s$ in the Hybrid phase, reducing 
further the parameter freedom to only $\a_{4,R}$, $\a_{s,R}$ and $v_\ast$. Moreover it is possible that the above will also
give some information about $\a_{s,R}$ and measurements of scalar Polyakov loop expectation values 
in the Higgs phase most likely can constrain $v_\ast$.

Let us close the current section with a comment on the possible effect on our boundary effective action 
\eq{SBHrd} of fermionic degrees of freedom and leave a more thorough analysis for a future work.
Part of the motivation behind such an extension is the fact that fermions play a catalytic role in the 
validity of both CW-like models and the Standard Model since, when dominant, they may render the effective potentials unstable.
Crucial for this to happen is the coupling of the top quark (the heaviest fermion of the SM) to the Higgs boson be stronger than both the self and the gauge-Higgs couplings.
Then, beyond a critical scale the former dominates the running of the quartic coupling which then takes large and negative values. 
A fair question is whether such a situation could be realized in our model.
In our case the starting point would be a 5d orbifold lattice with an $SU(2)$ gauge field minimally coupled to massless fermions.    
Even though the introduction of fermionic degrees of freedom on the lattice is non-trivial, let us assume that 
we have followed analogous steps as described in this work and ended up with a classical continuum Lagrangian.
It is then expected that in the continuum, the boundary action will 
inherit extra derivative operators some of which generate fermion-scalar vertices.
These extra vertices, being fermionic, will have the usual negative-sign contribution to the 
1-loop effective potential and this could be the root of an instability.
However, in our prescription there is only one 5d gauge coupling, $g_5$, which appears in the form of
$g_4$ on the boundary for both the gauge-scalar and the fermion-scalar vertices.
Due to this universality the Yukawa couplings will share the same order of magnitude and the same running properties 
with the gauge coupling so, their contribution to the effective potential will not be dominant and the latter will stay stable.
On the other hand, under the presence of fermions there would be changes in the constraints 
\eq{mRhast} and \eq{SMspec} which, however, are not expected to affect the current picture drastically. 
Suppose finally that, for an unforeseen reason, the above scenario fails and the top quark dominates 
violently the running of our quartic coupling when a critical scale is reached.
The SM prediction for this scale is $\m \sim 10^{10}$ GeV and lies far beyond, say, the cut-off scale $\m = \m_\ast \simeq 209.1$ GeV of our prime example, where 
our boundary model exhibits a first order phase transition.
In other words, the boundary action \eq{SBHrd} enhanced with fermions will never reach the instability 
scale hence the potential as well as the associated SSB mechanism will remain valid.

\begin{appendices}

\section{Review of the NPGHU model}\label{review}

In part I, the Non-Perturbative Gauge-Higgs Unification model was computed at leading order in the
lattice spacing expansion in \cite{IrgesFotis2} such that only operators of classical dimension less or equal to $d$ were inserted in the effective action.
Here we perform an extensive review of the main characteristics of this calculation and we move one step further, including next to leading order effects.
We show that like this, Higher Dimensional Operators are introduced and we recalculate the extended continuum effective action.   

Let us start with a short reminder of the NPGHU model.
This is a non-perturbative construction whose simplest realization is a five-dimensional pure 
Yang--Mills gauge theory defined on a periodic hyper-cubic orbifold lattice \cite{IrgesK1, IrgesK2, IrgesK3} with an anisotropy in its fifth dimension.
The four-dimensional planes, in which the lattice is infinite, have lattice spacing $a_4$,
while the anisotropic fifth dimension, on which the orbifold boundary conditions are implemented, has lattice spacing $a_5$.
The above discrepancy results in a geometry which has a 5d bulk and two 4d boundaries located at the endpoints of the anisotropic dimension.
In its simplest realization in the bulk an $SU(2)$ gauge symmetry is embedded and the boundary 
conditions leave at the boundaries a $U(1)$ gauge field and a complex scalar field.
At the end, the lattice can be folded about its midpoint in the extra dimension and the infinite dimensional limit can be taken.
Our notation follows \cite{IrgesFotis2} with small exceptions. 

\subsection{The orbifold lattice action}\label{orb.lat}

We consider a Euclidean hyper-cubic 5d lattice which is periodic in all dimensions.
The extra dimension is a circle with radius $R$ and the orbifold lattice is constructed by projecting the circle by the discrete group $\mathbb Z_2$.
This action identifies the upper with the lower semicircle and results in a discretized interval with two fixed points.
On each of these fixed points lives a 4d slice and the produced geometry is that of $\mathbb R^4 \times S^1/\mathbb Z_2$.
The projection acts also on the gauge group.
If there is a 5d $SU(2)$ gauge theory in the bulk, the action of $\mathbb Z_2$ leaves at the endpoints of the interval a $U(1)$ subgroup and a complex scalar field.
In what follows capital Latin letters 
$M,\, N, \cdots=0,1,2,3,5$ denote the 5d Euclidean or Minkowski index while small Greek letters $\m,\, \n, \cdots = 0,1,2,3$
denote the four-dimensional part.
Moreover, the lattice coordinates are defined as $n_M\equiv(n_\m,n_5)$ with $n_\m = 1,\cdots,\, L_\m$ and $n_5 = 0,\cdots, \, N_5$. 
The orbifold fixed points are located at $n_5=0,\, N_5$.
Recall that the four-dimensional part of the lattice is taken infinite, a fact that is mirrored by sending $L_\m \to \infty$.
On the other hand, the extra dimension is allowed to be either finite or infinite.
Since the number of lattice nodes in this direction is given by $N_5 = \pi R/a_5$, the choice $R \to \infty$ or $R \equiv \rm finite$ determines the magnitude of the fifth dimension.  
The next step is the determination of the lattice gauge variables. The needed ingredients for that are the gauge links $U(n_M,N) \in SU(2)$. Their form is
\be\label{Ugen}
U(n_M,N) = e^{i a_N g_5 {\bf A}_M (n_M)}\, ,
\ee
where $n_M$ indicates their location while $N$ the direction in which they point.
The above exponential includes the lattice spacings $a_\n$ and $a_5$ given by $a_N$ with $N=\n$ and $N=5$ respectively.
The former is identified as $a_\n = a_4$, for every $\n$, while in general $a_4 \ne a_5$.
In addition we get the continuum 5d dimensionfull coupling $g_5$ and the Lie algebra valued gauge field
(always written in bold character) ${\bf A}_M \equiv A^A_M T^A$ which carries the adjoint index $A$. The generators are normalized using ${\rm tr} \{ T^A T^B \} = \d^{AB}/2$. 
Dimensional analysis of the exponent in \eq{Ugen} indicates that $[{\bf A}_M] = 3/2$, $[a_N] = -1$ and $[g_5 ]= -1/2$, where $[\cdots]$ represents the classical dimension of its argument.
Since the structure of the lattice and the associated gauge variables are settled down we can now consider the orbifold projection on both of them.
We define the reflection operator ${\cal R}$ that acts as a group conjugation, ${\cal T}_g$, on the corresponding generators.
The orbifold condition on the links is implemented on the lattice through 
\be\label{orb.UnM}
{\cal R} U(n_M,N) = {\cal T}_g U(n_M,N) \, .
\ee
Regarding the lattice nodes $n_M = (n_\m, n_5)$, the reflection operator acts on them as
\be
{\cal R} (n_\m, n_5) = (n_\m, -n_5) \equiv \bar n_M
\ee
while on the gauge links as
\be
{\cal R} U(n_M,\n) = U(\bar n_M,\n) \,\,\, {\rm and} \,\,\, {\cal R} U(n_M, 5) = U^{\dagger}(\bar n_M - a_5 \hat 5, 5)\, .
\ee
On the other hand, ${\cal T}_g$ acts solely on the gauge links and the associated action yields
\be
{\cal T}_g \, U(n_M,N) = g U(n_M,N) g^{-1} \, ,
\ee
with $g^2$ an element in the centre of $SU(2)$ which we can choose to be $g = -i \s^3$, the third Pauli matrix.
Notice that ${\cal R}$ commutes with ${\cal T}_g$ and also that ${\cal R}^2 = {\cal T}_g^2 = 1$.
The orbifolding becomes interesting when we look at the fixed points.
As a small comment note that the orbifold has a mirror symmetry around the midpoint of the fifth dimension.
Then, without loss of generality, we can fold it and consider only one of the two boundaries multiplied now with a factor of 2. 
Here we choose to work with the boundary located at $n_5=0$.
Now, the reflection operation has a trivial effect on the boundary nodes, $n_M = (n_\m,0)$, and hence \eq{orb.UnM} becomes
\be\label{UnTU}
U((n_\m,0),\n) = {\cal T}_g \, U((n_\m,0),\n) \equiv g U((n_\m,0),\n) g^{-1} \, .
\ee
The above relation is satisfied only for the gauge links that commute with $\s^3$.
Therefore on the boundary the generators of the bulk group $\cal G$ separate in $T^a$ and $T^{\hat a}$.
The former belong to the unbroken group ${\cal H}$ with ${\cal H} \subset {\cal G}$ while the latter to the broken ones.
Since here $U(n_M,N)$ is an $SU(2)$ element with generators $T^A = \s^A/2$, the three Pauli matrices, the only acceptable choice is $a=3$ and $\hat a =1,2$.
In other words only a $ U(1)$ gauge symmetry remains unbroken on the boundary.
The lattice action is constructed using the Wilson plaquettes which are gauge invariant objects consisting of gauge links.
The above discussion indicates that there are $U(1)$ and $SU(2)$ links on the boundary and in the bulk respectively, 
as well as, hybrid links which have one end on the boundary and the other in the bulk.
The gauge transformation for the latter is given by $U \to \Omega^{U(1)} U (\Omega^{SU(2)})^\dagger $.
With these in our hand we get the following two plaquette categories: $1)$ the boundary-hybrid plaquette and $2)$ the bulk plaquette.
In $1)$ we define as $U^{b}_{\m\n}$ the 4d boundary plaquettes, including links lying only on the boundary, 
and as $U^{h}_{\m5}$ the hybrid plaquettes with two links lying on the fifth dimension with one end on the boundary and the other on the bulk.
In $2)$ we define $U_{\m\n}$ and $U_{\m5}$ plaquettes with gauge links lying exclusively in the bulk.
Hence the anisotropic orbifold Wilson action, $S_{S^1/\mathbb{Z}_2} \equiv S^{\rm orb}$, reads
\be\label{SorA}
S^{\rm orb} = S^{\rm b-h} + S^{B}
\ee 
where orb stands for orbifold, $S^{\rm b-h}$ is the boundary-hybrid action 
\be\label{SbhA}
S^{\rm b-h} = \frac{1}{2 N} \sum_{n_\m} \Biggl[ \frac{\b_4}{2} \sum_{\m<\n} \tr \Bigl \{  1- U^{b}_{\m\n}(n_\m,0) \Bigr\}  + \b_5 \sum_{\m} \tr \Bigl \{  1- U^{h}_{\m5} (n_\m,0) \Bigr\}  \Biggr]
\ee
and $S^{B}$ corresponds to the bulk action
\be\label{SBA}
S^{B} = \frac{1}{2 N} \sum_{n_\m,n_5} \Biggl[ \b_4 \sum_{\m<\n} \tr \Bigl \{  1- U_{\m\n}(n_\m,n_5) \Bigr\}  + \b_5 \sum_{\m} \tr \Bigl \{  1- U_{\m5}(n_\m,n_5) \Bigr\}  \Biggr] \, .
\ee
$\b_4$ and $\b_5$ are the dimensionless lattice couplings and $N$ is the degree of the group $\cal G$. 
Since here ${\cal G} \equiv SU(2)$ then $N = 2$. Notice that we sum only over plaquettes with counterclockwise orientation.

Few comments are in order.
Keep in mind that the extra $1/2$ factor in the boundary lattice action will be canceled due to the folding of the orbifold around the midpoint of the fifth dimension.
Moreover notice that $S^{\rm orb}$ is the same with the one developed in \cite{IrgesFotis2} (apart from some modifications in the notation, trivial to be synchronized).
Finally the anisotropy, which at classical level is given by $\g = a_4/a_5$, is explicitly introduced in the above framework when we switch to the equivalent pair of dimensionless couplings:
\be\label{b45}
\b_4 = \frac{\b}{\g}\, , \,\, \b_5 = \b\g
\ee
which holds for both the boundary and the bulk.
$\b$ here should not be confused with the usual definition of a $\b$-function.
Since we know the anisotropic orbifold lattice action our next step is to construct its continuum version 
so we first need to explicitly compute $S^{\rm b-h}$ and $S^{B}$. Towards that direction we consider 
the expansion in small lattice spacing for \eq{SbhA} and \eq{SBA} according to the program of \cite{IrgesFotis2}.
There the expansion was truncated to the lowest non-trivial order in $a_4, a_5$ which led to a disconnected boundary-bulk 
system\footnote{Trailing back the computational details regarding the small lattice spacing expansion, which are 
presented in \cite{IrgesFotis2}, will be a comprehensive guide for the current work.}.
Here we move a step further in our computation so that the boundary is no longer disconnected from the bulk, 
a property which should be mirrored in the continuum effective action. The truncation in $a_4, a_5$ is such that it reveals 
the lowest higher dimensional operators which respect the symmetries.

\subsection{The Boundary-Hybrid action with higher order terms}\label{orb.bh}

Let us start our computation with the pure boundary part of \eq{SbhA} which we denote here as
\be\label{Sb}
S^b = \frac{1}{2 N} \sum_{n_\m} \frac{\b_4}{2} \sum_{\m<\n} \tr \Bigl \{  1- U^{b}_{\m\n}(n_\m,0) \Bigr\} \, .
\ee
As we have already mentioned the above plaquette includes only boundary-lying gauge links and its explicit form reads
\bea\label{Ubmn2}
U^{b}_{\m\n}(n_\m) &=& U^{b}(n_\m,\m) \, U^{b}(n_\m+a_4 \hat \m,\n)\, (U^{b}(n_\m+a_4 \hat \n,\m))^\dagger \, (U^{b}(n_\m,\n))^\dagger \nonumber\\
 &=& e^{i a_4 g_5 {\bf A}_\m(n_\m)} \,e^{i a_4 g_5 {\bf A}_\n(n_\m+a_4 \hat \m)}\,e^{-i a_4 g_5 {\bf A}_\m(n_\m+a_4 \hat \n)}\,e^{-i a_4 g_5 {\bf A}_\n(n_\m)} \, ,
\eea
where we have used \eq{Ugen} for $M,N=\m,\n$. For the calculation of the plaquette we use the Baker--Campbell--Housdorff (BCH) formula.
To be more specific when there are four exponentials, with exponents $X,Y,Z,W$ respectively, the complete basis for the formula is the following:
\bea\label{BCH}
e^X e^Y e^Z e^W &=& \exp \Bigl [ X+ Y+Z+W + \frac{1}{2} \Bigl(  [X,Y] +[X,Z] + [X,W] + [Y,Z] + [Y,W] + [Z,W]  \Bigr) \nonumber\\
&+& \frac{1}{4} \Bigl(   [X,[Z,W]] + [Y,[Z,W]] + [[X,Y],Z] + [[X,Y,],W]  \Bigr) + \frac{1}{8} [[X,Y],[Z,W]]  \Bigr] \nonumber\\
\eea
where here the exponents of the gauge links are represented by $X,Y,Z$ and $W$.
Regarding the boundary gauge field ${\bf A}_\m \equiv A_\m^3 T^3$, BCH reduces to the usual product of exponentials since $[A_\m^3 T^3, A_\n^3 T^3] = 0$.
As a small side note, we would like to clarify a bit more the arguments of the above exponentials. In particular, the corresponding exponents are dimensionless only if
\be
 [ a_4 ] + [ g_5 ] + [ {\bf A}_\m ] =0 \nonumber 
\ee
which, according to the dimensional analysis of Sect. \ref{orb.lat} gives $[ {\bf A}_\m ] = 3/2$.
This is an expected result regarding the 5d bulk but on the 4d boundary the dimensionality of the gauge field should be different.
One way to overcome this obstacle, in agreement with \cite{IrgesFotis2}, is to note that the boundary-hybrid action does not include a summation over $a_5$ which then becomes a free parameter.
Since here we consider $R \to \infty$ and we are allowed to choose $N_5 \to \infty$ then, from $a_5 = \pi R/N_5$, we get that $a_5 \to a_{5,f} $ which is finite.
Then we can define $g_5 \equiv g_4 \sqrt{a_{5,f}} $, with $ g_4$ a dimensionless coupling, and reabsorb $\sqrt{a_{5,f} }$ in the gauge field.
In that sense $ \sqrt{a_{5,f} } {\bf A}_\m \to {\bf A}_\m $ on the exponentials and now $[{\bf A}_\m] = 1$ which is the usual dimension of a 4d gauge potential.
Another way of seeing the above argument is that we want the fields and currents of $S^{\rm b-h}$ to be localized at the lattice boundary.
This happens when the anisotropy is small, \cite{Itou}, which is true when $a_5>a_4$. So since here we consider the $a_4 \to 0$ limit then $a_5$ cannot approach zero but it can take any finite value without losing generality.
With the above in mind \eq{Ubmn2} becomes
\be\label{Ubmng4}
U^{b}_{\m\n}(n_\m) = e^{i a_4 g_4 [ {\bf A}_\m(n_\m) +  {\bf A}_\n(n_\m+a_4 \hat \m) - {\bf A}_\m(n_\m+a_4 \hat \n) - {\bf A}_\n(n_\m) ]} 
\ee
and we are ready to consider the series of $ {\bf A}_\n(n_\m+a_4 \hat \m) $ and $ {\bf A}_\m(n_\m+a_4 \hat \n) $ for small $a_4$ while we keep terms up to order ${\cal O}(a_4^2)$.
In other words the expansion gives
\bea
{\bf A}_\n(n_\m+a_4 \hat \m) &\equiv& {\bf A}_\n(n_\m) + a_4 \hat \D_\m {\bf A}_\n(n_\m) + \frac{a_4^2}{2}  \hat \D_\m ( \hat \D_\m {\bf A}_\n(n_\m) ) 
\nonumber \\
{\bf A}_\m(n_\m+a_4 \hat \n) &\equiv& {\bf A}_\m(n_\m) + a_4 \hat \D_\n {\bf A}_\m(n_\m) + \frac{a_4^2}{2}  \hat \D_\n ( \hat \D_\n {\bf A}_\m(n_\m) ) 
\nonumber
\eea
with $\hat \D_\m {\bf A}_\n(n_\m) = (1/a) [ {\bf A}_\n(n_\m +a \hat \m) - {\bf A}_\n(n_\m) ]$ a discretized derivative.
Therefore the above plaquette yields
\be\label{Ubmn3}
U^{b}_{\m\n}(n_\m) = e^{ i a_4^2 g_4 {\bf F}_{\m\n}(n_\m) + \frac{i a_4^3 g_4}{2} [ \hat\D_\m {\bf F}_{\m\n}(n_\m) + ( \hat\D_\m -\hat\D_\n )\hat\D_\n  {\bf A}_\m(n_\m)  ]  
+ {\cal O}(a_4^4) }\, ,
\ee
where we have defined that ${\bf F}_{\m\n}(n_\m) =  \hat \D_\m {\bf A}_\n(n_\m) - \hat \D_\n {\bf A}_\m(n_\m)$ with ${\bf F}_{\m\n} \equiv F_{\m\n}^3 T^3$.
Recall that a plaquette is designed to be a gauge and rotational (Lorentz) invariant object on the Euclidean (Minkowski) lattice. 
Nevertheless, due to the truncation level of the lattice-spacing expansion several terms, inconsistent with these symmetries, appear.
This mismatch is an artifact of the lattice in a sense that adding higher in $a_4$ terms to the expansion the gauge and rotational symmetry will eventually be restored.
Nevertheless, a delicate point here is the assumption that the continuum action, no matter the truncation level, should respect these symmetries.
Hence, in the following we will expand all the formulae up to a specific order at the lattice-spacing and then neglect every contribution which is not gauge and rotationally invariant.
Then implying \eq{Ubmn3} back to \eq{Sb}, performing one more expansion for small $a_4$ and taking the trace the boundary action becomes
\bea\label{Sb2}
S^b &=& 2 \frac{1}{2 N} \sum_{n_\m} \frac{\b_4}{2} \sum_{\m<\n} \tr \Bigl \{ - i a_4^2 g_4 {\bf F}_{\m\n} - i \frac{a_4^3}{2} g_4 [ \hat\D_\m {\bf F}_{\m\n} + ( \hat\D_\m -\hat\D_\n )\hat\D_\n  {\bf A}_\m  ] \nonumber\\
&+& \frac{a_4^4}{2} g_4^2  {\bf F}_{\m\n}^2 + \frac{a_4^5}{2} g_4^2 {\bf F}_{\m\n} [ \hat\D_\m {\bf F}_{\m\n}(n_\m) + ( \hat\D_\m -\hat\D_\n )\hat\D_\n  {\bf A}_\m  ] \nonumber\\
&+& \frac{a_4^6}{8} g_4^2 [ \hat\D_\m {\bf F}_{\m\n} + ( \hat\D_\m -\hat\D_\n )\hat\D_\n  {\bf A}_\m  ]^2 + i \frac{a_4^6 }{6} g_4^3 {\bf F}_{\m\n} {\bf F}_{\n\r} {\bf F}_{\r\m}
  \Bigr\} + {\cal O}(a_4^7)  \nonumber\\
 &=& \sum_{n_\m} a_4^4 \sum_{\m<\n} \Bigl \{ \frac{1}{4} F^3_{\m\n} F^3_{\m\n} + \frac{a_4^2}{16}  ( \hat\D_\m F^3_{\m\n} ) ( \hat\D_\m F^3_{\m\n} ) \Bigr\} + {\cal O}(a_4^7)  \, ,
\eea
where we added the factor of 2 in front of the sum due to the folding of the orbifold and kept operators up to dimension 6 which are gauge and rotationally invariant. 
To give an example the above action originally includes the dimension-3 operators
\be
 \hat\D_\m {\bf F}_{\m\n} + ( \hat\D_\m -\hat\D_\n )\hat\D_\n  {\bf A}_\m \, , \nonumber
 \ee
which are zero when the trace is taken and the dimension-5 operators 
\be
{\bf F}_{\m\n} \Bigl (  \hat\D_\m {\bf F}_{\m\n} + ( \hat\D_\m -\hat\D_\n )\hat\D_\n  {\bf A}_\m \Bigr ) \nonumber
\ee
which are not rotationally and gauge invariant. Note moreover that the lattice coupling is fixed by
\be\label{lb4}
\b_4 = \frac{2 N a_{5,f}}{g_5^2} = \frac{2 N}{g_4^2} \, .
\ee
Let us make a small pause here and comment on the distribution of $a_4$'s in \eq{Sb2}. In particular considering the naive continuum limit, $a_4 \to 0$, four of them will be sacrificed so as to form the 4d integral
\be\label{a4to0}
\sum_{n_0} a_4 \sum_{n_1} a_4 \sum_{n_2} a_4 \sum_{n_3} a_4 = (n_0 a_4) (n_1 a_4) (n_2 a_4) (n_3 a_4) \to \int dx_0 dx_1 dx_2 dx_3 \, .
\ee
Nevertheless the above is consistent regarding a dimension-4 operator, when the action includes HDO the situation is different.
In \eq{Sb2} for example there is also a dimension-6 operator which comes with two $a_4$'s extra. 
So the rising question is what happens to that term when the $a_4 \to 0$ limit is taken.
To answer this recall that lattice-spacing works both as a definition of the lattice length and as a natural theory regulator which, in addition, respects gauge invariance.
On the other hand, the same theory defined at the continuum is regularized in a gauge invariant manner using dimensional regularization.
Moreover, DR introduces the intrinsic scale $\mu$ which plays a crucial role in the quantum behaviour of the theory.
Matching the situation on the lattice with that of the continuum and since $a_4$ has a dual role, it is legal to assume that there is a non-trivial connection between the lattice-spacing and the intrinsic scale.
Back to \eq{Sb2} the above yields that we are allowed to replace the remnant $a_4$ with  $\m(a_4)$ and get
\be\label{SbfA}
S^b = \sum_{n_\m} a_4^4 \sum_{\m<\n} \Biggl (  \frac{1}{4} F^3_{\m\n} F^3_{\m\n} + \frac{1}{16}  \frac{ ( \hat\D_\m F^3_{\m\n} ) ( \hat\D_\m F^3_{\m\n} ) }{\m^2(a_4)}  \Biggr ) + {\cal O}(a_4^7)  \, .
\ee
The other part of \eq{SbhA} regards the hybrid action defined here as 
\be\label{Sh}
S^h = \frac{1}{2 N} \sum_{n_\m} \b_5 \sum_{\m} \tr \Bigl \{  1- U^{h}_{\m 5}(n_\m,0) \Bigr\} 
\ee
and includes the hybrid plaquette $U^{h}_{\m 5}$.
Recall that this plaquette is constructed from a $U(1)$ gauge link, two $SU(2)$ hybrid gauge links and one $SU(2)$ gauge link.
Therefore its explicit form using \eq{Ugen} for $M,N = \m, 5$ reads
\bea\label{Uhm51}
U^{h}_{\m 5} &=& U^{b}(n_\m,\m) \, U^{h}(n_\m+a_4 \hat \m,\hat 5)\, (U(n_\m+a_5 \hat 5,\m))^\dagger \, (U^{h}(n_\m,\hat 5))^\dagger \nonumber\\
 &=& e^{i a_4 g_4 {\bf A}^b_\m(n_\m,0)} \,e^{i a_5 g_4 {\bf A}_5(n_\m+a_4 \hat \m,0)}\,e^{-i a_4 g_4 {\bf A}_\m(n_\m, a_5 \hat 5)}\,e^{-i a_5 g_4 {\bf A}_5(n_\m,0)}  \, ,
\eea
where the superscript $b$ on the gauge field refers to the boundary.
More precisely we get ${\bf A}^b_\m \equiv A^3_\m T^3$ and $( {\bf A}_\m, {\bf A}_5 ) \equiv ( A^A_\m T^A, A^A_5 T^A ) $.
Note that here we have directly followed the arguments above \eq{Ubmng4} so the gauge potential has already mass dimension one.
The hybrid plaquette has been set up and the next step is to consider the lattice-spacing expansion.
For that purpose we exploit the BCH formula given in \eq{BCH} for 
\be
X = i a_4 g_4 {\bf A}^b_\m(n_\m,0)\, , \,\, Y = i a_5 g_4 {\bf A}_5(n_\m+a_4 \hat \m,0)\, , \,\, Z = -i a_4 g_4 {\bf A}_\m(n_\m, a_5 \hat 5)\, , \,\, W = -i a_5 g_4 {\bf A}_5(n_\m,0) \, , \nonumber
\ee
where now it will not be reduced to the usual case since the commutators are not necessarily zero.
Recall that the current work deals with a higher in $a_4$ and $a_5$ order compared to \cite{IrgesFotis2} so in the following we will keep terms up to ${\cal O}(a_4^3, a_5^3)$.
As a consequence, non-linear in gauge fields commutators are expected to play a crucial role in the calculation of the hybrid action.
With the above in mind the desired plaquette yields
\bea\label{Uhm52}
U^{h}_{\m 5} &=& \exp \Biggl [  i a_5 g_4 \Biggl ( \g {\bf A}^b_\m(n_\m,0) + {\bf A}_5(n_\m+a_4 \hat \m,0) - \g {\bf A}_\m(n_\m, a_5 \hat 5) - {\bf A}_5(n_\m,0)   \Biggr ) \nonumber\\
&-& \frac{g_4^2}{2} \Biggl ( a_4 a_5 [ {\bf A}^b_\m(n_\m,0), {\bf A}_5(n_\m+a_4 \hat \m,0)  ] - a_4^2 [  {\bf A}^b_\m(n_\m,0),  {\bf A}_\m(n_\m, a_5 \hat 5) ] \nonumber\\
&-& a_4 a_5 [  {\bf A}^b_\m(n_\m,0),  {\bf A}_5(n_\m,0)  ] - a_4 a_5 [  {\bf A}_5(n_\m+a_4 \hat \m,0) , {\bf A}_\m(n_\m, a_5 \hat 5) ] \nonumber\\
&-& a_5^2 [  {\bf A}_5(n_\m+a_4 \hat \m,0) ,  {\bf A}_5(n_\m,0)  ] + a_4 a_5 [ {\bf A}_\m(n_\m, a_5 \hat 5),  {\bf A}_5(n_\m,0) ]     \Biggr ) \nonumber\\
&-&\frac{i g_4^3}{4} \Biggl \{  a_4^2 a_5  [ {\bf A}^b_\m(n_\m,0)  ,[ {\bf A}_\m(n_\m, a_5 \hat 5),  {\bf A}_5(n_\m,0) ] ]  + a_4 a_5^2  [  {\bf A}_5(n_\m+a_4 \hat \m,0)  ,[ {\bf A}_\m(n_\m, a_5 \hat 5),  {\bf A}_5(n_\m,0) ] ]   \nonumber\\
&-&  a_4^2 a_5  [ [ {\bf A}^b_\m(n_\m,0), {\bf A}_5(n_\m+a_4 \hat \m,0)  ],  {\bf A}_\m(n_\m, a_5 \hat 5)  ]  - a_4 a_5^2  [ [ {\bf A}^b_\m(n_\m,0), {\bf A}_5(n_\m+a_4 \hat \m,0)  ],   {\bf A}_5(n_\m,0)  ]  
 \Biggr \}  \nonumber\\ 
 &+& {\cal O}(a^4)  \Biggr ], 
\eea
where ${\cal O}(a^k) \equiv {\cal O}(a_4^{k_1}, a_5^{k_2}) $ $\forall$ $k_1$, $k_2$ that satisfy $k_1 + k_2 = k$.
Note here that the terms in the parenthesis had also been found in \cite{IrgesFotis2} while the ones in the curly bracket denote a new, higher order, contribution.
Next we deal with the expansion of the gauge potentials for small lattice-spacing while we keep in mind that a next-to-next-to leading order truncation is needed.
Then, the desired expressions yield
\bea\label{A5Amexp1}
{\bf A}_5(n_\m+a_4 \hat \m,0) &\equiv& {\bf A}_5(n_\m,0) + a_4 \hat \D_\m {\bf A}_5(n_\m,0) + \frac{a_4^2}{2}  \hat \D_\m ( \hat \D_\m {\bf A}_5(n_\m,0) )   \nonumber \\
{\bf A}_\m(n_\m, a_5 \hat 5) &\equiv& {\bf A}_\m(n_\m,0) + a_5 \hat \D_5 {\bf A}_\m(n_\m,0) + \frac{a_5^2}{2}  \hat \D_5 ( \hat \D_5 {\bf A}_\m(n_\m,0) )   \, ,
\eea
with $\hat \D_5$ a discretized derivative in accordance with $\hat \D_\m$.
Now the above relations should be combined with \eq{Uhm52}, an attempt that will give back a rather messy result.
Nevertheless, we can avoid the undesired mess if we first consider \eq{UnTU}, at a gauge-field level, 
which will inherit our expressions with the Dirichlet and Neumann orbifold boundary conditions.
To be more specific these conditions demand the following:
\bea
{\cal R} A^{\hat a}_\m T^{\hat a} = \a_\m A^{\hat a}_\m T^{\hat a} = \eta^{\hat a} A^{\hat a}_\m T^{\hat a}  &\Rightarrow& A^{\hat a}_\m T^{\hat a} = - A^{\hat a}_\m T^{\hat a} \Rightarrow A^{1,2}_\m = 0 \nonumber\\
{\cal R} A^{ a}_5 T^{ a} = \a_5 A^{ a}_5 T^{ a} = \eta^{a} A^{ a}_5 T^{ a} &\Rightarrow& - A^{ a}_5 T^{ a} =  A^{ a}_5 T^{ a} \Rightarrow A^{ 3}_5  = 0 \nonumber\\
{\cal R} \hat \D_5 A^{a}_\m T^{ a} = \a_5 \a_\m \hat \D_5 A^{ a}_\m T^{ a} = \eta^{ a} \hat \D_5 A^{ a}_\m T^{ a}  &\Rightarrow& - \hat \D_5 A^{ a}_\m T^{ a} = \hat \D_5 A^{ a}_\m T^{ a} \Rightarrow \hat \D_5 A^3_\m = 0 \nonumber\\
{\cal R} \hat \D_5 A^{\hat a}_5 T^{\hat a} = \a_5 \a_5 \hat \D_5 A^{\hat a}_5 T^{\hat a} = \eta^{\hat a} \hat \D_5 A^{\hat a}_5 T^{\hat a}  &\Rightarrow& \hat \D_5 A^{\hat a}_5 T^{\hat a} = - \hat \D_5 A^{\hat a}_5 T^{\hat a} \Rightarrow \hat \D_5 A^{1,2}_5 = 0 \nonumber
\eea
where $\a_\m$ and $\a_5 $ correspond to the parity of the gauge fields ${\bf A}_\m$ and ${\bf A}_5$ respectively, while $\eta^A$ with $\eta^a = +1 $ and $\eta^{\hat a} = -1$ is the parity of the generators coming from the relation $g T^A g^{-1} = \eta^A T^A$. 
As a small comment notice that the hybrid plaquette includes also the term $ \hat \D_5 ( \hat \D_5  A^{3}_\m(n_\m,0) )$ which survives from the Neumann boundary conditions since
\be
{\cal R} \hat \D_5 ( \hat \D_5 A^{ a}_\m T^{ a} ) = \a_5 \a_5 \a_\m \hat \D_5 ( \hat \D_5 A^{a}_\m T^{ a} ) = \eta^{ a} \hat \D_5 ( \hat \D_5 A^{ a}_\m T^{ a} )  \Rightarrow \hat \D_5 ( \hat \D_5 A^{ a}_\m T^{ a} ) \ne 0 \, .\nonumber 
\ee
However, after the lattice-spacing expansion is taken all the gauge fields are defined on the boundary, which is located at $n_5=0$, so there is no evolution along the fifth dimension. In other words the derivative of $\hat \D_5$, of any given order, on the boundary fields should vanish.
In that sense the gauge potentials become $A_5^A(n_\m, 0) \to A_5^{1,2}(n_\m, 0)$ and $  A^A_\m(n_\m, 0) \to  A^3_\m(n_\m, 0)$ so \eq{A5Amexp1} now gives
\bea\label{A5Amexp2}
{\bf A}_5(n_\m+a_4 \hat \m,0) &\to& {\bf A}^b_5(n_\m,0) + a_4 \hat \D_\m {\bf A}^b_5(n_\m,0) + \frac{a_4^2}{2}  \hat \D_\m ( \hat \D_\m {\bf A}^b_5(n_\m,0) )   \nonumber \\
{\bf A}_\m(n_\m, a_5 \hat 5) &\to& {\bf A}^b_\m(n_\m,0) 
\, ,
\eea
where we have defined that ${\bf A}^b_5 \equiv  A_5^{\hat a} T^{\hat a}$.
With these in hand our calculation is simplified and the hybrid palquette reads
\bea\label{Uhm53}
U^{h}_{\m 5} &=& \exp \Biggl [  i g_4 \Biggl (   a_4 a_5 \hat \D_\m {\bf A}^b_5(n_\m,0) + \frac{a_4^2 a_5}{2} \hat \D_\m ( \hat \D_\m {\bf A}^b_5(n_\m,0) ) 
   \Biggr )  \nonumber\\
&-&g_4^2 \Biggl ( a_4 a_5 [ {\bf A}^b_\m(n_\m,0), {\bf A}^b_5(n_\m,0)  ] + a_4^2 a_5 [  {\bf A}^b_\m(n_\m,0),  \hat \D_\m {\bf A}^b_5(n_\m,0)  ]   \Biggr ) \nonumber\\
&-& i g_4^3 \Biggl \{  \frac{a_4^2 a_5}{2}  [ {\bf A}^b_\m(n_\m,0)  ,[ {\bf A}^b_\m(n_\m, 0),  {\bf A}^b_5(n_\m,0) ] ]  + \frac{a_4 a_5^2}{2}  [  {\bf A}^b_5(n_\m, 0)  ,[ {\bf A}^b_\m(n_\m, 0),  {\bf A}^ b_5(n_\m,0) ] ]   \Biggr \}  + {\cal O}(a^4)  \Biggr ] \nonumber\\
  &=& \exp \Biggl [  i g_4 \Biggl ( a_4 a_5 \Biggl [  \hat \D_\m {\bf A}^b_5(n_\m,0)  + i g_4 [ {\bf A}^b_\m(n_\m,0), {\bf A}^b_5(n_\m,0)  ]   \Biggr ]  \nonumber\\
&+& \frac{a_4^2 a_5}{2}  \Biggl [ \hat \D_\m ( \hat \D_\m {\bf A}^b_5(n_\m,0) ) + 2i g_4 [  {\bf A}^b_\m(n_\m,0),  \hat \D_\m {\bf A}^b_5(n_\m,0)  ]  - g_4^2  [ {\bf A}^b_\m(n_\m,0)  ,[ {\bf A}^b_\m(n_\m, 0),  {\bf A}^b_5(n_\m,0) ] ]   \Biggr ] \nonumber\\
&-& \frac{a_4 a_5^2}{2} g_4^2 [  {\bf A}^b_5(n_\m, 0)  ,[ {\bf A}^b_\m(n_\m, 0),  {\bf A}^ b_5(n_\m,0) ] ] 
 + {\cal O}(a^4)  \Biggr )  \Biggr ] \, ,
\eea
and now each bracket in the last exponential includes terms of the same dimension.
Actually this is not the end since there are some extra manipulations which will lead to an even more simple relation for \eq{Uhm53}.
In particular we can give up $(n_\m,0)$ from now on, since we are exclusively on the boundary, and do the following:
The first bracket can be rewritten defining the hybrid field-strength ${\bf F}_{\m5}$ as
\be
{\bf F}_{\m5} =  \hat \D_\m {\bf A}^b_5 + i g_4 [{\bf A}^b_\m , {\bf A}^b_5 ] \, . \nonumber 
\ee
Then the second bracket, after using the identity
\be
\hat \D_\m [{\bf A}^b_\m,  {\bf A}^ b_5] = [\hat \D_\m {\bf A}^b_\m,  {\bf A}^ b_5] + [ {\bf A}^b_\m,  \hat \D_\m  {\bf A}^ b_5] \nonumber
\ee
to rewrite one of the two $[  {\bf A}^b_\m,  \hat \D_\m {\bf A}^b_5  ]$ and collecting terms which form ${\bf F}_{\m5}$, reads
\be
{\bf \hat D}_\m {\bf F}_{\m5} - i g_4 [\hat \D_\m {\bf A}^b_\m,  {\bf A}^ b_5 ] \, , \nonumber 
\ee
with ${\bf \hat D}_\m =  \hat \D_\m  + i g_4 [{\bf A}^b_\m ,  ] $ the discretized version of the covariant derivative.
In that sense \eq{Uhm53} admits the final form
\be\label{Uhm5f}
U^{h}_{\m 5} = \exp \Biggl [  i g_4 \Biggl ( a_4 a_5 {\bf F}_{\m5} + \frac{a_4^2 a_5}{2} {\bf \hat D}_\m {\bf F}_{\m5} - i \frac{g_4 a_4^2 a_5}{2} [\hat \D_\m {\bf A}^b_\m,  {\bf A}^ b_5 ] 
- \frac{g_4^2 a_4 a_5^2}{2} [  {\bf A}^b_5 , [ {\bf A}^b_\m,  {\bf A}^ b_5 ] ]  + {\cal O}(a^4)  \Biggr )  \Biggr ] 
\ee
and the next step is to insert the above relation in \eq{Sh} and expand the exponential for small lattice-spacing.
This is a path which needs caution since choosing to truncate the expansion at a specific order inherits the action with terms which do not respect the symmetries of the theory, in the same manner with the boundary action.
In particular if we set $U^{h}_{\m 5}$ back in \eq{Sh}, perform the trace and keep terms up to ${\cal O}(a^6)$, the hybrid action reads
\bea\label{Sh2}
S^h &=& \sum_{n_\m} a_4^4 \sum_{\m} \Biggl(  (\overline{\hat D_\m \phi}) \hat D_\m \phi + \frac{a_4 }{2}  (\overline{\hat D_\m \phi}) [ \overline{\hat D_\m} +  \hat D_\m ]  \hat D_\m \phi  + \frac{i g_4 a_4 }{2} \hat \D_\m A^3_\m [  (\overline{\hat D_\m \phi}) \phi - \bar \phi \hat D_\m \phi  ]  \nonumber\\
&+& \frac{a_4^2}{4} (\overline{\hat D_\m \hat D_\m \phi}) \hat D_\m \hat D_\m \phi  + \frac{g_4^2 a_4^2}{4}  \bar \phi \phi (\hat \D_\m A^3_\m)^2  +  i g4 \frac{ a_4^2}{4} \hat \D_\m A^3_\m \Bigl [ ( \overline{\hat D_\m \hat D_\m \phi} ) \phi - \bar \phi ( \hat D_\m \hat D_\m \phi )  \Bigr ]   \nonumber\\
&+& \frac{a_5^2}{2} (\bar \phi \phi)^2 A^3_\m A^3_\m   \Biggr ) + {\cal O}(a^7) \, ,
\eea
defining first the lattice coupling $\b_5 = 2 N a_4^2/ a_5^2 g_4^2 $ and in addition the complex scalar field $\phi = ( A_5^1 + i A_5^2 )/2$.
Moreover the covariant derivative has been set as $\hat D_\m = \hat \D_\m - i g_4 A^3_\m $.
However, notice in \eq{Sh2} that the dimension-five and six operators, except from the case of $(\overline{\hat D_\m \hat D_\m \phi}) \hat D_\m \hat D_\m \phi$, lack of rotational and gauge invariance.
Of course the situation is changed and symmetry restoration is achieved when someone keeps higher and higher order terms at the lattice-spacing expansion and the BCH formula.
Nevertheless, at the expansion order of the current work these terms can be neglected without affecting the consistency of the theory.
Therefore, the desired expression for the hybrid action yields
\bea\label{Shf}
S^h &=& \sum_{n_\m} a_4^4 \sum_{\m} \Biggl(  | \hat D_\m \phi |^2 +  \frac{a_4^2}{4} | \hat D_\m \hat D_\m \phi |^2  \Biggr )  + {\cal O}(a^7) \nonumber\\
&=&  \sum_{n_\m} a_4^4 \sum_{\m} \Biggl(  | \hat D_\m \phi |^2 +  \frac{1}{4} \frac{ | \hat D_\m \hat D_\m \phi |^2 }{\m^2}   \Biggr )   + {\cal O}(a^7)  \, ,
\eea
where we have identified the remnant $a_4$ of the dim-6 operator as part of the regularization scale, $\m \equiv \m(a_4)$, in accordance with the pure boundary action.
So now the combination of \eq{SbfA} and \eq{Shf}, according to \eq{SbhA}, gives the boundary-hybrid action
\be\label{SbhfA}
S^{\rm b-h} = \sum_{n_\m} a_4^4 \sum_{\m} \Biggl [ \sum_{\n} \Biggl (  \frac{1}{4} F^3_{\m\n} F^3_{\m\n} + \frac{1}{16}  \frac{ ( \hat\D_\m F^3_{\m\n} ) ( \hat\D_\m F^3_{\m\n} ) }{\m^2}  \Biggr ) +   | \hat D_\m \phi |^2 +  \frac{1}{4} \frac{ | \hat D_\m \hat D_\m \phi |^2 }{\m^2}   \Biggr ] \, ,
\ee
where all the information of the anisotropy is hidden in the coupling between the fields.
Recall that the boundary action includes $g_4$ which however is not an independent coupling since
\be\label{defg4}
g_4 = \frac{g_5}{\sqrt{a_{5,f}}} = \frac{g_5}{\sqrt{a_4}}  \sqrt{\g} \equiv  g \sqrt{\g}
\ee 
with $g = \frac{g_5}{\sqrt{a_4}}$ a dimensionless coupling.
Therefore, now the covariant derivative is expressed with the help of the new coupling as a function of the anisotropy and becomes $\hat D_\m = \hat \D_\m - i g \sqrt{\g} A^3_\m $.
Of course \eq{SbhfA} is defined on the lattice and it can be useful only after considering the naive continuum limit according to \eq{a4to0}.
Nevertheless, before following this path we complete the picture constructing the bulk lattice action which will include terms coming from a higher order truncation of the lattice-spacing.



\subsection{The Bulk action with higher order terms}\label{orb.b}
Here the main ingredient is the pure bulk action, given by \eq{SBA}, which can be broken into a four-dimensional and a five-dimensional part.
The former includes plaquettes which lie exclusively on the 4d bulk and reads
\be\label{SBm1}
S^{4d} = \frac{1}{2 N} \sum_{n_\m,n_5}  \b_4 \sum_{\m<\n} \tr \Bigl \{  1- U_{\m\n}(n_\m,n_5) \Bigr\}  \, ,
\ee
while the latter includes plaquettes with gauge links leaving on the full 5d bulk and reads
\be\label{SB51}
S^{5d} = \frac{1}{2 N} \sum_{n_\m,n_5}  \b_5 \sum_{\m} \tr \Bigl \{  1- U_{\m5}(n_\m,n_5) \Bigr\} 
\ee  
with
$S^B = S^{4d} + S^{5d}$, $\b_4 = 2 N a_{5}/g_5^2$ and $\b_5 = 2 N a_4^2/a_5 g_5^2$.
Notice that the two sets of couplings, $\{ \b'_4, \b'_5\}$ and $\{ \b_4, \b_5\}$, are essentially the same except that in the bulk case $a_5$ is not an undetermined parameter any longer.
An important difference between the boundary and bulk case is that the latter includes fields with mass dimension $[ A_M ] = 3/2$, with $M= \m,5$.
This is a crucial fact regarding the calculations of the following sections. 
Now one of the arguments in \cite{IrgesFotis2} was that the actions $S^{4d}$ and $S^{5d}$, at leading order, were 5d covariant so their sum reconstructed the 5d bulk action.
Here the next-to-leading order is considered so there would be higher derivative terms in the action which do not allow us to expect the same conclusion regarding the covariance of \eq{SBm1} and \eq{SB51}.
Actually if this is the case then the breaking of the 5d covariance should be related with the existence of the anisotropy.
Of course the above argument can be proved following the path of the previous section.
For that purpose let us draw the calculation, without many details, starting from $S^{4d}$ whose plaquette yields
\bea\label{UMN1}
U_{\m\n}(n_\m,n_5) &=& U(n_\m,n_5;\m) \, U(n_\m+a_4 \hat \m,n_5;\n)\, (U(n_\m+a_4 \hat \n,n_5;\m))^\dagger \, (U(n_\m,n_5;\n))^\dagger \nonumber\\
 &=& e^{i a_4 g_5 {\bf A}_\m(n_\m,n_5)} \,e^{i a_4 g_5 {\bf A}_\n(n_\m+a_4 \hat \m,n_5)}\,e^{-i a_4 g_5 {\bf A}_\m(n_\m+a_4 \hat \n,n_5)}\,e^{-i a_4 g_5 {\bf A}_\n(n_\m,n_5)} \, , \nonumber
\eea
with ${\bf A}_\m \equiv  A_\m^A T^A$.
Then the algorithm that we follow suggests that the 4d bulk palquette is properly rewritten using the BCH formula.
In particular we need \eq{BCH} whose exponents here admit that
\bea
X &=& i a_4 g_5 {\bf A}_\m(n_\m,n_5)\, , \,\, Y = i a_4 g_5 {\bf A}_\n(n_\m+a_4 \hat \m,n_5)\nonumber\\
Z &=& -i a_4 g_5 {\bf A}_\m(n_\m+a_4 \hat \n,n_5)\, , \,\, W = -i a_4 g_5 {\bf A}_\n(n_\m,n_5) \, . \nonumber
\eea
So now we employ the lattice-spacing expansion, at the next-to-next-to leading order in the expansion, exactly as we did for the boundary.
In that sense, after collecting and massaging terms of the same dimension and keeping terms up to ${\cal O}(a_4^3)$, the plaquette becomes
\bea\label{UMN2}
U_{\m\n}(n_\m,n_5) &=& \exp \Biggl [   i a_4^2 g_5 {\bf F}_{\m\n}(n_\m,n_5) + \frac{i a_4^3 g_5}{2} \Bigl( \hat {\bf D}'_\m {\bf F}_{\m\n}(n_\m,n_5) + ( \hat\D_\m -\hat\D_\n )\hat\D_\n  {\bf A}_\m(n_\m,n_5) \nonumber\\
&+& i g_5 [{\bf A}_\n(n_\m,n_5), {\bf F}_{\m\n}(n_\m,n_5) ]  + i g_5 [ {\bf A}_\m(n_\m,n_5) - {\bf A}_\n(n_\m,n_5), \hat\D_\n  {\bf A}_\m(n_\m,n_5) ]  \Bigr )   \nonumber\\
&+& {\cal O}(a_4^4)  \Biggr ]  \, , \nonumber
\eea
where we have defined
\be
{\bf F}_{\m\n} \equiv F_{\m\n}^A T^A = \hat\D_\m  {\bf A}_\n - \hat\D_\n  {\bf A}_\m + i g_5 [{\bf A}_\m,{\bf A}_\n ] \,\,\, {\rm and} \,\,\, \hat {\bf D}'_\m = \hat \D_\m  + i g_5 [{\bf A}_\m ,  ] \, .
\ee
The final step of the calculation is to set $U_{\m\n}$ back in \eq{SBm1}, expand, perform the trace and keep terms up to ${\cal O}(a_4^6)$.
Keep in mind that we neglect terms which do not respect the symmetries of the theory, the lattice artifacts induced from the truncation, in the same manner with the previous section.
Hence the 4d bulk action now reads\footnote{Notice that in principle the complete base of 7-dimensional terms includes also two more entries, that of $( \hat D_\m F_{\n\l})^2$ and $( \hat D_\m F_{\m\l})( \hat D_\n F_{\n\l})$. Nevertheless these are not independent since, using the Bianchi identity, they reduce to the already existing dim-7 terms of \eq{SBmf}. Considering for example $( \hat D_\m F_{\n\l})^2 $ where we can see through the Bianchi identity
\be
 \hat D_\m F_{\n\l} +  \hat D_\n F_{\l\m} +  \hat D_\l F_{\m\n} =0 \nonumber
\ee
and the relation $[\hat D_\m, \hat D_\n] \equiv F_{\m\n}$ that
\be
( \hat D_\m F_{\n\l})^2 \sim  ( \hat D_\m F_{\m\n} )^2 - F_{\m\n} F_{\n\r} F_{\r\m} \, . \nonumber
\ee
This is the reason why the action that we get after the lattice spacing expansion does not include these redundant terms.}
\be\label{SBmf}
S^{4d} =  \sum_{n_\m} a_4^4 \sum_{n_5}  a_5 \sum_{\m<\n}  \Biggl (  \frac{1}{4} F^A_{\m\n} F^A_{\m\n} + \frac{a_4^2}{16}  ( \hat D_\m F^A_{\m\n} ) ( \hat D_\m F^A_{\m\n} )  - a_4^2 \frac{ g_5 f_{ABC} }{24} F^A_{\m\n} F^B_{\n\r} F^C_{\r\m}   \Biggr ) + {\cal O}(a_4^7) \, ,
\ee
where $a_4^2$ has not been identified as the regulator $\m^2$ yet for reasons which will become clear in the following.
The evaluation of the full bulk action ends with the determination of $S^{5d}$, which is given in \eq{SB51}.
For that purpose the explicit form of the corresponding plaquette, $U_{\m5}$, is exploited and reads
\bea\label{UM51}
U_{\m5}(n_\m,n_5) &=& U(n_\m,n_5;\m) \, U(n_\m+a_4 \hat \m,n_5;5)\, (U(n_\m ,n_5 + a_5 \hat 5;\m))^\dagger \, (U(n_\m,n_5;5))^\dagger \nonumber\\
 &=& e^{i a_4 g_5 {\bf A}_\m(n_\m,n_5)} \,e^{i a_5 g_5 {\bf A}_5(n_\m+a_4 \hat \m,n_5)}\,e^{-i a_4 g_5 {\bf A}_\m(n_\m,n_5 + a_5 \hat 5)}\,e^{-i a_5 g_5 {\bf A}_5(n_\m,n_5)} \, , \nonumber\\
\eea
with the Lie element ${\bf A}_5 \equiv  A_5^A T^A$.
Similarly with the previous case the above relation can be rewritten using \eq{BCH} but now for
\bea
X &=& i a_4 g_5 {\bf A}_\m(n_\m,n_5)\, , \,\, Y = i a_5 g_5 {\bf A}_5(n_\m+a_4 \hat \m,n_5)\nonumber\\
Z &=& -i a_4 g_5 {\bf A}_\m(n_\m,n_5 + a_5 \hat 5)\, , \,\, W = -i a_5 g_5 {\bf A}_5(n_\m,n_5) \, . \nonumber
\eea
Before we perform the lattice-spacing expansion and present the results for the action let us make a short comment.  
To begin note that $U_{\m5}$ is similar with the hybrid plaquette, given in \eq{Uhm51}, which was used in the boundary action.
Nevertheless, this is a delicate point since the information that each of the two plaquettes includes is completely different and so we expect that the corresponding actions will not match.
In particular $U_{\m5}$ includes only bulk fields whose mass dimension is different than the boundary fields, as we have already explained. In addition there is no need to respect any boundary conditions here since the corresponding gauge links live exclusively on the bulk and no dimensional reduction is implemented.
Therefore \eq{UM51} will contain terms which are forbidden on the boundary.
With these in mind and after the expansion of the lattice spacings the desired plaquette becomes
\bea\label{UMN2}
U_{\m5}(n_\m,n_5) &=& \exp \Biggl [   i a_4 a_5 g_5 {\bf F}_{\m5}(n_\m,n_5) - i a_4 a_5 g_5 \hat {\bf D}_5 {\bf A}_{\m}(n_\m,n_5) - a_4 a_5 g_5^2 [{\bf A}_5(n_\m,n_5), {\bf A}_{\m}(n_\m,n_5) ] \nonumber\\
&+& \frac{i a_4^2 a_5 g_5}{2} \hat {\bf D}'_\m {\bf F}_{\m5}(n_\m,n_5) + \frac{ a_4^2 a_5 g_5^2}{2} [\hat \D_\m {\bf A}_\m(n_\m,n_5),  {\bf A}_5(n_\m,n_5) ] \nonumber\\
&-& \frac{i a_4 a_5^2 g_5}{2} \hat {\bf D}_5(\hat {\bf D}_5 {\bf A}_{\m}(n_\m,n_5) ) - \frac{ a_4 a_5^2 g_5^2}{2} [ \hat \D_5 {\bf A}_5(n_\m,n_5) , {\bf A}_\m(n_\m,n_5) ] + {\cal O}(a^4)  \Biggr ]  \, , \nonumber
\eea
where we have collected and massaged terms of the same dimension.
In addition we kept terms up to ${\cal O}(a^3) \equiv {\cal O}(a_4^3, a_5^3)$, as we did for the hybrid action, while we defined that
\be
{\bf F}_{\m5} =  \hat \D_\m {\bf A}_5 + i g_5 [{\bf A}_\m , {\bf A}_5 ] \,\,\, {\rm and} \,\,\, \hat {\bf D}_5 = \hat \D_5  + i g_5 [{\bf A}_5 ,  ] \, . \nonumber
\ee
Finally the last step is to set $U_{\m5}$ back to \eq{SB51} which, after considering the trace, reads
\be\label{SB5f}
S^{5d} =  \sum_{n_\m} a_4^4 \sum_{n_5}  a_5 \sum_{\m}  \Biggl (  \frac{1}{4} F^A_{\m5} F^A_{\m5} + \frac{a_4^2}{16}  ( \hat D_\m F^A_{\m5} ) ( \hat D_\m F^A_{\m5} )  - i a_4 a_5 \frac{g_5 }{6} \tr \{ F^3_{\m5} \}   \Biggr ) + {\cal O}(a^7) \, ,
\ee
where only terms up to dimension 7 have been kept while we neglected as usual those which do not respect the symmetries of the theory.
The calculation of the bulk action is almost done since the only thing left is to add $S^{4d}$ and $S^{5d}$ so as to form $S^B$.
Nevertheless there is a tricky point here which should be clarified. 
To be more specific both \eq{SBmf} and \eq{SB5f} lack of regularization scale in contrast with the boundary case.
In the latter recall that the leftover lattice-spacing for each higher-order term was the only available physical scale, $a_4$, which regularized the theory.
On the other hand $S^B$ is defined on the full 5d orbifold lattice and because of the anisotropy there are two physical scales, $a_4$ and $a_5$, which both participate in the regularization game.
Nevertheless, these two scales depend upon each other through the anisotropy factor $\g$ so there is some freedom left to choose one of them so as to regularize both actions.
Hence if $a_4$ is chosen then both $S^{4d}$ and $S^{5d}$ inherit the scale $\m \equiv \m(a_4)$ while the anisotropy enters only in the latter action. 
With the above in our mind $S^{4d}$ and $S^{5d}$ become
\be
S^{4d} =  \sum_{n_\m} a_4^4 \sum_{n_5}  a_5 \sum_{\m<\n}  \Biggl (  \frac{1}{4} F^A_{\m\n} F^A_{\m\n} + \frac{1}{16} \frac{ ( \hat D_\m F^A_{\m\n} ) ( \hat D_\m F^A_{\m\n} )}{\m^2}  - \frac{ g_5 f_{ABC} }{24} \frac{ F^A_{\m\n} F^B_{\n\r} F^C_{\r\m} }{\m^2}   \Biggr ) \nonumber
\ee
and
\be
S^{5d} =  \sum_{n_\m} a_4^4 \sum_{n_5}  a_5 \sum_{\m}  \Biggl (  \frac{1}{4} F^A_{\m5} F^A_{\m5} + \frac{1}{16} \frac{ ( \hat D_\m F^A_{\m5} ) ( \hat D_\m F^A_{\m5} ) }{\m^2}  -  i \frac{g_5 }{6} \frac{ \tr \{ F^3_{\m5} \} }{\g \m^2}  \Biggr ) \nonumber
\ee
respectively, with $\hat D_\m$ the $SU(2)$ covariant derivative.
Indeed the above reflects that due to the anisotropy the 5d covariance of the theory is broken, already at the classical level, when all the higher-order terms, at the truncation level that we work, are taken into account.
In particular, responsible for the breaking are only the $F^3$ terms since all the rest restore the covariance in accordance with \cite{IrgesFotis2}.
There is no clear evidence why the breaking of the 5d covariance becomes obvious only when the cubic HDO appears while this is not true regarding the quadratic one.
Answering that question is beyond the scope of the current work, however there is a chance to have a glance at the solution through the quantization of the bulk action.
Now there are two possible ways to describe the corresponding quantum effects.
The first one is to consider $S^{5d}$ as an adjoint-scalar action, so then $S^B$ will correspond to a 5d cartoon-version of the Lee-Wick(LW) model\footnote{An extensive analysis of higher-derivative Yang-Mills theories coupled to an adjoint-scalar or spinor matter Lagrangian can be found 
at \cite{Tseytlin, Wise, Grinstein, Schuster, Casarin} and references therein.} which has extensively been studied.
The second one is to take advantage of the lattice periodicity so as to compactify the fifth dimension.
This step allows us to rewrite $S^{5d}$ introducing the Kaluza-Klein states into the spectrum.
However, since the 5d covariance is broken, the most profitable way to describe the system is through the LW approach.
In that sense after recalculating \eq{SB5f} we arrive, regarding the bulk action, at the following result 
\bea\label{SBf}
S^B &=&  \sum_{n_\m} a_4^4 \sum_{n_5}  a_5 \sum_{\m}  \Biggl [ \sum_{\n}  \Bigl ( \frac{1}{4} F^A_{\m\n} F^A_{\m\n} + \frac{1}{16 \m^2} ( \hat D_\m F^A_{\m\n} ) ( \hat D_\m F^A_{\m\n} ) - \frac{ g_5 }{24 \m^2} f_{ABC} F^A_{\m\n} F^B_{\n\r} F^C_{\r\m}  \Bigr )  \nonumber\\
&+& ( \overline{ \hat D_\m \Phi^A}) (\hat D_\m \Phi^A ) + \frac{1}{4 \m^2} ( \overline{ \hat D^2 \Phi^A}) (\hat D^2 \Phi^A )  \Biggr ] 
\eea
where the capital $\Phi^A$ corresponds to a scalar doublet given by
\be
\Phi^A = 
\begin{pmatrix}
\frac{A_5^1 + i A_5^2}{2} 
\\
\frac{A_5^3}{2} 
\\
\end{pmatrix}
\equiv
\begin{pmatrix}
\phi 
\\
\phi^0 
\\
\end{pmatrix}
\ee
while $\phi$ corresponds exactly to the scalar d.o.f which leaves at the boundary.
Moreover notice that the cubic term $F_{\m5}^3$ is absent, after considering the trace, a fact that seems to be characteristic of the Lee-Wick description which is exploited here.
As a last comment we emphasize the fact that \eq{SBf} is not the most general 5d LW theory which can be made.
This becomes obvious looking at the interacting terms which all of them depend on the same coupling $g_5$.
In addition, even though $S^B$ seems like a 4d LW model and $\sum a_5$ seems irrelevant, $A^A_\m$ and $\Phi^A$ are both functions of the anisotropic dimension, $n_5$.
Actually this is characteristic of the 5d orbifold lattice where the fields correspond to different components of the original 5d gauge field $A^A_M$.

\section{Field redefinition and gauge invariance}\label{FRGI}

There are several ways to perform a field redefinition but not all of them lead to an expression which respects 
the symmetries of the theory, especially when gauge symmetry is present.
Here the derivation of the appropriate conditions so as to get a gauge invariant and consistent theory after a field redefinition is at work.
Some useful and relevant comments in this respect are given in \cite{Criado}. Here we present a self contained discussion of the issue with some additional comments.
We focus on the Boundary-Hybrid action, \eq{SBH1} that corresponds to a particular version of 
SQED so we are concerned about the proper way to redefine the Abelian-gauge theory while respecting gauge invariance.
We perform an analysis showing the criteria which the proper field redefinition should fulfill.
We recall that, according to \cite{FotisLetter1,FotisLetter2}, the Jacobean after the field redefinition plays a crucial role since renders a theory with O-ghosts consistent.
This happens with the insertion of the R-ghosts in the spectrum and therefore they should also enter in a gauge-invariant manner.

Let us begin the analysis with the path-integral of \eq{SBH1} in the absence of sources
\be
{\cal Z} [0] = \int  {\cal D} A  {\cal D} \bar\phi {\cal D} \phi \, e^{i  S[A, \bar \phi, \phi]}\, , \hskip 0.5 cm  
S[A, \bar \phi, \phi] \equiv S[A^3_\m, \bar \phi, \phi] = \int d^4 x \, {\cal L}[A^3_\m, \bar \phi, \phi] = S^{\rm b-h} \, , \nonumber
\ee
where $A$ stands for the $U(1)$ gauge field $A^3_\m$.
In the following both notations are used interchangeably.
In the presence of sources the terms $J \, \phi$, $J \, \bar \phi$ and $J \, A^3_\m$ 
should also be properly transformed, otherwise the $S$-matrix can not be kept unchanged.
If we perform a finite gauge transformation to the path-integral through 
\be\label{gatr}
\phi' = V(x) \phi \, ,\,\,\,\bar \phi' = \bar \phi \bar V(x) \,\,\,{\rm and}\,\,\, A' = V(x) ( A + \frac{1}{g_4} \pa ) \bar V(x) \, ,
\ee
with $V(x) = e^{i \a(x)}$ and $\bar V(x) = e^{- i \a(x)} = V^{-1}(x)$, then we get
\bea
{\cal Z}' [0] &=& \int  {\cal D} A'  {\cal D} \bar\phi' {\cal D} \phi' \, e^{i  S'[A', \bar \phi', \phi']} \nonumber\\
 &=& \int  {\cal D} A  {\cal D} \bar\phi {\cal D} \phi\, \left| \frac{d  A'}{d A}\right| \left| \frac{d \bar \phi'}
 {d  \bar \phi}\right| \left| \frac{d  \phi'}{d  \phi}\right| \, e^{i  S[A, \bar \phi, \phi]} \nonumber\\
 &=& {\cal Z} [0] \, .
\eea
The action is by definition gauge invariant while the functional measure inherits the Jacobeans.
For the transformation of \eq{gatr} these are however trivial and the l.h.s matches the original partition function.
The question is how can this argument be combined with our intention to redefine the fields.
To see this, we perform a general field redefinition of the form
\bea\label{fred.2}
\phi &\rightarrow& \hat \phi = \phi + \frac{x}{\L^2} F_\phi(\phi) \nonumber \\
A^3_\m &\rightarrow& \hat A^3_\m = A^3_\m + \frac{x_\a}{\L^2} F_{A,\m}(A^3_\m)
\eea
and after that gauge transform the path-integral so as to see whether it will stay invariant.
Let us forget for a moment about the gauge field and deal only with the complex scalar.
We apply the first of \eq{fred.2} to the path-integral ${\cal Z} [0]$ to get
\bea\label{Z0phi1}
\hat {\cal Z} [0] &=& \int  {\cal D} A  {\cal D} \bar{\hat \phi} {\cal D} \hat \phi \, e^{i  S[A, \bar {\hat \phi}, \hat \phi]} \nonumber\\
 &=& \int  {\cal D} A  {\cal D} \bar\phi {\cal D} \phi\, \left| \frac{d \bar{\hat \phi}}{d  \bar \phi}\right| \left| \frac{d \hat \phi}{d  \phi}\right| \, e^{i  S[A, \bar {\hat \phi}, \hat \phi]} \, ,
 \eea
where now the Jacobeans are not trivial and they must be accounted for.
Since everything is valid up to total derivatives there is a freedom in the choice of the independent field variables after the redefinition.
We can fix this gauge by demanding pole cancellation\footnote{See \cite{FotisLetter2} for more details.} between the O- and R-ghosts.
This is the diagonal gauge so the Jacobeans are diagonal matrices with determinant
\bea
D(\phi) &=& \det \left[ \frac{d \hat \phi}{d  \phi} \right] = \left( 1 + \frac{x}{\L^2} 
\frac{d F_\phi(\phi) }{d  \phi} \right)^\b \simeq 1+ \frac{\b x}{\L^2}  \frac{d F_\phi(\phi) }{d  \phi} \nonumber\\
\overline{ D(  \phi)} &=& \det \left[ \frac{d \bar{ \hat \phi} }{d \bar \phi} \right] = 
\left( 1 + \frac{x}{\L^2}  \overline{ \Bigl [ \frac{d F_\phi( \phi)  }{d \phi} \Bigr ]} \right)^\b \simeq 1+ \frac{\b x}{\L^2}  \overline{ \Bigl [ \frac{d F_\phi( \phi)  }{d \phi} \Bigr ]}
\eea
where only terms up to ${\cal O}(1/\L^2)$ have been kept.
$\b$ refers to the dimension of the Jacobean matrices.
Therefore the product of the determinants in \eq{Z0phi1} would read
\be
\overline{ D(  \phi)} D(\phi) \equiv 1+ \frac{\b x}{\L^2}  \frac{d  F_\phi( \phi)  }{d \phi} + \frac{\b x}{\L^2}  \overline{ \Bigl [ \frac{d F_\phi( \phi)  }{d \phi} \Bigr ]} = D(\bar \phi, \phi) \, ,
\ee
indicating that the R-ghost should be given by the path-integral of the form
\be
D(\bar \phi, \phi) =  \int {\cal D} \bar \chi {\cal D} \chi\, e^{- i \int d^4 x \bar \chi D(\bar \phi, \phi)  \chi} \, . \nonumber
\ee
Combining now \eq{Z0phi1} with the above relation the resulting path-integral reads
\bea\label{Z0phi2}
{\cal Z} [0] &=& \int  {\cal D} A  {\cal D} \bar\phi {\cal D} \phi {\cal D} \bar \chi {\cal D} \chi\, e^{i S[ A, \bar \phi + \frac{x}{\L^2} \overline{F_\phi(\phi)}, \, \phi + \frac{x}{\L^2}  F_\phi(\phi)  ] - i \int d^4 x \bar \chi D(\bar \phi, \phi)  \chi } \nonumber\\
 &=&  \int  {\cal D} A  {\cal D} \bar\phi {\cal D} \phi {\cal D} \bar \chi {\cal D} \chi\, e^{i S[ A, \bar \phi + \frac{x}{\L^2} \overline{F_\phi(\phi)}, \, \phi + \frac{x}{\L^2} F_\phi(\phi)  ] + i S_R(\bar \chi, \chi) } \, ,
\eea
where we have defined the action of the scalar R-ghost through
\be
S_R(\bar \chi, \chi) = - \int d^4 x \bar \chi D(\bar \phi, \phi)  \chi  \, . \nonumber
\ee
There are two more steps that we should take.
The first is to expand the action with respect to $F_\phi$, $\overline F_\phi$ and the second is to gauge transform the result.
However, before we get into this note that for $S_R(\bar \chi, \chi) $ to contribute to the pole cancelation, the R-ghost should be charged under the $U(1)$ group.
Then after a gauge transformation also the ghost action should be invariant.
As a consequence, since $\chi$ transforms as $\chi \to \chi' = V(x) \chi$, the operator $D(\bar \phi, \phi) $ should let the transformation pass through it showing that $D'(\bar \phi, \phi) \chi' = V(x )D(\bar \phi, \phi) \chi $ and more importantly that $F_\phi$ is gauge covariant.
In particular the gauge transformation of the scalar field in \eq{fred.2} gives that
\be
F'_\phi(\phi') = V(x)  F_\phi(\phi) \,\,\,\, {\rm and} \,\,\,\,  \overline{F'_\phi(\phi')} =  \overline{F_\phi(\phi)} V^{-1}(x) \nonumber
\ee
rendering, indeed, $F_\phi$ a covariant operator which is in the same representation with $\phi$.
Let us now return to our algorithm and expand the action in \eq{Z0phi2} to get
\be\label{Z0phi3}
{\cal Z} [0] = \int  {\cal D} A  {\cal D} \bar\phi {\cal D} \phi {\cal D} \bar \chi {\cal D} \chi\, \exp \Bigl [ i \sum^{\infty}_{m,n=0} \frac{x^{m+n}}{\L^{2(m+n)}(m+n)!} \overline{F^m_\phi } F^n_\phi \frac{\d^{(m+n)} S[ A, \bar \phi, \, \phi ]}{ (\d \bar \phi)^m (\d \phi)^n} + i S_R(\bar \chi, \chi) \Bigr ] \, ,
\ee
which then is gauge transformed to
\bea
{\cal Z}' [0] &=& \int  {\cal D} A  {\cal D} \bar\phi {\cal D} \phi {\cal D} \bar \chi {\cal D} \chi\, \exp \Bigl [ i \sum^{\infty}_{m,n=0} \frac{x^{m+n}}{\L^{2(m+n)}(m+n)!} \overline{(F'_\phi)^m } (F'_\phi)^n \frac{\d^{(m+n)} S[ A, \bar \phi, \, \phi ]}{(\d \bar \phi')^m (\d \phi')^n} + i S_R(\bar \chi, \chi) \Bigr ] \nonumber\\
 &=& {\cal Z} [0] \, . \nonumber
\eea
The actions $S[ A, \bar \phi, \, \phi ]$, $S_R(\bar \chi, \chi)$ and the functional integrals are invariant, as it was shown above, so the gauge covariance of $F_\phi$ makes the partition function gauge invariant.
Following the same strategy for the redefinition of the gauge field leads to the same outcome indicating that the path-integral is manifestly gauge invariant under proper field redefinition.
To be more specific we sketch the calculation starting with \eq{Z0phi3} where we use \eq{fred.2} to redefine $A^3_\m$.
In that sense the path-integral yields
\bea\label{Z0Aphi1}
{\cal Z} [0] &=& \int  {\cal D} \hat A  {\cal D} \bar\phi {\cal D} \phi {\cal D} \bar \chi {\cal D} \chi\, \exp \Bigl [ i \sum^{\infty}_{m,n=0} \frac{x^{m+n}}{\L^{2(m+n)}(m+n)!} \overline{F^m_\phi } F^n_\phi \frac{\d^{(m+n)} S[ \hat A^3_\m, \bar \phi, \, \phi ]}{(\d \bar \phi)^m (\d \phi)^n} + i S_R(\bar \chi, \chi) \Bigr ] \nonumber\\
 &=& \int  {\cal D} A \left| \frac{d \hat A}{d A }\right| {\cal D} \bar\phi {\cal D} \phi {\cal D} \bar \chi {\cal D} \chi\, \exp \Bigl [ i \sum^{\infty}_{m,n=0} \frac{x^{m+n}}{\L^{2(m+n)}(m+n)!} \overline{F^m_\phi } F^n_\phi \frac{\d^{(m+n)} S[ A^3_\m + \frac{x_\a}{\L^2} F_{A,\m}, \bar \phi, \, \phi ]}{(\d \bar \phi)^m (\d \phi)^n} \nonumber\\
 &+& i S_R(\bar \chi, \chi) \Bigr ] 
\eea
and the Jacobean, keeping in mind the pole cancelation condition also for the gauge field, is a diagonal matrix with determinant
\be
D^{\m\n}(A^3_\m) = \det \left[  \frac{d \hat A^3_\m}{d A^3_\n } \right] = \left( \d^{\m\n} + \frac{x_\a}{\L^2} \frac{d F_{A,\m}(A^3_\m) }{d  A^3_\n } \right)^{\b'} \equiv \d^{\m\n}+ \frac{\b' x_\a}{\L^2} \frac{d F_{A,\m}(A^3_\m) }{d A^3_\n} \, ,
\ee
where $\b'$ indicates the dimension of the Jacobean matrix.
Again we have kept terms up to ${\cal O}(1/\L^2)$.
The generated R-ghost for $A^3_\m$ is defined as $B\equiv B^3_\m$ and it is inserted in the path-integral by the relation
\be
D^{\m\n}(A^3_\m) =  \int {\cal D} B\, e^{- i \int d^4 x B^3_\m D^{\m\n}(A^3_\m) B^3_\n } \nonumber
\ee
leading to 
\bea\label{Z0Aphi2}
{\cal Z} [0] &=& \int  {\cal D} A {\cal D} \bar\phi {\cal D} \phi {\cal D} \bar \chi {\cal D} \chi {\cal D} B\, \exp \Bigl [ i \sum^{\infty}_{m,n=0} \frac{x^{m+n}}{\L^{2(m+n)}(m+n)!} \overline{F^m_\phi } F^n_\phi \frac{\d^{(m+n)} S[ A^3_\m + \frac{x_\a}{\L^2} F_{A,\m}, \bar \phi, \, \phi ]}{ (\d \bar \phi)^m (\d \phi)^n} \nonumber\\
&+& i S_R(\bar \chi, \chi) + i S_R(B) \Bigr ] \, ,
\eea
with
\be
S_R(B)  = -  \int d^4 x B^3_\m D^{\m\n}(A) B^3_\n \equiv \int d^4 x B_{\m\n} B^{\m\n} \, . \nonumber
\ee
Here $S_R(B)$ as well as $D^{\m\n}(A^3_\m)$ should be invariant under gauge transformations which is also true for $F_{A,\m}(A^3_\m)$ since the gauge field is Abelian. If the gauge group is non-Abelian then the corresponding $F_{A,\m}$ should be a gauge covariant operator in the same representation with the associated gauge field.
Finally we expand around $A^3_\m$ and gauge transform \eq{Z0Aphi2} to get
\bea
{\cal Z}' [0] &=& \int  {\cal D} A {\cal D} \bar\phi {\cal D} \phi {\cal D} \bar \chi {\cal D} \chi {\cal D} B\, \exp \Bigl [ i \sum^{\infty}_{m,n,p=0} \frac{x^{m+n} x_\a^p}{\L^{2(m+n+p)}(m+n+p)!} \overline{F^m_\phi } F^n_\phi  (F'_A)^p \frac{\d^{(m+n+p)} S[ A, \bar \phi, \, \phi ]}{ (\d \bar \phi)^m (\d \phi)^n (\d A')^p} \nonumber\\
&+& i S_R(\bar \chi, \chi) + i S_R(B) \Bigr ] \nonumber\\
 &=& {\cal Z} [0] \nonumber
\eea
which shows that using the proper field redefinitions a gauge invariant and, due to the pole cancelation at work, O-ghost-free path-integral can be constructed yielding
\be
{\cal Z} [0] = \int  {\cal D} A {\cal D} \bar\phi {\cal D} \phi {\cal D} \bar \chi {\cal D} \chi {\cal D} B\, e^{ i S[\hat A, \bar {\hat \phi}, \hat \phi] + i S_R(\bar \chi, \chi) + i S_R(B) } \, .
\ee
Note that specifically for our case an appropriate choice, regarding $F_\phi$ and $F_{A,\m}$, is the following
\be\label{FphiFA}
F_\phi(\phi) =  \frac{x}{\L^2} D^\m D_\m \phi + \frac{y}{\L^2} (\bar \phi \phi) \phi \,\,\,\, {\rm and} \,\,\,\, F_{A,\m}(A^3_\m) = \frac{x_\a}{\L^2} ( \eta_{\m\n} \Box - \pa_\m \pa_\n ) A^{3,\n} \, ,
\ee  
with $D_\m$ the usual covariant derivative and $x$, $y$, $x_\a$ yet undefined parameters.

\section{Feynman Rules for the Boundary-Hybrid action}\label{FRBHA}
In this Appendix the derivation of the Feynman rules for the field-redefined Boundary-Hybrid action, \eq{SBHrd}, is in order.
The procedure that we follow is the usual one keeping in mind that we split the action into its kinetic and interaction part.
Let us start with the former which after the redefinition reads
\be
S^{\rm b-h}_{\rm Kin, 0}  = \int d^4x \Biggl [ - \frac{1}{4} F^3_{\m\n,0} F_0^{3,\m\n} + \frac{1}{2\xi} A^3_{\m,0} \pa^\m \pa_\n A_0^{3,\n} - \bar \phi_0 \Box \phi_0 - \bar c_0^3 \Box c_0^3   \Biggr ] 
\ee
while using the fact that $F^3_{\m\n,0} = \pa_\m A^3_{\m,0} - \pa_\n A^3_{\n,0}$ and performing some massaging becomes
\bea
S^{\rm b-h}_{\rm Kin, 0}  &=& \int d^4x \Biggl [ - \frac{1}{4} \Bigl ( \pa_\m A^3_{\n,0} - \pa_\n A^3_{\m,0}  \Bigr) \Bigl ( \pa^\m A_0^{3,\n}  - \pa^\n A_0^{3,\m}   \Bigr) + \frac{1}{2\xi} A^3_{\m,0} \pa^\m \pa_\n A_0^{3,\n} - \bar \phi_0 \Box \phi_0 - \bar c_0^3 \Box c_0^3   \Biggr ] \nonumber\\
 &=& \int d^4x \int d^4y \d_{xy} \Biggl [ - \frac{1}{4} \Bigl ( \pa_\m A^3_{\n,0}(x) - \pa_\n A^3_{\m,0} (x) \Bigr) \Bigl ( \pa^\m A_0^{3,\n}(y)  - \pa^\n A_0^{3,\m}(y)   \Bigr) + \frac{1}{2\xi} A^3_{\m,0}(x) \pa^\m \pa_\n A_0^{3,\n}(y) \nonumber\\
 &-& \bar \phi_0(x) \Box \phi_0(y) - \bar c_0^3(x) \Box c_0^3(y)   \Biggr ] \nonumber\\
 &=& \int d^4x \int d^4y \d_{xy} \Biggl [  \frac{1}{2} A^3_{\m,0}(x) \Bigl ( \eta^{\m\n} \Box_y + (\frac{1}{\xi} - 1 ) \pa_y^\m \pa_y^\n  \Bigr)A^3_{\n,0} (y) - \bar \phi_0(x) \Box_y \phi_0(y) - \bar c_0^3(x) \Box_y c_0^3(y)   \Biggr ] \, . \nonumber\\
\eea
The next step is to Fourier transform to momentum space using that
\be\label{FT}
A^3_\m (x) = \int \frac{d^4 p}{(2 \pi)^4} A^3_\m (p) e^{-i p \cdot x} \, , \,\,\, \phi_0(x) = \int \frac{d^4 p}{(2 \pi)^4}\phi_0 (p) e^{-i p \cdot x}
\ee
so the above equation returns the following
\bea
S^{\rm b-h}_{\rm Kin, 0}  &=& \int d^4x \int d^4y  \frac{d^4 p}{(2 \pi)^4}  \frac{d^4 q}{(2 \pi)^4}  \d_{xy}  e^{-i p \cdot x}\Biggl [  \frac{1}{2} A^3_{\m,0}(p) \Bigl ( \eta^{\m\n} \Box_y + (\frac{1}{\xi} - 1 ) \pa_y^\m \pa_y^\n  \Bigr) e^{-i q \cdot y} A^3_{\n,0} (q) \nonumber\\
 &-& \bar \phi_0(p) \Box_y e^{-i q \cdot y} \phi_0(q) - \bar c_0^3(p) \Box_y e^{-i q \cdot y} c_0^3(q)   \Biggr ]  \nonumber\\
 &=&  \int \frac{d^4 p}{(2 \pi)^4} \Biggl [  \frac{1}{2} A^3_{\m,0}(p) \Bigl ( - \eta^{\m\n} p^2 + (  1 - \frac{1}{\xi} ) p^\m p^\n  \Bigr) A^3_{\n,0} (-p) + \bar \phi_0(p) p^2  \phi_0(-p) - \bar c_0^3(p) p^2 c_0^3(-p)   \Biggr ]  \nonumber\\
 &=&  \int \frac{d^4 p}{(2 \pi)^4} \Biggl [  \frac{1}{2} A^3_{\m,0}(p) M_A^{\m\n} A^3_{\n,0} (-p) + \bar \phi_0(p) M_\phi  \phi_0(-p) - \bar c_0^3(p) M_c c_0^3(-p)   \Biggr ]  \nonumber\\
\eea
where we have used that $ \int d^4x \int d^4y \d_{xy}  e^{-i p \cdot x}e^{-i q \cdot y} = \int d^4x  e^{-i (p +q) \cdot x} = (2 \pi)^4 \int d^4x \d(p +q)$.
Then we used the $\d$-function to eliminate the $q$-integral setting $q=-p$ and defined the matrices $ M_A^{\m\n} = - \eta^{\m\n} p^2 + (  1 - \frac{1}{\xi} ) p^\m p^\n  $, $M_\phi = p^2$ and $M_c = p^2$.
So inverting the above matrices we end up with the corresponding propagators.
Finally the Feynman rules for the interaction part of \eq{SBHrd} will be in our hand after performing the Fourier transform given in \eq{FT} and calculating the various partial derivatives.
Hence, collectively, the Feynman rules that $S^{b-h}$ generates are:
\begin{itemize}
\item Gauge Propagator
\begin{center}
\begin{tikzpicture}[scale=0.8]
\draw[photon] (0,-0.19)--(2.5,-0.19) ;
\node at (0,0.2) {$\m$};
\node at (2.5,0.2) {$\n$};
\node at (6,-0.15) {$= \hskip .1 cm \displaystyle
 \frac{i}{q^2} \Bigl( - \eta_{\m\n}  +(1-\xi ) \frac{ q_\m q_\n }{q^2} \Bigr) $};
\end{tikzpicture}
\end{center}
\item Scalar Propagator
\begin{center}
\begin{tikzpicture}[scale=0.8]
\draw[dashed] (0,-0.19)--(2.5,-0.19) ;
\draw [->, very thick]  (1.25,-0.19)--(1.3,-0.19);
\node at (4,-0.2) {$= \hskip .1 cm \displaystyle
 \frac{i}{p^2} $};
\end{tikzpicture}
\end{center}
\item Scalar-Scalar-gauge vertex
\begin{center}
\begin{tikzpicture}[scale=0.7]
\draw [dashed] (-2.5,1.5)--(-1,0);
\draw [dashed] (-2.5,-1.5)--(-1,0);
\draw [<-] (-1.5,0.8)--(-2.2,1.5);
\node at (-2.7,1.1) {$p_2$};
\draw [<-] (-1.5,-0.8)--(-2.2,-1.5);
\draw [->, very thick] (-1.9,-0.9)--(-1.7,-0.7);
\draw [<-, very thick] (-1.9,0.9)--(-1.7,0.7);
\node at (-2.5,-1) {$p_1$};
\draw [<-] (-0.5,0.3)--(0.3,0.3);
\node at (0,-0.5) {$q$};
\node at (1.6,0) {$\m$};
\draw[photon] (-1,0)--(1,0) ;
\node at (5,0) {$=  \displaystyle  i g_{4,0} Q_{\m}(p,q)  \, . $};
\end{tikzpicture}
\end{center}
\item Four-point self interaction vertex
\begin{center}
\begin{tikzpicture}[scale=0.7]
\draw [dashed] (0,0)--(1.5,1.4);
\draw [dashed] (0,0)--(1.5,-1.4);
\draw [dashed] (-1.5,1.4)--(0,0);
\draw [dashed] (-1.5,-1.4)--(0,0);
\draw [<-]  (0.7,0.3)--(1.3,0.9);
\node at (0.9,1.5) {$p_3$};
\draw [->, very thick]  (0.7,0.7)--(0.8,0.8);
\draw [<-, very thick]  (0.7,-0.7)--(0.8,-0.8);
\draw [->, very thick] (-0.9,-0.9)--(-0.7,-0.7);
\draw [<-, very thick] (-0.7,0.7)--(-0.5,0.5);
\draw [<-]  (0.9,-0.4)--(1.5,-1);
\node at (1.1,-1.6) {$p_4$};
\draw [<-] (-0.9,0.3)--(-1.6,1);
\node at (-1,1.5) {$p_2$};
\draw [<-] (-0.8,-0.4)--(-1.5,-1);
\node at (-0.8,-1.5) {$p_1$};
\node at (4.6,0) {$= \hskip .1 cm \displaystyle  i c_{1,0}^{(6)} \frac{p^2} {\L^2}  \, .$};
\end{tikzpicture}
\end{center}
\item Scalar-Scalar-gauge-gauge vertex
\begin{center}
\begin{tikzpicture}[scale=0.7]
\draw [photon] (0,0)--(1.5,1.4);
\draw [photon] (0,0)--(1.5,-1.4);
\draw [dashed] (-1.5,1.4)--(0,0);
\draw [dashed] (-1.5,-1.4)--(0,0);
\draw [<-]  (0.7,0.3)--(1.3,0.9);
\node at (0.8,1.4) {$q_1$};
\node at (2,1.6) {$\m$};
\draw [<-]  (0.9,-0.4)--(1.5,-1);
\draw [->, very thick] (-0.9,-0.9)--(-0.7,-0.7);
\draw [<-, very thick] (-0.7,0.7)--(-0.5,0.5);
\node at (1.2,-1.7) {$q_2$};
\node at (2.2,-1.6) {$\n$};
\draw [<-] (-0.9,0.3)--(-1.6,1);
\node at (-1,1.4) {$p_2$};
\draw [<-] (-0.8,-0.4)--(-1.5,-1);
\node at (-0.8,-1.4) {$p_1$};
\node at (5,0) {$= \,\,\,\, \displaystyle 2 i  g^2_{4,0}  Q_{\m\n}(p,q) \, .$};
\end{tikzpicture}
\end{center}
\item Four-scalars one-photon vertex
\begin{center}
\begin{tikzpicture}[scale=0.7]
\draw [dashed] (0,0)--(1.5,1.4);
\draw [dashed] (0,0)--(1.5,-1.4);
\draw [dashed] (-1.5,1.4)--(0,0);
\draw [dashed] (-1.5,-1.4)--(0,0);
\draw [photon] (0.02,1.4)--(0.02,0);
\draw [<-]  (0.7,0.3)--(1.3,0.9);
\node at (0.9,1.5) {$p_3$};
\draw [->, very thick]  (0.7,0.7)--(0.8,0.8);
\draw [<-, very thick]  (0.7,-0.7)--(0.8,-0.8);
\draw [->, very thick] (-0.9,-0.9)--(-0.7,-0.7);
\draw [<-, very thick] (-0.7,0.7)--(-0.5,0.5);
\draw [<-]  (0.9,-0.4)--(1.5,-1);
\node at (1.1,-1.6) {$p_4$};
\draw [<-] (-0.9,0.3)--(-1.6,1);
\node at (-1,1.5) {$p_2$};
\draw [<-] (-0.8,-0.4)--(-1.5,-1);
\node at (-0.8,-1.5) {$p_1$};
\node at (0,1.75) {$q, \m$};
\node at (4.6,0) {$= \hskip .1 cm \displaystyle i g_{4,0} \, c_{1,0}^{(6)} \frac{  (p_1 + p_2)_\m }{\L^2}  \, .$};
\end{tikzpicture}
\end{center}
\item Four-scalars two-photons vertex
\begin{center}
\begin{tikzpicture}[scale=0.7]
\draw [dashed] (0,0)--(1.5,1.4);
\draw [dashed] (0,0)--(1.5,-1.4);
\draw [dashed] (-1.5,1.4)--(0,0);
\draw [dashed] (-1.5,-1.4)--(0,0);
\draw [photon] (0.02,-1.4)--(0.02,0);
\draw [photon] (0.02,1.4)--(0.02,0);
\draw [<-]  (0.7,0.3)--(1.3,0.9);
\node at (0.9,1.5) {$p_3$};
\draw [<-]  (0.9,-0.4)--(1.5,-1);
\node at (1.1,-1.6) {$p_4$};
\draw [<-] (-0.9,0.3)--(-1.6,1);
\node at (-1,1.5) {$p_2$};
\draw [->, very thick]  (0.7,0.7)--(0.8,0.8);
\draw [<-, very thick]  (0.7,-0.7)--(0.8,-0.8);
\draw [->, very thick] (-0.9,-0.9)--(-0.7,-0.7);
\draw [<-, very thick] (-0.7,0.7)--(-0.5,0.5);
\draw [<-] (-0.8,-0.4)--(-1.5,-1);
\node at (-0.9,-1.5) {$p_1$};
\draw [<-] (-0.2,0.5)--(-0.2,1.2);
\node at (0,1.75) {$q_1$};
\draw [<-] (0.25,-0.5)--(0.25,-1.2);
\node at (0,-1.75) {$q_2$};
\node at (4.2,0) {$= \hskip .1 cm \displaystyle  2 i  g^2_{4,0} \frac{ c_{1,0}^{(6)} }{\L^2} \eta_{\m\n} \, ,$};
\end{tikzpicture}
\end{center}
\end{itemize}
where we have used the $p$'s and $q$'s for the scalar-field and the gauge-field momenta respectively and in particular $p_1$ corresponds to $\phi$ while $p_2$ to $\bar \phi$.
Moreover we have define the functions
\bea\label{QmQmn}
Q_{\m}(p,q) &=& ( \eta_{\m\n} + \frac{ \eta_{\m\n} q^2- q_\m q_\n }{\L^2}  ) (p_1 + p_2)^\n \nonumber\\
Q_{\m\n}(p,q) &=&  \eta_{\m\n} + \frac{ \eta_{\m\n} p_1 \cdot p_2 + q_{1,\m} ( p_{2,\n} - p_{1,\n} )  } {2 \L^2}  +  \frac{2 ( \eta_{\m\n} q_1^2- q_{1,\m} q_{1,\n} )} { \L^2}  
\eea
while notice that the propagator of the Faddeev-Popov ghosts is absent here since they are completely decoupled from the theory.
A final comment is that the diagram which includes four scalar-fields and one photon-field will not contribute at the 1-loop renormalization of the theory.

\section{Review of the Coleman-Weinberg model}\label{CWm}
This part of the Appendix works as a short review of the main characteristics of the Coleman-Weinberg model \cite{Coleman}.
The way that it is constructed follows the calculation conducted at \sect{cCWm} so as to elucidate the differences between the physical results of the Boundary-Hybrid model, \eq{SBHrd}, and that of the CW model.
The analysis here is interested in the part of \cite{Coleman} which refers to the mSQED whose effective potential and scalar-to-gauge mass ratio have been analytically evaluated.
   
Let us start with the fact that all the calculations along with the regularization of the CW model are performed exclusively at $d=4$ dimensions.
On the other hand, recall that our strategy here is to keep the dimensions arbitrary and specify only when this is necessary.

Regarding the renormalization program notice that in \cite{Coleman}, even though the model of interest is massless, a counter-term for both the mass term and the scalar-field is in use.
According to the authors this is a legal step since there is no symmetry preventing the production of a bare mass in the limit of vanishing renormalized mass.
Actually, as they mention, there is scale symmetry at work but they do not pay much attention to that since, in general, it could be anomalous.
Therefore they brake it already at the classical level using their mass counter-term.
Recall that this was not the path that we followed here where we introduced a counterterm for the scalar field but not for the mass term.
The reason for that is hidden behind our intention to show how the breaking of scale invariance, inserting the HDO at quantum level, is connected with the spontaneous breaking of the internal symmetry.

Now the loop calculation of the CW model is done, at 1-loop level, through the effective action which led to the determination of the effective potential.
In particular the renormalization of the model using the above arguments gives the effective potential
\be\label{VCW1}
V_{CW} =  \frac{\l}{4!} \phi_c^4 + \Bigl ( \frac{5 \l^2}{1152 \pi^2} + \frac{3 e^4}{64 \pi^2} \Bigr ) \phi_c^4 \Bigl( \ln\frac{\phi_c^2}{M^2} - \frac{25}{6} \Bigr) \, ,
\ee
where $\phi_c^2 \equiv \bar \phi \phi$ while $\l$ and $e$ correspond to the quartic and gauge coupling respectively. 
According to \cite{Coleman} this effective potential has a local minimum away from the origin which is made clear by supposing that $\l$ and $e^4$ share the same order of magnitude.
This claim, connecting the two couplings, seems a little bit suspicious at the CW analysis while it follows as a natural implementation regarding the Boundary-Hybrid model developed here.
Notice however that the authors show the validity of the SSB through radiative corrections for arbitrary, but still small, couplings.

Then, the minimization condition of \eq{VCW1}, neglecting the terms proportional to $\l^2$, indicates that
\be\label{le4}
\l = \frac{33}{8\pi^2} e^4
\ee 
and therefore the effective potential now becomes
\be\label{VCW2}
V_{CW} =  \frac{3 e^4}{64 \pi^2} \, \phi_c^4 \Bigl( \ln\frac{\phi_c^2}{\langle \phi \rangle^2} - \frac{1}{2} \Bigr) \, ,
\ee
with $\langle \phi \rangle$ the vacuum expectation value of $\phi_c$.
Keep in mind that during the above procedure the number of the independent parameters stays fixed.
In particular the explicit form of $V_{CW}$ depends on two parameters, $\l$ and $e$, which is also true for the potential of \eq{VCW2} but now the parameters are $e$ and $\langle \phi \rangle$.
This was defined by the authors as dimensional transmutation.

Since SSB does occur the authors sift the field $\phi_c$ around its vev which inherits the scalar field with the mass
\be\label{SmCW}
m^2(S) = \frac{3}{8\pi^2} e^4 \langle \phi \rangle^2
\ee
and the gauge field with the mass
\be\label{VmCW}
m^2(V) = e^2 \langle \phi \rangle^2
\ee
at leading loop order.
Then the scalar-to-gauge mass ration is constructed and given by
\bea\label{mSmV}
\frac{m^2(S)}{m^2(V)} \equiv \r^2_{CW} &=& \frac{3}{8\pi^2} e^2  \Rightarrow \nonumber\\
\r_{CW} &=& \sqrt{ \frac{3}{8\pi^2}} e \, .
\eea
Finally, notice that in \cite{Coleman} the authors act with the Callan-Symanzik equation on the fourth derivative of the effective potential \eq{VCW1} and on the wave function renormalization.
Then a system of equations is constructed through which the scalar-field anomalous dimension and the $\b$-functions of $e$ and $\l$ are evaluated and give
\bea
\g &=& \frac{3 e^2}{16\pi^2} \label{gCW} \\
\b_e &=& \frac{e^3}{48\pi^2} \,\,\,\, {\rm or} \,\,\,\, \b_{\a_e} = \frac{2 \a_e^2}{3} \label{beCW} \\
\b_\l &=& \frac{\frac{10}{3} \l^2 - 12 \l e^2 + 36 e^4}{16\pi^2} \label{blCW}
\eea
respectively.
Recall that the calculations are performed in $d=4$, so the above results represent the loop-part of the corresponding $\b$-functions.

\end{appendices}



\end{document}